# Critical Exponents, Hyperscaling and Universal Amplitude Ratios for Two- and Three-Dimensional Self-Avoiding Walks

Bin Li*
*Debt and Equity Markets Group*
*Merrill Lynch*
*World Financial Center*
*New York, NY 10281-1315 USA*
Internet: `BLI@ML.COM`

Neal Madras
*Department of Mathematics and Statistics*
*York University*
*4700 Keele Street*
*North York, Ontario M3J 1P3*
*CANADA*
Internet: `MADRAS@NEXUS.YORKU.CA`

Alan D. Sokal
*Department of Physics*
*New York University*
*4 Washington Place*
*New York, NY 10003 USA*
Internet: `SOKAL@NYU.EDU`

September 6, 1994

---

*Formerly with the Department of Physics, New York University.

## Abstract

We make a high-precision Monte Carlo study of two- and three-dimensional self-avoiding walks (SAWs) of length up to 80000 steps, using the pivot algorithm and the Karp-Luby algorithm. We study the critical exponents $\nu$ and $2\Delta_4 - \gamma$ as well as several universal amplitude ratios; in particular, we make an extremely sensitive test of the hyperscaling relation $d\nu = 2\Delta_4 - \gamma$. In two dimensions, we confirm the predicted exponent $\nu = 3/4$ and the hyperscaling relation; we estimate the universal ratios $\langle R_g^2 \rangle / \langle R_e^2 \rangle = 0.14026 \pm 0.00007$, $\langle R_m^2 \rangle / \langle R_e^2 \rangle = 0.43961 \pm 0.00034$ and $\Psi^* = 0.66296 \pm 0.00043$ (68% confidence limits). In three dimensions, we estimate $\nu = 0.5877 \pm 0.0006$ with a correction-to-scaling exponent $\Delta_1 = 0.56 \pm 0.03$ (subjective 68% confidence limits). This value for $\nu$ agrees excellently with the field-theoretic renormalization-group prediction, but there is some discrepancy for $\Delta_1$. Earlier Monte Carlo estimates of $\nu$, which were $\approx 0.592$, are now seen to be biased by corrections to scaling. We estimate the universal ratios $\langle R_g^2 \rangle / \langle R_e^2 \rangle = 0.1599 \pm 0.0002$ and $\Psi^* = 0.2471 \pm 0.0003$; since $\Psi^* > 0$, hyperscaling holds. The approach to $\Psi^*$ is from above, contrary to the prediction of the two-parameter renormalization-group theory. We critically reexamine this theory, and explain where the error lies.

# Contents

# 1 Introduction

The self-avoiding walk (SAW) is a well-known lattice model of a polymer molecule in a good solvent [1]. Its equivalence to the $n = 0$ limit of the $n$-vector model [1,2,3,4,5,6,7] has also made it an important test-case in the theory of critical phenomena.

In this paper we report the results of an extensive Monte Carlo study of two- and three-dimensional SAWs of length up to 80000 steps, using the pivot algorithm [8,9,10].[1] We make a high-precision determination of the critical exponents $\nu$ and $2\Delta_4 - \gamma$ as well as several universal amplitude ratios. In particular, we make an extremely sensitive test of the hyperscaling relation $d\nu = 2\Delta_4 - \gamma$, which plays a central role in the general theory of critical phenomena (Section 1.1).

Our results have also led us to reexamine critically the conventional theory of polymer molecules, the so-called "two-parameter renormalization-group theory" [11,12]. Indeed, such a reexamination is unavoidable, as our Monte Carlo data are *inconsistent* with this theory as it has been heretofore applied. But this is because — as we explain in Section 1.2 — the theory has heretofore been applied *incorrectly*!

---

[1]These computations were carried out over a 4-year period on a variety of RISC workstations. The total CPU time was several years, but we have by now lost track of exactly how many!

These points were first made three years ago by Nickel [13], in an important but apparently under-appreciated paper; they have recently been extended by one of us [14,15].

## 1.1 The Problem of Hyperscaling

One of the key unsolved problems in the theory of critical phenomena has been the status of the so-called *hyperscaling* relations (scaling laws in which the spatial dimension $d$ appears explicitly). These relations have long been known to rest on a much more tenuous physical basis than the other scaling laws [16,17,18,19,20,21,22]. Indeed, it has been understood since the early 1970's that hyperscaling should *not* hold for systems above their upper critical dimension $d_u$: for $d > d_u$ the critical exponents are expected to be those of mean-field theory, and these exponents satisfy the hyperscaling relations only *at* $d = d_u$.[2] (For models in an $n$-vector universality class, including the SAW, $d_u$ equals 4.) It has generally been believed that hyperscaling *should* hold in dimensions $d < d_u$, but in our opinion there is no particularly compelling justification for such a belief (although the claim itself is probably correct). Hyperscaling in the three-dimensional Ising model is the subject of a controversy that has been raging for 25 years, and which is still not completely settled.[3] We remark that hyperscaling is also of interest in quantum field theory, where it is equivalent to the nontriviality of the continuum limit for a strongly-coupled $\varphi^4$ field theory [48,49,50,37,51,52,7].

Although the hyperscaling relations appear naively to be ineluctable consequences of the renormalization-group approach to critical phenomena, closer examination reveals a mechanism by which hyperscaling can fail: the so-called "dangerous irrelevant variables".[4] But the much more difficult question of whether this violation actually *occurs* in a given model can be resolved only by detailed calculation. Unfortunately, a direct analytical test of hyperscaling appears to be possible only at or in the immediate neighborhood of a Gaussian fixed point, that is, for asymptotically free theories [59,60,61,62] or for small $\epsilon = d_u - d$ or large $n$ [63,64,65,66]. We note that the real-space RG [67,68] and field-theoretic RG [69] frameworks, as typically used in approximate calculations, implicitly *assume* the hyperscaling relations, so they cannot be used to *test* hyperscaling.

It is therefore of interest to make an unbiased numerical test of hyperscaling, working directly from first principles. One approach is series extrapolation [70,71], which affords a direct test of universality and scaling laws, including hyperscaling.

---

[2] This belief has now been confirmed by rigorous proofs of the failure of hyperscaling for the Ising model in dimension $d > 4$ [7], the self-avoiding walk in dimension $d \geq 5$ [1,23,24], and spread-out percolation in dimension $d > 6$ [25,26].

[3] See [17,27,28,29,30,31,32,33,34,35,36,37,38,39,40,41,42] for series-extrapolation work, and [43,44,45,46,47] for Monte Carlo work.

[4] This mechanism was proposed independently by Fisher [53] and Wegner and Riedel [54, p. 250 and footnote 8] in the early 1970's. For further discussion, see also Ma [55, Section VII.4], Amit and Peliti [56], Fisher [57, Appendix D] and van Enter, Fernández and Sokal [58, Section 5.2].

It gives numerical results of apparently very high accuracy: the claimed (subjective) error bars on critical exponents are on the order of ±0.001–0.005 [71], which is comparable to the best alternative calculational schemes. However, as is inherent in any extrapolation method, the results obtained depend critically on the assumptions made about the singularity structure of the exact function, notably the nature of the confluent singularity, if any [33,38,41,72,73,71]. Indeed, estimates by different methods from the same series sometimes differ among themselves by several times their claimed error bars. This, together with systematic differences between lattices of the same dimension, accounts for much of the controversy over hyperscaling. Quite a few extra terms would be needed to resolve these discrepancies in a convincing manner [33]. Unfortunately, the computer time required to evaluate the series coefficients grows exponentially with the number of terms desired, while the extrapolation error is proportional to some inverse power of the number of terms (the power depends on the details of the correction-to-scaling terms and the extrapolation method).

In a Monte Carlo study, by contrast, one aims to probe directly the regime where the correlation length $\xi$ is $\gg 1$. (For SAWs this corresponds to a chain length $N \gg 1$.) The method affords a direct test of universality and scaling laws, including hyperscaling. In practice, however, it has been extremely difficult to obtain good data in the neighborhood of the critical point. There are two essential difficulties: *finite system size* [74,75,76] and *critical slowing-down* [77,78,79]. For spin models and lattice field theories, these two factors together imply that the CPU time needed to obtain one "effectively independent" sample grows as $\sim L^d \xi^z \gtrsim \xi^{d+z}$, where $d$ is the spatial dimension of the system and $z$ is the dynamic critical exponent of the Monte Carlo algorithm.[5] (This situation may be alleviated somewhat by a new finite-size-scaling technique [80] that yields accurate estimates of infinite-volume quantities from Monte Carlo data on lattice sizes $L \ll \xi$.) The situation for SAWs is rather more favorable: one can simulate a SAW directly in infinite space, with no finite-size corrections or $L^d$ factor in the CPU time. There is, to be sure, critical slowing-down; but vast progress has been made over the last decade or so in inventing new and more efficient algorithms for simulating the SAW [81]. In particular, using the pivot algorithm [8,9,10] one can generate an "effectively independent" $N$-step SAW (at least as regards global observables) in a CPU time of order $\sim N$ [10,81]. (This is the best possible order of magnitude, since it takes a time of order $N$ merely to *write down* an $N$-step walk!) Since $\xi \sim N^\nu$, this corresponds to a CPU time $\sim \xi^{1/\nu}$ in a spin system — which (if $d \geq 2$) is better than $\sim \xi^{d+z}$ even if $z = 0$! So the SAW is a uniquely favorable "laboratory" for studying the problem of hyperscaling.

---

[5]Conventional local algorithms have $z \approx 2$, while the new collective-mode algorithms [77,78,79] can have $z \ll 2$ and in some cases even $z = 0$.

## 1.2 Which Quantities are Universal?

Over the past four decades, various mathematical models have been employed to describe the behavior of linear polymer molecules in a good solvent.[6] Among these models are the self-avoiding walk [1], the bead-rod model [82], and the continuum Edwards model [83,84,85,86,87,11,12]. The detailed behavior depends on the specific model chosen, just as the detailed behavior of real polymer molecules depends on the particular chemical structure of the polymer and solvent (and on the temperature). However, it has long been understood that *some* aspects of polymer behavior become *universal* in the long-chain limit $N \to \infty$ (where $N$ is the number of monomers in the chain). Unfortunately, there has been considerable confusion about which quantities are universal and which are not. In this subsection we summarize recent work of Nickel [13] and one of us [14,15] which clarifies this issue. (For further discussion, see Section 5.2 below.)

Standard renormalization-group (RG) arguments predict [13] that the mean-square end-to-end distance $\langle R_e^2 \rangle$, the mean-square radius of gyration $\langle R_g^2 \rangle$ and the second virial coefficient $A_2^{(mol)} \equiv (N^2 M_{monomer}^2 / N_{Avogadro}) A_2$ of any real or model polymer chain should have the asymptotic behavior

$$\langle R_e^2 \rangle = A_{R_e} N^{2\nu} (1 + b_{R_e}^{(1)} N^{-\Delta_1} + \ldots) \qquad (1.1)$$

$$\langle R_g^2 \rangle = A_{R_g} N^{2\nu} (1 + b_{R_g}^{(1)} N^{-\Delta_1} + \ldots) \qquad (1.2)$$

$$A_2^{(mol)} = A_A N^{d\nu} (1 + b_A^{(1)} N^{-\Delta_1} + \ldots) \qquad (1.3)$$

as $N \to \infty$, where $d$ is the spatial dimension.[7] The critical exponents $\nu$ and $\Delta_1$ are universal. The amplitudes $A_{R_e}, A_{R_g}, A_A, b_{R_e}^{(1)}, b_{R_g}^{(1)}, b_A^{(1)}$ are nonuniversal; in fact, even the *signs* of the correction-to-scaling amplitudes $b_{R_e}^{(1)}, b_{R_g}^{(1)}, b_A^{(1)}$ [and their various combinations such as $b_\Psi^{(1)} \equiv b_A^{(1)} - (d/2) b_{R_g}^{(1)}$] are nonuniversal. However, the RG theory also predicts that the dimensionless amplitude *ratios* $A_{R_g}/A_{R_e}$, $A_A/A_{R_e}^{d/2}$, $b_{R_g}^{(1)}/b_{R_e}^{(1)}$ and $b_A^{(1)}/b_{R_e}^{(1)}$ are universal [13,88].

So there is no reason why the correction-to-scaling amplitudes should have any particular sign. In the continuum Edwards model, the effective exponents $\nu_{eff,R_e} \equiv \frac{1}{2} d \log \langle R_e^2 \rangle / d \log N$ and $\nu_{eff,R_g} \equiv \frac{1}{2} d \log \langle R_g^2 \rangle / d \log N$ and the interpenetration ratio $\Psi \equiv 2(d/12\pi)^{d/2} A_2^{(mol)} / \langle R_g^2 \rangle^{d/2}$ all approach their asymptotic values *from below* [11,12,89,90,91]: that is, $b_{R_e}^{(1)}, b_{R_g}^{(1)} > 0$ and $b_\Psi^{(1)} < 0$. On the other hand, high-precision Monte Carlo data on lattice self-avoiding walks — see Section 4 below, as well as Ref. [13] — show clearly that these quantities approach their asymptotic values *from above*; and the same occurs in the bead-rod model with sufficiently large bead diameter [92]. Indeed, this behavior is almost obvious qualitatively: short self-avoiding walks behave roughly like hard spheres; only at larger $N$ does one see the softer excluded volume (smaller $\Psi$) characteristic of a fractal object. In any case, all these models are in excellent agreement for the leading *universal* quantities $\nu$, $A_{R_g}/A_{R_e}$ and $\Psi^* \equiv 2(d/12\pi)^{d/2} A_A/A_{R_g}^{d/2}$, and they are in rough agreement for the universal correction-to-scaling quantities $\Delta_1$, $b_{R_g}^{(1)}/b_{R_e}^{(1)}$ and $b_A^{(1)}/b_{R_e}^{(1)}$.

---

[6] Here "good solvent" means that we work at any fixed temperature strictly above the theta temperature for the given polymer-solvent pair.

[7] In (1.3) we have assumed for simplicity that the hyperscaling relation $d\nu = 2\Delta_4 - \gamma$ is valid.

It is thus misguided to analyze the experimental data in the good-solvent regime by attempting to match the real polymer molecules to the continuum Edwards model via the correspondence $z_{Edwards} = aN^{1/2}$ (where $a$ is an empirically determined scale factor depending on the polymer, solvent and temperature)[8]: the continuum Edwards model can predict *only* the universal quantities. Indeed, there is evidence [93,94,95,96] that real polymers in a sufficiently good solvent behave like self-avoiding walks, i.e. they approach $\Psi^*$ from above; in this case they *cannot* be matched to *any* value of $z_{Edwards}$. This behavior has heretofore been considered paradoxical; in fact, it is quite natural. (Huber and Stockmayer [94] attributed this behavior to the effects of chain stiffness. In fact, as pointed out by Nickel [13], chain stiffness is quite irrelevant here, as the effect occurs also for *perfectly flexible* chains, such as self-avoiding walks or the bead-rod model.)

These points have been made previously by Nickel [13]. Similar comments have been made with regard to liquid-gas critical points by Liu and Fisher [97].

In summary, the error of all two-parameter theories is to fail to distinguish correctly which quantities are universal and which are non-universal. In particular, the modern two-parameter theory begins from one special model — the continuum Edwards model — and assumes (incorrectly) that it can describe certain aspects of polymer behavior (e.g. the sign of approach to $\Psi^*$) which in reality are non-universal.

**Remark.** A very different limiting behavior is obtained (in dimension $d < 4$) if we take simultaneously $N \to \infty$ and $T \to T_\theta$ such that $x \equiv N^\phi (T - T_\theta)$ remains fixed, for a suitable crossover exponent $\phi$. In a separate work [14,15], one of us has argued that it is precisely this *universal crossover scaling behavior* in an infinitesimal region just above the theta temperature that is described by the continuum Edwards model.

## 1.3 Plan of this Paper

The plan of this paper is as follows: In Section 2 we review the needed background information about the self-avoiding walk and the pivot algorithm. In Section 3 we analyze several algorithms for computing the second virial coefficient; this section can be skipped by readers whose main interest is in the results. In Section 4 we present and analyze our Monte Carlo data for self-avoiding walks in two and three dimensions. In Section 5 we compare our results with previous work, discuss further the interpretation of the sign of approach to $\Psi^*$, and discuss prospects for the future. In Appendix A we prove some geometric bounds for subsets of $\mathbb{Z}^d$; as a corollary, we prove hyperscaling for SAWs in dimension $d = 2$. In Appendix B we discuss the problem of ensuring adequate thermalization for the pivot algorithm.

---

[8]See, for example, the comparisons between theory and experiment in [11, Section 10.F] and [12, Section 15.3].

In Appendix C we discuss some subtleties involved in the statistical analysis of our data. In Appendix D we make a few remarks on the field-theoretic calculations of universal amplitude ratios.

## 2 Background and Notation

### 2.1 The Self-Avoiding Walk (SAW): A Review

In this section we review briefly the basic facts and conjectures about the SAW that will be used in the remainder of the paper. Let $\mathcal{L}$ be some regular $d$-dimensional lattice. Then an $N$-*step self-avoiding walk* (SAW) $\omega$ on $\mathcal{L}$ is a sequence of *distinct* points $\omega_0, \omega_1, ..., \omega_N$ in $\mathcal{L}$ such that each point is a nearest neighbor of its predecessor. For simplicity we shall restrict attention to the simple (hyper-)cubic lattice $\mathbb{Z}^d$; similar ideas apply with minor alterations to other regular lattices. We assume all walks to begin at the origin ($\omega_0 = 0$) unless stated otherwise, and we let $\mathcal{S}_N$ be the set of all $N$-step SAWs starting at the origin (and ending anywhere).

First we define the quantities relating to the *number* (or "entropy") of SAWs: Let $c_N$ [resp. $c_N(x)$] be the number of $N$-step SAWs on $\mathbb{Z}^d$ starting at the origin and ending anywhere [resp. ending at $x$]. Then $c_N$ and $c_N(x)$ are believed to have the asymptotic behavior

$$c_N \sim \mu^N N^{\gamma-1} \tag{2.1}$$

$$c_N(x) \sim \mu^N N^{\alpha_{sing}-2} \qquad (x \text{ fixed} \neq 0) \tag{2.2}$$

as $N \to \infty$; here $\mu$ is called the connective constant of the lattice, and $\gamma$ and $\alpha_{sing}$ are *critical exponents*. The critical exponents are believed to be universal among lattices of a given dimension $d$. For rigorous results concerning the asymptotic behavior of $c_N$ and $c_N(x)$, see [1,23,24,98].

Next we define several measures of the *size* of an $N$-step SAW[9]:

- The *squared end-to-end distance*

$$R_e^2 = \omega_N^2 . \tag{2.3}$$

- The *squared radius of gyration*

$$R_g^2 = \frac{1}{N+1} \sum_{i=0}^{N} \left( \omega_i - \frac{1}{N+1} \sum_{j=0}^{N} \omega_j \right)^2 \tag{2.4a}$$

$$= \frac{1}{N+1} \sum_{i=0}^{N} \omega_i^2 - \left( \frac{1}{N+1} \sum_{i=0}^{N} \omega_i \right)^2 \tag{2.4b}$$

$$= \frac{1}{2(N+1)^2} \sum_{i,j=0}^{N} (\omega_i - \omega_j)^2 . \tag{2.4c}$$

---

[9]Some other measures of the size of a SAW will be defined in Appendix A.

- The *mean-square distance of a monomer from the endpoints*

$$R_m^2 = \frac{1}{2(N+1)} \sum_{i=0}^{N} \left[ \omega_i^2 + (\omega_i - \omega_N)^2 \right] . \qquad (2.5)$$

We then consider the mean values $\langle R_e^2 \rangle_N$, $\langle R_g^2 \rangle_N$ and $\langle R_m^2 \rangle_N$ in the probability distribution which gives equal weight to each $N$-step SAW. Very little has been proven rigorously about these mean values, but they are believed to have the asymptotic behavior

$$\langle R_e^2 \rangle_N, \ \langle R_g^2 \rangle_N, \ \langle R_m^2 \rangle_N \ \sim \ N^{2\nu} \qquad (2.6)$$

as $N \to \infty$, where $\nu$ is another (universal) critical exponent. Moreover, the amplitude ratios

$$A_N = \frac{\langle R_g^2 \rangle_N}{\langle R_e^2 \rangle_N} \qquad (2.7)$$

$$B_N = \frac{\langle R_m^2 \rangle_N}{\langle R_e^2 \rangle_N} \qquad (2.8)$$

are expected to approach universal values in the limit $N \to \infty$.[10] Indeed, the full probability distribution of $\omega_N$ is expected to scale as

$$\frac{c_N(x)}{c_N} \ \sim \ N^{-d\nu} f(x/N^\nu) \qquad (2.9)$$

as $N \to \infty$, for a suitable *scaling function f* (also universal modulo a single nonuniversal scale factor); and $f$ is expected to be rotation-invariant.[11] All these beliefs can be subsumed in the even more general assertion that the probability distribution of the SAW, with lengths rescaled by $N^\nu$, converges weakly as $N \to \infty$ to some well-defined probability measure on a space of continuum chains.[12]

Finally, let $c_{N_1,N_2}$ be the number of pairs $(\omega^{(1)}, \omega^{(2)})$ such that $\omega^{(1)}$ is an $N_1$-step SAW starting at the origin, $\omega^{(2)}$ is an $N_2$-step SAW starting *anywhere*, and $\omega^{(1)}$ and $\omega^{(2)}$ have at least one point in common (i.e. $\omega^{(1)} \cap \omega^{(2)} \neq \emptyset$). Equivalently, we can write $c_{N_1,N_2}$ in terms of walks that both start at the origin:

$$c_{N_1,N_2} = \sum_{\omega^{(1)} \in \mathcal{S}_{N_1}} \sum_{\omega^{(2)} \in \mathcal{S}_{N_2}} T(\omega^{(1)}, \omega^{(2)}) , \qquad (2.10)$$

---

[10] Sometimes the notation $\aleph = 6 \langle R_g^2 \rangle / \langle R_e^2 \rangle$ is used instead.

[11] Actually, (2.9) is claimed to hold only for $|x|$ of order $N^\nu$. The precise statement of (2.9) is therefore that the limit

$$f(y) \equiv \lim_{N \to \infty} N^{d\nu} \frac{c_N(N^\nu y)}{c_N}$$

exists for each $y \neq 0$, and that $0 < f(y) < \infty$.

[12] Very recently, Hara and Slade [23,24] have proven that the SAW in dimension $d \geq 5$ converges weakly to Brownian motion when $N \to \infty$ with lengths rescaled by $CN^{1/2}$ for a suitable (nonuniversal) constant $C$. It follows from this that (2.6) holds with $\nu = \frac{1}{2}$, and also that (2.7)/(2.8) have the limiting values $A_\infty = \frac{1}{6}$, $B_\infty = \frac{1}{2}$. Earlier, Slade [99,100,101] had proven these results for sufficiently high dimension $d$. See also [1].

where
$$T(\omega^{(1)},\omega^{(2)}) \equiv \#\{x \in \mathbb{Z}^d \colon \omega^{(1)} \cap (\omega^{(2)} + x) \neq \varnothing\} \qquad (2.11)$$
is the number of translates of $\omega^{(2)}$ that somewhere intersect $\omega^{(1)}$. It is believed that
$$c_{N_1,N_2} \sim \mu^{N_1+N_2} (N_1 N_2)^{(2\Delta_4+\gamma-2)/2} g(N_1/N_2) \qquad (2.12)$$
as $N_1, N_2 \to \infty$, where $\Delta_4$ is yet another (universal) critical exponent and $g$ is a (universal) scaling function.

The quantity $c_{N_1,N_2}$ is closely related to the second virial coefficient. To see this, consider a rather general theory in which "molecules" of various types interact. Let the molecules of type $i$ have a set $S_i$ of "internal states", so that the complete state of such a molecule is given by a pair $(x,s)$ where $x \in \mathbb{Z}^d$ is its position and $s \in S_i$ is its internal state. Let us assign Boltzmann weights (or "fugacities") $W_i(s)$ [$s \in S_i$] to the internal states, normalized so that $\sum_{s \in S_i} W_i(s) = 1$; and let us assign an interaction energy $\mathcal{V}_{ij}\big((x,s),(x',s')\big)$ [$x, x' \in \mathbb{Z}^d$, $s \in S_i$, $s' \in S_j$] to a molecule of type $i$ at $(x,s)$ interacting with one of type $j$ at $(x',s')$. Then the second virial coefficient between a molecule of type $i$ and one of type $j$ is [102]
$$B_2^{(ij)} = \frac{1}{2} \sum_{\substack{s \in S_i \\ s' \in S_j}} \sum_{x' \in \mathbb{Z}^d} W_i(s) W_j(s') \left[1 - e^{-\mathcal{V}_{ij}((0,s),(x',s'))}\right]. \qquad (2.13)$$

In the SAW case, the types are the different lengths $N$, the internal states are the conformations $\omega \in \mathcal{S}_N$ starting at the origin, the Boltzmann weights are $W_N(\omega) = 1/c_N$ for each $\omega \in \mathcal{S}_N$, and the interaction energies are hard-core repulsions
$$\mathcal{V}_{NN'}\big((x,\omega),(x',\omega')\big) = \begin{cases} +\infty & \text{if } (\omega + x) \cap (\omega' + x') \neq \varnothing \\ 0 & \text{otherwise} \end{cases} \qquad (2.14)$$
It follows immediately that
$$B_2^{(N_1,N_2)} = \frac{c_{N_1,N_2}}{2 c_{N_1} c_{N_2}}. \qquad (2.15)$$

The second virial coefficient $B_2^{(N_1,N_2)}$ is a measure of the "excluded volume" between a pair of SAWs. It is useful to define a *dimensionless* quantity by normalizing $B_2^{(N_1,N_2)}$ by some measure of the "size" of these SAWs. Theorists prefer $\langle R_e^2 \rangle$ as the measure of size, while experimentalists prefer $\langle R_g^2 \rangle$ since it can be measured by light scattering. We thus define the *theorists' interpenetration ratio*
$$\Psi_{R,N} \equiv 2(d/2\pi)^{d/2} \frac{B_2^{(N,N)}}{\langle R_e^2 \rangle_N^{d/2}} = (d/2\pi)^{d/2} \frac{c_{N,N}}{c_N^2 \langle R_e^2 \rangle_N^{d/2}} \qquad (2.16)$$
and the (usual) *interpenetration ratio*
$$\Psi_N \equiv 2(d/12\pi)^{d/2} \frac{B_2^{(N,N)}}{\langle R_g^2 \rangle_N^{d/2}} = (d/12\pi)^{d/2} \frac{c_{N,N}}{c_N^2 \langle R_g^2 \rangle_N^{d/2}} \qquad (2.17)$$

(for simplicity we consider only $N_1 = N_2 = N$). The numerical prefactors are a convention that arose historically for reasons not worth explaining here. Crudely speaking, $\Psi$ measures the degree of "hardness" of a SAW in its interactions with other SAWs. A useful standard of comparison is the hard sphere of radius $r$ and constant density:

$$B_2 = \frac{\pi^{d/2}}{d\,\Gamma(d/2)} (2r)^d \qquad (2.18)$$

$$R_g^2 = \frac{d}{d+2} r^2 \qquad (2.19)$$

and hence

$$\Psi_{hard-sphere} = \frac{2}{d\,\Gamma(d/2)} \left(\frac{d+2}{3}\right)^{d/2} = \begin{cases} \approx 1.12838 & \text{in } d=1 \\ 4/3 & \text{in } d=2 \\ \approx 1.61859 & \text{in } d=3 \\ 2 & \text{in } d=4 \end{cases} \qquad (2.20)$$

Inserting (2.1), (2.6) and (2.12) into (2.17), we see that

$$\Psi_N \sim N^{2\Delta_4 - \gamma - d\nu} \qquad (2.21)$$

as $N \to \infty$. We can therefore distinguish three *a priori* possibilities:

(a) $2\Delta_4 - \gamma - d\nu > 0$, so that $\Psi_N \to \infty$ as $N \to \infty$. This behavior cannot occur unless typical SAWs have a very strange (porcupine-like) shape, which is quite implausible.[13]

(b) $2\Delta_4 - \gamma - d\nu = 0$. In the simplest case this means that $\Psi_N \to \Psi^* > 0$ as $N \to \infty$, i.e. typical SAWs exclude each other "within a constant factor like hard spheres". This behavior is called *hyperscaling*. However, the relation $2\Delta_4 - \gamma - d\nu = 0$ is also consistent with a logarithmic violation of hyperscaling, i.e. $\Psi_N \sim (\log N)^{-p} \to 0$ as $N \to \infty$ for some power $p > 0$.

(c) $2\Delta_4 - \gamma - d\nu < 0$, so that $\Psi_N \to 0$ as $N \to \infty$. This is a power-law violation of hyperscaling; typical SAWs exclude each other infinitely more weakly than hard spheres.

A very beautiful heuristic argument concerning hyperscaling for SAWs was given by des Cloizeaux [103]. Note first from (2.17) that $\Psi$ measures, roughly speaking, the probability of intersection of two independent SAWs that start a distance of order $\langle R_g^2 \rangle^{1/2} \sim N^\nu$ apart. Now, by (2.6), we can interpret a long SAW as an object with "fractal dimension" $1/\nu$. Two independent such objects will "generically" intersect if and only if the sum of their fractal dimensions is at least as large as the

---

[13]Modulo some reasonable assumptions, this behavior can in fact be *rigorously* excluded: see Theorem A.1 and equations (A.20)–(A.22) in Appendix A.

dimension of the ambient space. So we expect $\Psi^* \equiv \lim_{N \to \infty} \Psi_N$ to be nonzero if and only if
$$\frac{1}{\nu} + \frac{1}{\nu} \geq d , \quad \text{i.e.} \quad d\nu \leq 2 . \tag{2.22}$$
Since it is believed that
$$\nu = \begin{cases} \frac{1}{2} & \text{for } d > 4 \\ \frac{1}{2} \times \log^{1/8} & \text{for } d = 4 \text{ [69]} \\ \frac{1}{2} + \frac{\epsilon}{16} + \frac{15\epsilon^2}{512} + \ldots & \text{for } d = 4 - \epsilon \text{ [69]} \\ \approx 0.588 & \text{for } d = 3 \text{ [104,105,106,90,91,107, this paper]} \\ \frac{3}{4} & \text{for } d = 2 \text{ [108,109]} \end{cases} \tag{2.23}$$
we see that
$$\begin{aligned} d\nu &< 2 & \text{for } d < 4 \\ d\nu &= \text{``2 + logs''} & \text{for } d = 4 \\ d\nu &> 2 & \text{for } d > 4 \end{aligned}$$
Therefore we expect

- hyperscaling for $d < 4$
- logarithmic violation of hyperscaling for $d = 4$
- power-law violation of hyperscaling for $d > 4$

One half of this heuristic argument can be proven rigorously. It is easy to see [110] that
$$c_{N_1,N_2} \leq (N_1 + 1)(N_2 + 1) c_{N_1} c_{N_2} , \tag{2.24}$$
so that
$$B_2^{(N_1,N_2)} \leq \frac{1}{2}(N_1 + 1)(N_2 + 1) , \tag{2.25}$$
or in terms of critical exponents,
$$2\Delta_4 - \gamma \leq 2 . \tag{2.26}$$
It follows that $d\nu > 2$ implies $2\Delta_4 - \gamma - d\nu < 0$, i.e. the power-law violation of hyperscaling. This is now proven rigorously to occur for $d \geq 5$ [1,23,24].

In the polymer-physics literature it is usually *taken for granted* that hyperscaling holds in dimension $d = 3$. But in our opinion hyperscaling is a deep property that *needs to be tested*.

We remark that dimension $d = 2$ is a different case: here hyperscaling can be *proven rigorously* (modulo some reasonable assumptions on the scaling of individual SAWs). We present this proof in Appendix A; it is the analogue for SAWs of Aizenman's proof [111, Section 8] of hyperscaling for two-dimensional Ising models with finite-range ferromagnetic interaction. The underlying geometric idea is that SAWs in the plane cannot avoid intersecting each other.

Finally, we need to make some comments about corrections to scaling. Clearly, (2.1)/(2.2)/(2.6)/(2.12) are only the leading term in a large-$N$ asymptotic expansion. According to renormalization-group theory [112], the mean value of any global observable $\mathcal{O}$ behaves as $N \to \infty$ as

$$\langle \mathcal{O} \rangle_N = A N^p \left[ 1 + \frac{a_1}{N} + \frac{a_2}{N^2} + \ldots + \frac{b_0}{N^{\Delta_1}} + \frac{b_1}{N^{\Delta_1+1}} + \frac{b_2}{N^{\Delta_1+2}} + \ldots \right.$$
$$\left. + \frac{c_0}{N^{\Delta_2}} + \frac{c_1}{N^{\Delta_2+1}} + \frac{c_2}{N^{\Delta_2+2}} + \ldots \right]. \quad (2.27)$$

Thus, in addition to "analytic" corrections to scaling of the form $a_k/N^k$, there are "non-analytic" corrections to scaling of the form $b_k/N^{\Delta_1+k}$, $c_k/N^{\Delta_2+k}$ and so forth, as well as more complicated terms [not shown in (2.27)] which have the general form const/$N^{k_1\Delta_1+k_2\Delta_2+\cdots+l}$ where $k_1, k_2, \ldots$ and $l$ are non-negative integers. The leading exponent $p$ and the correction-to-scaling exponents $\Delta_1 < \Delta_2 < \ldots$ are universal; $p$ of course depends on the observable in question, but the $\Delta_i$ do not. [Please note that the exponents $\Delta_1 < \Delta_2 < \ldots$ have no relation whatsoever to the gap exponent $\Delta_4$ defined in (2.12). The notation used here is standard but unfortunate.] The various amplitudes (both leading and subleading) are all nonuniversal. However, *ratios* of the corresponding amplitudes $A$, $b_0$ and $c_0$ (but not $a_k$ or the higher $b_k, c_k$) for different observables are universal [88,13].

**Remark.** The names of the critical exponents $\gamma$, $\alpha_{sing}$, $\nu$ and $\Delta_4$ are chosen by analogy with the corresponding exponents in ferromagnetic spin systems [17,113]. Indeed, the generating functions of self-avoiding walks,

$$\chi(\beta) \equiv \sum_{N=0}^{\infty} \beta^N c_N \quad (2.28)$$

$$G(x;\beta) \equiv \sum_{N=0}^{\infty} \beta^N c_N(x) \quad (2.29)$$

$$\bar{u}_4(\beta) \equiv -3 \sum_{N_1,N_2=0}^{\infty} \beta^{N_1+N_2} c_{N_1,N_2} \quad (2.30)$$

are *equal* to the susceptibility, spin-spin correlation function and fourth cumulant in the $n$-vector model analytically continued to $n = 0$ [1,4,5,6,7]. In particular, if $x$ is a nearest neighbor of the origin, then $G(x;\beta)$ is essentially the energy $E$ (up to an additive and multiplicative constant). The quantity

$$\xi(\beta) \equiv \left( \frac{\sum_x |x|^2 G(x;\beta)}{\sum_x G(x;\beta)} \right)^{1/2} \quad (2.31)$$

is the second-moment correlation length. Inserting (2.1)/(2.2)/(2.6)/(2.12) into (2.28)–(2.31), we obtain the leading behavior

$$\chi(\beta) \sim (\beta_c - \beta)^{-\gamma} \quad (2.32)$$

$$G(x;\beta) \sim (\beta_c - \beta)^{1-\alpha_{sing}} + \text{regular terms} \qquad (2.33)$$
$$\xi(\beta) \sim (\beta_c - \beta)^{-\nu} \qquad (2.34)$$
$$\bar{u}_4(\beta) \sim (\beta_c - \beta)^{-\gamma - 2\Delta_4} \qquad (2.35)$$

as $\beta$ approaches the *critical point* $\beta_c \equiv 1/\mu$. Note, in particular, that $\alpha_{sing}$ is the exponent for the *singular part* of the specific heat $C_H \sim \partial E/\partial \beta$; the exponent for the full specific heat is $\alpha = \max(\alpha_{sing}, 0)$. If the hyperscaling relation $d\nu = 2\Delta_4 - \gamma$ holds (without multiplicative logarithmic corrections), then the renormalized coupling constant $g \equiv -\bar{u}_4/\chi^2\xi^d$ tends to a nonzero limiting value $g^*$ as $\beta \to \beta_c$; so hyperscaling (without multiplicative logarithmic corrections) can be interpreted as the non-Gaussianness (nontriviality) of the scaling-limit quantum field theory [7, p. 281].

## 2.2 The Pivot Algorithm: A Review

The *pivot algorithm* was invented in 1969 by Lal [8], reinvented in 1985 by MacDonald *et al.* [9], and again reinvented a short time later by Madras [10]. The pivot algorithm is the most efficient algorithm currently known for simulating SAWs in the fixed-$N$, variable-$x$ ensemble. Here we summarize briefly the relevant features of the algorithm; more details can be found in [10,81].

The elementary move of the pivot algorithm is as follows: Choose at random a pivot point $k$ along the walk ($0 \le k \le N-1$); choose at random an element $g$ of the symmetry group of the lattice (rotation or reflection or a combination thereof); then apply $g$ to the part of the walk subsequent to the pivot point (namely $\omega_{k+1}, \ldots, \omega_N$), using $\omega_k$ as the temporary "origin". That is, the proposed new walk $\omega'$ is

$$\omega'_i = \begin{cases} \omega_i & \text{for } 0 \le i \le k \\ \omega_k + g(\omega_i - \omega_k) & \text{for } k+1 \le i \le N \end{cases} \qquad (2.36)$$

The walk $\omega'$ is accepted if it is self-avoiding; otherwise, it is rejected, and the old walk $\omega$ is counted once more in the sample. It is easy to see that this algorithm satisfies detailed balance for the standard equal-weight SAW distribution. Ergodicity is less obvious, but it can be proven [10,114].

At first thought this seems to be a terrible algorithm: for $N$ large, nearly all the proposed moves will get rejected. In fact, this latter statement is true, but the hasty conclusion drawn from it is radically false! The acceptance fraction $f$ does indeed go to zero as $N \to \infty$, roughly like $N^{-p}$; empirically, it is found that the exponent $p$ is $\approx 0.19$ in $d = 2$ [10] and $\approx 0.11$ in $d = 3$ [10,115,116]. But this means that roughly once every $N^p$ moves one gets an acceptance. And the pivot moves are very radical: one might surmise that after very few accepted moves (say, 5 or 10) the SAW will have reached an "essentially new" configuration. One conjectures, therefore, that the autocorrelation time $\tau$ of the pivot algorithm behaves as $\sim N^p$. Things are in fact somewhat more subtle (see the next paragraph), but roughly speaking (and modulo a possible logarithm) this conjecture appears to be true. On the other hand, a careful analysis of the computational complexity of the pivot algorithm (see also below) shows that one accepted move can be produced in a computer time of order $N$. Combining these two facts, we conclude that one "effectively independent" sample can be produced in a computer time of order $N$ (or perhaps $N \log N$).

Let's look more closely: Suppose we know that the acceptance fraction $f$ in the pivot algorithm behaves as $f \sim N^{-p}$ as $N \to \infty$. Then, as argued above, after a few successful pivots — i.e. a time of order $1/f \sim N^p$ — the *global* conformation of the walk should have reached an "essentially new" state. Thus, we expect that for observables $A$ which measure the *global* properties of the walk — such as $R_e^2$, $R_g^2$ or $R_m^2$ — the autocorrelation time $\tau_{int,A}$ [see (C.4)] should be a few times $1/f$. This is confirmed numerically [10, Section 4.3]. On the other hand, it is important to recognize that *local* observables — such as the angle between the $17^{th}$ and $18^{th}$ steps of the walk — may evolve a factor of $N$ more slowly than global observables. For example, the observable mentioned in the preceding sentence changes only when $\omega_{17}$ serves as a successful pivot point; and this happens, on average, only once every $N/f$ attempted moves. Thus, for *local* observables $A$ we expect $\tau_{int,A}$ to be of order $N/f$. General properties of reversible Markov chains [77] then imply that the exponential autocorrelation time $\tau_{exp}$ must be of at least this order; and if we have not overlooked any slow modes in the system, then $\tau_{exp}$ should be of exactly this order. Finally, even the global observables are unlikely to be precisely orthogonal to the slowest mode; so it is reasonable to expect that $\tau_{exp,A}$ be of order $N/f$ for these observables too. In other words, for global observables $A$ we expect the autocorrelation function $\rho_{AA}(t)$ to have an extremely-slowly-decaying tail which, however, contributes little to the area under the curve. This behavior is illustrated by the exact solution of the pivot dynamics for the case of ordinary random walk [10, Section 3.3], and by numerical calculations for the SAW (see Appendix C).

*Computational complexity.* A very important issue in any algorithm — but especially in a non-local one — is the CPU time per iteration. By using a hash table [117,118], the self-avoidance of a proposed new walk can be checked in a time of order $N$. But one can do even better: by starting the checking at the pivot point and working outwards, *failures* can be detected in a mean time of order $N^{1-p}$ [10, Sections 3.4 and 4.4]. The mean CPU time *per successful pivot* is therefore $\sim N^{1-p}$ for each of $\sim N^p$ failures, plus $\sim N$ for one success, or $\sim N$ in all. Combining this with the observations made previously, we conclude that one "effectively independent" sample — as regards *global* observables — can be produced in a computer time of order $N$.

*Initialization.* There are two main approaches:

1) *Equilibrium start.* Generate the initial configuration by dimerization [119,1,81]; then the Markov chain is in equilibrium from the beginning, and no data need be discarded. This approach is feasible (and recommended) at least up to $N$ of order a few thousand. There is no harm in spending even days of CPU time on this step, provided that this time is small compared to the rest of the run; after all, the algorithm need only be initialized once.

2) *"Thermalization".* Start in an arbitrary initial configuration, and then discard the first $n_{disc} \gg \tau_{exp} \sim N/f$ iterations. This is painful, because $\tau_{exp}$ is a factor $\sim N$ larger than $\tau_{int,A}$ for global observables $A$; thus, for very large $N$ ($\gtrsim 10^5$), the CPU time of the algorithm could end up being dominated by the thermalization. Nevertheless, one must resist the temptation to cut corners here, as even a small initialization bias can lead to systematically erroneous results, especially if the statistical error is small: see Appendix B for striking evidence of this. Some modest gain can probably be obtained by using closer-to-equilibrium initial configurations (e.g. [120]), but it is still prudent to take $n_{disc}$ at least several times $N/f$.

Initialization will become a more important issue in the future, as faster computers permit simulations at ever-larger chain lengths.

# 3 Algorithms for Counting Overlaps

In this section we discuss algorithms for computing the excluded volume between a given pair of SAWs; this is the key step in a Monte Carlo study of the second virial coefficient. This section can be skipped by readers whose main interest is in the results rather than the algorithms.

## 3.1 Generalities

Let $\omega^{(1)}$ and $\omega^{(2)}$ be, respectively, $N_1$-step and $N_2$-step SAWs, and define $T(\omega^{(1)}, \omega^{(2)})$ to be the number of translates of $\omega^{(2)}$ which somewhere intersect $\omega^{(1)}$:

$$T(\omega^{(1)}, \omega^{(2)}) = \#\{x \in \mathbb{Z}^d \colon \omega^{(1)} \cap (\omega^{(2)} + x) \neq \varnothing\} \quad (3.1a)$$
$$= \#(\omega^{(1)} - \omega^{(2)}), \quad (3.1b)$$

where $A - B \equiv \{y - z \colon y \in A, z \in B\}$. The expected value of $T(\omega^{(1)}, \omega^{(2)})$, averaging over independent walks $\omega^{(1)} \in \mathcal{S}_{N_1}$ and $\omega^{(2)} \in \mathcal{S}_{N_2}$, is $c_{N_1,N_2}/c_{N_1}c_{N_2}$. This quantity has the asymptotic behavior

$$c_{N_1,N_2}/c_{N_1}c_{N_2} \sim (N_1 N_2)^{(2\Delta_4 - \gamma)/2} g(N_1/N_2), \quad (3.2)$$

where $g$ is a scaling function [cf. (2.1)/(2.12)]. It is thus possible to estimate the critical exponent $2\Delta_4 - \gamma$ by running two independent pivot algorithms and measuring $T(\omega^{(1)}, \omega^{(2)})$. [Typically one would run at $N_1 = N_2 = N$ for a sequence of values of $N$.] In particular, this allows a direct Monte Carlo test of the hyperscaling relation $d\nu = 2\Delta_4 - \gamma$. Note that an independent measurement of $\gamma$ is *not* needed.

The efficient determination of $T(A_1, A_2) \equiv \#(A_1 - A_2)$ for a specified pair of sets $A_1, A_2 \subset \mathbb{Z}^d$ is a very interesting and nontrivial problem in computer science. We see two broad approaches:

1) Deterministic algorithms which compute $T(A_1, A_2)$ exactly.

2) Monte Carlo algorithms which produce an unbiased (or almost unbiased) estimate of $T(A_1, A_2)$.

In the latter case, the statistical fluctuations in the auxiliary (inner-loop) Monte Carlo process would be added to those in the main Monte Carlo program; but this is acceptable provided that the former are not too large compared to the latter (see Section 3.4).

We shall discuss the deterministic algorithms for computing $T(A_1, A_2)$ in Section 3.3, and the Monte Carlo algorithms in Sections 3.4–3.9.

## 3.2 Notation

We shall denote by $\mathcal{N}_1$ (resp. $\mathcal{N}_2$) the number of points in the set $A_1$ (resp. $A_2$). We also write $\mathcal{N}_{min} = \min(\mathcal{N}_1, \mathcal{N}_2)$ and $\mathcal{N}_{max} = \max(\mathcal{N}_1, \mathcal{N}_2)$. An $N$-step self-avoiding walk has $\mathcal{N} = N + 1$ points.

Now fix a pair of sets $A_1, A_2 \subset \mathbb{Z}^d$, and write

$$S \equiv A_1 - A_2 \equiv \{y - z \colon y \in A_1,\, z \in A_2\}. \tag{3.3}$$

Our goal is to compute $T(A_1, A_2) \equiv \#(S)$. An important role is played by the function

$$\rho(x) = \#\{(y,z) \colon y \in A_1,\, z \in A_2,\, y - z = x\} \tag{3.4a}$$
$$= (\chi_{A_1} * \check{\chi}_{A_2})(x), \tag{3.4b}$$

where $\chi_{A_1}$ and $\chi_{A_2}$ are the indicator functions of the sets $A_1$ and $A_2$, respectively, and $\check{\chi}_{A_2}(z) \equiv \chi_{A_2}(-z)$, and $*$ denotes convolution. Clearly

$$\rho(x) = 0 \quad \text{for } x \notin S \tag{3.5a}$$
$$1 \leq \rho(x) \leq \mathcal{N}_{min} \quad \text{for } x \in S \tag{3.5b}$$

We also write

$$I(x) = \left\{ \begin{array}{ll} 1 & \text{if } \rho(x) > 0 \\ 0 & \text{if } \rho(x) = 0 \end{array} \right\} = \left\{ \begin{array}{ll} 1 & \text{if } x \in S \\ 0 & \text{if } x \notin S \end{array} \right\}. \tag{3.6}$$

Note that

$$\sum_{x \in \mathbb{Z}^d} I(x) = \sum_{x \in S} 1 = T(A_1, A_2) \tag{3.7}$$

$$\sum_{x \in \mathbb{Z}^d} \rho(x) = \sum_{x \in S} \rho(x) = \mathcal{N}_1 \mathcal{N}_2 \tag{3.8}$$

For future reference we define also

$$U(A_1, A_2) \equiv \sum_{x \in S} \frac{1}{\rho(x)}. \tag{3.9}$$

This observable has little *intrinsic* interest, but it will play an important role in two of the Monte Carlo algorithms. It is not hard to see that[14]

$$\mathcal{N}_1 + \mathcal{N}_2 - 1 \leq T(A_1, A_2) \leq \mathcal{N}_1 \mathcal{N}_2 \tag{3.10}$$

$$\frac{T(A_1, A_2)^2}{\mathcal{N}_1 \mathcal{N}_2} \leq U(A_1, A_2) \leq T(A_1, A_2) \tag{3.11}$$

Moreover, we shall prove in Appendix A (Theorem A.3) that

$$U(A_1, A_2) \geq \log(\mathcal{N}_1 \mathcal{N}_2) . \tag{3.12}$$

We shall use $\langle \cdot \rangle$ to denote expectation with respect to some probability distribution on pairs of sets $A_1, A_2$ (e.g. the equal-weight ensemble on the space $\mathcal{S}_{N_1} \times \mathcal{S}_{N_2}$ of pairs of SAWs). We shall use $E(\cdot)$ to denote expectation with respect to some "inner-loop" Monte Carlo algorithm.

## 3.3 Deterministic Algorithms

Here we introduce some deterministic algorithms for computing $\rho(x)$, $I(x)$, $T(A_1, A_2)$ and/or $U(A_1, A_2)$. These algorithms can be employed either "stand-alone" or as building blocks for the Monte Carlo algorithms to be introduced later.

1) Suppose first that we want to compute $\rho(x)$ for a *single* value of $x$. This can be done as follows: write all the points of $A_1$ into a hash table [118,117]; then examine sequentially each of the points $z \in A_2$, inquiring whether $y \equiv x + z$ belongs to $A_1$, and incrementing a counter if it does. Clearly this requires a CPU time of order $\mathcal{N}_1 + \mathcal{N}_2$.

For computing $I(x)$, the algorithm can be streamlined by stopping as soon as one finds $\rho(x) > 0$.

2) Suppose next that we want to compute $\rho(x)$ for *several* different values of $x$, say $r$ of them. Then the foregoing algorithm requires a CPU time of order $\mathcal{N}_1 + r\mathcal{N}_2$. But since the answer is invariant under the interchange $A_1 \leftrightarrow A_2$, $x \to -x$, it now pays to choose $A_2$ to be the *smaller* of the two sets. As a result, the CPU time is of order $\mathcal{N}_{max} + r\mathcal{N}_{min}$.

3) Suppose, finally, that we want to compute $\rho(x)$ or $I(x)$ for *all* $x$. [A special case of this is to compute $T(A_1, A_2)$ or $U(A_1, A_2)$.] This can be done by examining sequentially each of the *pairs* $y \in A_1$, $z \in A_2$, and writing the points $x \equiv y - z$ into a hash table equipped with an auxiliary count field. Clearly this requires a CPU time of order $\mathcal{N}_1 \mathcal{N}_2$. [The CPU time for subsequently *utilizing* the counts $\{\rho(x)\}$ is of order $T(A_1, A_2) \leq \mathcal{N}_1 \mathcal{N}_2$.] If all one wants is $\{I(x)\}$, then the count field can be dispensed with; it suffices to know which sites $x$ are hit *at least once*.

---

[14]The two upper bounds are trivial. The lower bound in (3.11) is the Schwarz inequality. The lower bound in (3.10) is proven as follows: Let $y_*$ (resp. $z_*$) be the lexicographically smallest element of $A_1$ (resp. $A_2$). Then $y_* - A_2 \preceq y_* - z_* \preceq A_1 - z_*$; so $y_* - A_2$ and $A_1 - z_*$ are subsets of $S$ (of cardinalities $\mathcal{N}_2$ and $\mathcal{N}_1$, respectively) with only one point in common (namely, $y_* - z_*$). Q.E.D.

4) An alternative algorithm for computing $\rho(x)$ for all $x$ can be based on the Fast Fourier Transform (FFT). Let $\alpha_j$ ($1 \leq j \leq d$) be the extension of $A_1$ in the $j^{th}$ coordinate direction (i.e. the difference between the maximum and minimum values of $y_j$ for $y \in A_1$). Let $\beta_j$ be the corresponding extension for $A_2$. Now let $k_j$ be the least integer such that $2^{k_j} \geq \alpha_j + \beta_j + 1$. Then if we place $A_1$ and $A_2$ in a periodic box $B$ of size $2^{k_1} \times 2^{k_2} \times \cdots \times 2^{k_d}$, we can compute the convolution of $\chi_{A_1}$ and $\check{\chi}_{A_2}$ in the box $B$, without distortion by the periodic boundary conditions; this gives the full set of counts $\{\rho(x)\}$ [and thus also $\{I(x)\}$, $T(A_1, A_2)$ and $U(A_1, A_2)$]. The convolution can be carried out by the FFT in a time $\sim V \log V$, where

$$V = 2^{k_1 + \ldots + k_d} \leq \prod_{j=1}^{d} 2(\alpha_j + \beta_j) \,. \qquad (3.13)$$

Let us now estimate the performance of these methods as "stand-alone" algorithms for computing $T(\omega^{(1)}, \omega^{(2)})$, where $\omega^{(1)}$ and $\omega^{(2)}$ are self-avoiding walks. Algorithm #3 computes $T(\omega^{(1)}, \omega^{(2)})$ in a CPU time of order $\mathcal{N}_1 \mathcal{N}_2 = (N_1 + 1)(N_2 + 1)$, i.e. of order $N^2$ if $N_1 \sim N_2 \sim N$. By contrast, we expect that one "effectively independent" sample of the pair $(\omega^{(1)}, \omega^{(2)})$ can be produced by the pivot algorithm in a CPU time of order $N_1 + N_2$ (if $\tau_{int,T} \sim N^p$) or in any case not much greater. So this algorithm would spend more time analyzing the data than producing it! — and the overall computational complexity per "effectively independent" sample would be increased from $N$ to $N^2$, thereby nullifying the advantage of the pivot algorithm over previous [121,122] algorithms.

Algorithm #4 computes $T(\omega^{(1)}, \omega^{(2)})$ in a CPU time of order $V \log V$, where $V$ is given by (3.13). For "typical" SAWs we have $V \sim (N_1 + N_2)^{d\nu}$, so presumably we have

$$\langle V \log V \rangle \sim (N_1 + N_2)^{d\nu} \log(N_1 + N_2) \,. \qquad (3.14)$$

In the usual situation $N_1 \sim N_2 \sim N$, this method is asymptotically better than algorithm #3 if $d\nu < 2$, i.e. if $d < 4$. However, it is unlikely to be better in practice except for *very* large $N$. And the behavior is still vastly worse than the time of order $N$ for generating the walks (except in $d = 1$).

It may be possible to devise deterministic algorithms which are more efficient than either of these elementary ones; we leave this as an exercise for interested computer scientists.

## 3.4 Monte Carlo Algorithms: Generalities

Fix a pair of sets $A_1, A_2$, and suppose that we use some Monte Carlo algorithm to provide an unbiased estimate $Z$ of $T(A_1, A_2)$. We will thus have

$$E(Z|A_1, A_2) = T(A_1, A_2) \qquad (3.15\text{a})$$

$$\text{var}(Z|A_1, A_2) \equiv E(Z^2|A_1, A_2) - E(Z|A_1, A_2)^2 \equiv V(A_1, A_2) \qquad (3.15\text{b})$$

Here (3.15a) expresses the unbiasedness of the inner-loop Monte Carlo algorithm, while (3.15b) defines its (conditional) variance. We shall compute the functional $V(A_1, A_2)$ for each of the Monte Carlo algorithms we introduce (Sections 3.5–3.7).

Now let us call the Monte Carlo subroutine $R$ times for the given pair $(A_1, A_2)$ and average the results: $\bar{Z} = R^{-1} \sum_{i=1}^{R} Z_i$. Obviously we have

$$E(\bar{Z}|A_1, A_2) = T(A_1, A_2) \qquad (3.16\text{a})$$

$$\text{var}(\bar{Z}|A_1, A_2) = R^{-1} V(A_1, A_2) \qquad (3.16\text{b})$$

Now suppose that we generate a *random* pair $(A_1, A_2)$ from some probability distribution [e.g. SAWs $(\omega^{(1)}, \omega^{(2)})$ from the equal-weight distribution on $\mathcal{S}_{N_1} \times \mathcal{S}_{N_2}$]. Clearly $\bar{Z}$ is an unbiased estimator of $\langle T \rangle$, i.e.

$$\langle \bar{Z} \rangle = \langle E(\bar{Z}|A_1, A_2) \rangle = \langle T \rangle , \qquad (3.17)$$

where $\langle \cdot \rangle$ denotes expectation in the given probability distribution. The variance of $\bar{Z}$ is a sum of two terms:

$$\begin{aligned} \text{var}(\bar{Z}) &= \text{var}(T) + \langle \text{var}(\bar{Z}|A_1, A_2) \rangle \\ &= \left[ \langle T^2 \rangle - \langle T \rangle^2 \right] + R^{-1} \langle V \rangle . \end{aligned} \qquad (3.18)$$

The first term is the fluctuation of $T(A_1, A_2)$ from one pair of sets to another; the second term is the mean over pairs of sets of the fluctuation (conditional variance) in the inner Monte Carlo subroutine.

The mean CPU time for the computation of $\bar{Z}$ is $\langle T_{CPU} \rangle = a + bR$: here $a$ is the mean CPU time for generating a pair of "effectively independent" sets $(A_1, A_2)$ from the desired ensemble, plus any "setup" time associated with the inner-loop Monte Carlo algorithm; while $b$ is the mean additional CPU time per iteration of the inner-loop Monte Carlo algorithm. The goal is to minimize the variance-time product

$$\langle T_{CPU} \rangle \text{var}(\bar{Z}) = b \, \text{var}(T) R + a \langle V \rangle R^{-1} + a \, \text{var}(T) + b \langle V \rangle , \qquad (3.19)$$

since this quantity divided by the total CPU time equals the variance of our final estimate. Hence the optimal choice of $R$ is

$$R_{opt} = \left( \frac{a \langle V \rangle}{b \, \text{var}(T)} \right)^{1/2} , \qquad (3.20)$$

and the variance-time product is then

$$[\langle T_{CPU} \rangle \text{var}(\bar{Z})]_{opt} = \left[ (a \, \text{var}(T))^{1/2} + (b \langle V \rangle)^{1/2} \right]^2 . \qquad (3.21)$$

Of course, $R$ must be a positive integer, and so the true $R_{opt}$ is obtained by rounding the right-hand side of (3.20) up or down. This subtlety can be ignored if $R_{opt}$ is large, but may be significant otherwise. In particular, the deterministic inner-loop algorithms (Section 3.3) have $V = 0$, but in this case $R_{opt} = 1$ [rather than 0 as (3.20) claims] and $[\langle T_{CPU} \rangle \text{var}(\bar{Z})]_{opt} = (a+b)\text{var}(T)$.

In the remainder of this section we will assume that the CPU time for generating the sets $A_1$ and $A_2$ is of order $\mathcal{N}_1 + \mathcal{N}_2$. Clearly this is a lower bound, since it takes a time of order $\mathcal{N}_1 + \mathcal{N}_2$ simply to *write down* the two sets. On the other hand, for our application to SAWs, $A_1$ and $A_2$ will be generated by the pivot algorithm, which generates an "effectively independent" SAW (as regards *global* observables) in a CPU time of order $N$ (see Section 2.2). We expect that the observable $T(\omega^{(1)}, \omega^{(2)})$ is indeed "global" in the sense that $\tau_{int,T} \sim N^p$, where $p$ is the acceptance-fraction exponent.

## 3.5 "Hit-or-Miss" Monte Carlo Algorithm

Let $\alpha_j^+$ (resp. $\alpha_j^-$) be the maximum (resp. minimum) value of the $j^{th}$ coordinate among the points in $A_1$, so that

$$B_1 \equiv [\alpha_1^-, \alpha_1^+] \times \ldots \times [\alpha_d^-, \alpha_d^+] \tag{3.22}$$

is the smallest rectangular parallelopiped containing $A_1$. Let $\beta_j^+$ and $\beta_j^-$ be the corresponding values for $A_2$, and $B_2$ the corresponding box. It follows that

$$B \equiv B_1 - B_2 \equiv [\alpha_1^- - \beta_1^+, \alpha_1^+ - \beta_1^-] \times \ldots \times [\alpha_d^- - \beta_d^+, \alpha_d^+ - \beta_d^-] \tag{3.23}$$

is a parallelopiped which is guaranteed to contain all the points of $S \equiv A_1 - A_2$.[15]

Therefore, $T(A_1, A_2) \equiv \#(S) \equiv \sum_{x \in B} I(x)$ can be computed by the trivial "hit or miss" Monte Carlo method: Pick a point $x \in B$ at random, and compute $I(x)$ by the deterministic algorithm #1 of Section 3.3; then output $Z = \#(B) I(x)$, where $\#(B) \equiv \prod_{j=1}^{d}(\alpha_j + \beta_j + 1)$. [Here $\alpha_j = \alpha_j^+ - \alpha_j^-$ and $\beta_j = \beta_j^+ - \beta_j^-$.] Clearly $I(x)$ is a binomial random variable of mean $p = \#(S)/\#(B)$, so that

$$E(I(x)) = \frac{\#(S)}{\#(B)} \tag{3.24}$$

$$\text{var}(I(x)) = \frac{\#(S)}{\#(B)}\left(1 - \frac{\#(S)}{\#(B)}\right) \tag{3.25}$$

Hence $Z = \#(B) I(x)$ is an unbiased estimator of $\#(S)$, and its variance is

$$V_{hit-or-miss}(A_1, A_2) \equiv \text{var}(Z) = \#(S)[\#(B) - \#(S)]. \tag{3.26}$$

The CPU time for $R$ iterations of this algorithm is of order $T_{CPU} = \mathcal{N}_{max} + R\mathcal{N}_{min}$: we put the larger of the two sets $A_1, A_2$ in the hash table *once*, and then each time we compute $\rho(x)$ by looping over the smaller of the two sets. The CPU time for generating the two sets is by assumption also of order $\mathcal{N}_{max}$. Therefore, in (3.19)–(3.21) we have $a \sim \mathcal{N}_{max}$ and $b \sim \mathcal{N}_{min}$.

---

[15] In fact it is the smallest such parallelopiped, since for each index $j$ there is a point in $A_1 - A_2$ with $j^{th}$ coordinate equal to $\alpha_j^- - \beta_j^+$, and another point with $j^{th}$ coordinate equal to $\alpha_j^+ - \beta_j^-$.

## 3.6 Barrett Algorithm: Theory

Barrett [123] has proposed the following Monte Carlo algorithm, which gives an unbiased estimate of $T(A_1, A_2)$:

1. Choose at random $y \in A_1$ and $z \in A_2$. Set $x = y - z$.

2. Compute $\rho(x)$ using the deterministic algorithm #1 described in Section 3.3. [Note that by construction we have $x \in S$ and hence $\rho(x) > 0$.]

3. Output $Y = \mathcal{N}_1 \mathcal{N}_2 / \rho(x)$.

The analysis of this algorithm is easy: In Step 1 we choose the vector $x$ with probability
$$\text{Prob}(x) = \frac{\rho(x)}{\mathcal{N}_1 \mathcal{N}_2} \ . \tag{3.27}$$
It follows that
$$E(Y) = \sum_{x \in S} \text{Prob}(x) Y(x) = \sum_{x \in S} 1 = T(A_1, A_2) \tag{3.28a}$$
$$E(Y^2) = \sum_{x \in S} \text{Prob}(x) Y(x)^2 = \sum_{x \in S} \frac{\mathcal{N}_1 \mathcal{N}_2}{\rho(x)} = \mathcal{N}_1 \mathcal{N}_2 U(A_1, A_2) \tag{3.28b}$$
and hence
$$V_{Barrett}(A_1, A_2) \equiv \text{var}(Y) = \mathcal{N}_1 \mathcal{N}_2 U(A_1, A_2) - T(A_1, A_2)^2 \ . \tag{3.29}$$

The CPU time for $R$ iterations of the Barrett algorithm is of order $T_{CPU} \sim \mathcal{N}_{max} + R \mathcal{N}_{min}$: we put the larger of the two sets $A_1, A_2$ in the hash table *once*, and then each time we compute $\rho(x)$ by looping over the smaller of the two sets. The CPU time for generating the two sets is by assumption also of order $\mathcal{N}_{max}$. Therefore, in (3.19)–(3.21) we have $a \sim \mathcal{N}_{max}$ and $b \sim \mathcal{N}_{min}$.

## 3.7 Karp-Luby Algorithm: Theory

Karp and Luby [124,125] have devised an elegant Monte Carlo algorithm for estimating $T(A_1, A_2)$ [and somewhat more general combinatorial problems]. The Karp-Luby algorithm goes as follows:

1. Choose at random $y \in A_1$ and $z \in A_2$. Set $x = y - z$. Set $t = 1$.

2. Choose at random $y' \in A_1$.

3. If $z' \equiv y' - x \in A_2$, then go to Step 4. Otherwise, increment $t$ by 1 and go to Step 2.

4. Output $Z = \mathcal{N}_2 t$.

This algorithm can be understood as a randomized version of the Barrett algorithm. Step 1 is identical in the two algorithms, and it selects the vector $x$ with probability

$$\text{Prob}(x) = \frac{\rho(x)}{\mathcal{N}_1 \mathcal{N}_2}. \tag{3.30}$$

Then $t$ (the number of trials of Steps 2 and 3 needed to find a $y'$ such that $y' - x \in A_2$) is, conditioned on $x$, a random variable with a geometric distribution:

$$\text{Prob}(t = k|x) = \frac{\rho(x)}{\mathcal{N}_1}\left(1 - \frac{\rho(x)}{\mathcal{N}_1}\right)^{k-1} \quad \text{for } k = 1, 2, 3, \ldots. \tag{3.31}$$

Hence the conditional expectations of $Z = \mathcal{N}_2 t$ are

$$E(Z|x) = \frac{\mathcal{N}_1 \mathcal{N}_2}{\rho(x)} \tag{3.32a}$$

$$E(Z^2|x) = \frac{\mathcal{N}_1 \mathcal{N}_2^2}{\rho(x)}\left(\frac{2\mathcal{N}_1}{\rho(x)} - 1\right) \tag{3.32b}$$

(Of course, this makes sense only for $x \in S$.) Thus, Steps 2–4 of the Karp-Luby algorithm produce a *random* quantity $Z$ whose *mean* value (conditional on $x$) is precisely the deterministic quantity $Y = \mathcal{N}_1 \mathcal{N}_2/\rho(x)$ of the Barrett algorithm. It follows that the unconditional expectations are

$$E(Z) = \sum_{x \in S} \text{Prob}(x) E(Z|x) = T(A_1, A_2) \tag{3.33a}$$

$$E(Z^2) = \sum_{x \in S} \text{Prob}(x) E(Z^2|x) = \mathcal{N}_2 \sum_{x \in S} \left(\frac{2\mathcal{N}_1}{\rho(x)} - 1\right)$$
$$= 2\mathcal{N}_1 \mathcal{N}_2 U(A_1, A_2) - \mathcal{N}_2 T(A_1, A_2) \tag{3.33b}$$

and hence

$$V_{Karp-Luby}(A_1, A_2) \equiv \text{var}(Z) = 2\mathcal{N}_1 \mathcal{N}_2 U(A_1, A_2) - \mathcal{N}_2 T(A_1, A_2) - T(A_1, A_2)^2. \tag{3.34}$$

For future reference, we note the following inequality:

$$\left(\frac{\mathcal{N}_1 - 1}{\mathcal{N}_1 + \mathcal{N}_2 - 1}\right) \mathcal{N}_1 \mathcal{N}_2 U(A_1, A_2) \leq V_{Karp-Luby}(A_1, A_2) \leq 2\mathcal{N}_1 \mathcal{N}_2 U(A_1, A_2). \tag{3.35}$$

[PROOF: The upper bound is trivial. To prove the lower bound, use (3.10) and (3.11) to deduce $(\mathcal{N}_1 + \mathcal{N}_2 - 1)T \leq T^2 \leq \mathcal{N}_1 \mathcal{N}_2 U$.] This means that $V_{Karp-Luby}$ is of the same order of magnitude as its first term, namely $\sim \mathcal{N}_1 \mathcal{N}_2 U$, except perhaps when $\mathcal{N}_1 \ll \mathcal{N}_2$. An alternative (and often sharper) lower bound on $V_{Karp-Luby}$ can be obtained by noting that

$$\mathcal{N}_2 T \leq \left(\frac{\mathcal{N}_1}{\mathcal{N}_2} \log \mathcal{N}_{max}\right)^{-1/2} \mathcal{N}_1 \mathcal{N}_2 U. \tag{3.36}$$

[PROOF: From (3.11) we have $T^2/\mathcal{N}_1\mathcal{N}_2 \leq U$, while from (3.12) we have $\log \mathcal{N}_{max} \leq U$. Now take the geometric mean of these two bounds.] Therefore we have

$$V_{Karp-Luby}(A_1, A_2) \geq \left[2 - \left(\frac{\mathcal{N}_1}{\mathcal{N}_2}\log \mathcal{N}_{max}\right)^{-1/2}\right]\mathcal{N}_1\mathcal{N}_2 U(A_1, A_2) - T(A_1, A_2)^2 \ . \tag{3.37}$$

The CPU time for one execution of Steps 2 and 3 is essentially $t$, so the expected CPU time for one iteration of the Karp-Luby algorithm is $E(t) = T(A_1, A_2)/\mathcal{N}_2$. In addition, there is an initial CPU time of order $\mathcal{N}_2$ to place the elements of $A_2$ in a hash table. Finally, we should remember the time of order $\mathcal{N}_1 + \mathcal{N}_2$ for generating $A_1$ and $A_2$ in the first place. The expected CPU time for $R$ iterations of the Karp-Luby algorithm (plus generating $A_1$ and $A_2$) is thus

$$T_{CPU} \sim \mathcal{N}_{max} + \frac{T(A_1, A_2)}{\mathcal{N}_2} R \ . \tag{3.38}$$

Therefore, in (3.19)–(3.21) we have $a \sim \mathcal{N}_{max}$ and $b \sim \langle T \rangle/\mathcal{N}_2$.

## 3.8 Scaling Theory

In this section we consider the scaling theory of the three Monte Carlo algorithms — hit-or-miss, Barrett and Karp-Luby — in the case where $A_1$ and $A_2$ are independent random SAWs of lengths $N_1$ and $N_2$, respectively, and $N_1 \sim N_2 \sim N \to \infty$. In each case we need to compute (or guess heuristically) the scaling behavior of $\text{var}(\bar{Z})$ and $\langle T_{CPU} \rangle$ as $N \to \infty$. A good figure of (de)merit for an algorithm is the mean CPU time needed to estimate $\langle T \rangle$ with a relative variance of order 1; this time is

$$\frac{[\langle T_{CPU} \rangle \text{var}(\bar{Z})]_{opt}}{\langle T \rangle^2} \ . \tag{3.39}$$

For the hit-or-miss algorithm, we must study the scaling of $\langle T \rangle$, $\langle T^2 \rangle$ and $\langle T \#(B) \rangle$. For the Barrett and Karp-Luby algorithms, we must study the scaling of $\langle T \rangle$, $\langle T^2 \rangle$ and $\langle U \rangle$ as well as their various combinations.

As discussed in Section 2.2, we expect that $\langle T \rangle$ scales as

$$\langle T \rangle \sim N^{p_T} \ , \tag{3.40}$$

where

$$p_T = 2\Delta_4 - \gamma \ . \tag{3.41}$$

We further expect that the probability distribution of $T$ will, after rescaling by $N^{p_T}$, approach a nontrivial limiting distribution (here "nontrivial" means that the distribution is not a delta function). Therefore, we expect that

$$\langle T^2 \rangle, \text{var}(T) \sim N^{2p_T} \ . \tag{3.42}$$

On the other hand, it is reasonable to expect that

$$\langle U \rangle \sim N^{p_U} \tag{3.43}$$

for some (*a priori* unknown) exponent $p_U$. From (3.11) we know that $2p_T - 2 \le p_U \le p_T$; and we will be interested in knowing whether these inequalities are strict or not.

We now look at the individual algorithms:

**Hit-or-Miss Algorithm:** As just discussed, we expect

$$\langle T \rangle \sim N^{p_T} \tag{3.44}$$
$$\text{var}(T) \sim N^{2p_T} \tag{3.45}$$

with $p_T = 2\Delta_4 - \gamma$. On the other hand, for "typical" pairs of SAWs we clearly have $\#(B) \sim N^{d\nu}$, so we expect

$$\langle T \#(B) \rangle \sim N^{d\nu + p_T} . \tag{3.46}$$

In general we have $d\nu \ge p_T$; but even when $d\nu = p_T$ (i.e. hyperscaling holds), it seems intuitively clear that the ratio $\langle T^2 \rangle / \langle T \#(B) \rangle$ will stay well below 1 (except in dimension $d = 1$). Therefore, radical cancellations in (3.26) are excluded, and we expect

$$\langle V_{hit-or-miss} \rangle \sim N^{d\nu + p_T} . \tag{3.47}$$

It follows from (3.45) and (3.47) that

$$\text{var}(\bar{Z}) \sim N^{2p_T} + R^{-1} N^{d\nu + p_T} . \tag{3.48}$$

On the other hand, the CPU time is obviously

$$\langle T_{CPU} \rangle \sim N + RN . \tag{3.49}$$

The variance-time product is thus

$$\langle T_{CPU} \rangle \text{var}(\bar{Z}) \sim N \left[ N^{2p_T} R + N^{d\nu + p_T} R^{-1} + N^{2p_T} + N^{d\nu + p_T} \right] . \tag{3.50}$$

There are two cases:

(a) If the hyperscaling relation $d\nu = p_T$ holds, then $R_{opt} \sim 1$. [We expect that this is the case for $d < 4$.] With this choice of $R$, we have $\text{var}(\bar{Z}) \sim \langle T \rangle^2$ and $\langle T_{CPU} \rangle \sim N$. In other words, we can achieve a relative variance of order 1 in a CPU time of order $N$.

(b) If $d\nu > p_T$, then $R_{opt} \sim N^{(d\nu - p_T)/2}$. [We expect that this is the case for $d > 4$. More precisely, for $d > 4$ we expect that $\nu = 1/2$ and $p_T = 2$, in which case $R_{opt} \sim N^{(d-4)/4}$.] With this choice of $R$, we have $\text{var}(\bar{Z}) \sim \langle T \rangle^{3/2} N^{d\nu/2}$ and $\langle T_{CPU} \rangle \sim N^{1+d\nu/2} \langle T \rangle^{-1/2}$, which implies a variance-time product of order $N^{1+d\nu} \langle T \rangle$. [In fact, a variance-time product of this order can be achieved with any $R$ in the range $1 \lesssim R \lesssim N^{d\nu - p_T}$; the optimal value $R_{opt}$ lies at the geometric mean of these two extremes.] Hence a relative variance of order 1 requires a CPU time of order $N^{1+d\nu - p_T}$ [i.e. order $N^{1+(d-4)/2}$ for $d > 4$].

**Barrett and Karp-Luby algorithms:** The behavior of the Barrett and Karp-Luby algorithms is determined by the scaling behavior of $\langle T \rangle$, $\langle T^2 \rangle$ and $\langle U \rangle$ as $N \to \infty$. From (3.29) and (3.35)–(3.37), we have

$$\langle V_{Barrett} \rangle = \mathcal{N}_1 \mathcal{N}_2 \langle U \rangle \left[ 1 - \frac{\langle T^2 \rangle}{\mathcal{N}_1 \mathcal{N}_2 \langle U \rangle} \right] \tag{3.51}$$

$$\langle V_{Karp-Luby} \rangle \approx \mathcal{N}_1 \mathcal{N}_2 \langle U \rangle \left[ 2 - \frac{\langle T^2 \rangle}{\mathcal{N}_1 \mathcal{N}_2 \langle U \rangle} \right] \tag{3.52}$$

[We write $a \approx b$ to denote that $a/b \to 1$ as $N \to \infty$.] This focusses attention on the ratio

$$\kappa(N_1, N_2) = \frac{\langle T^2 \rangle_{N_1, N_2}}{(N_1+1)(N_2+1)\langle U \rangle_{N_1, N_2}} \tag{3.53}$$

[which by (3.11) satisfies $0 \leq \kappa(N_1, N_2) \leq 1$] and in particular on

$$\kappa = \lim_{N \to \infty} \kappa(N, N) . \tag{3.54}$$

(For simplicity we consider here only $N_1 = N_2 = N$, but the same principles obviously apply at any fixed ratio $N_1/N_2 \neq 0, \infty$.) There are then three possible cases:

(a) $\kappa = 0$. In this case $\langle V_{Barrett} \rangle \approx N^2 \langle U \rangle$ and $\langle V_{Karp-Luby} \rangle \approx 2N^2 \langle U \rangle$.

(b) $0 < \kappa < 1$. In this case $\langle V_{Barrett} \rangle \approx (1-\kappa)N^2 \langle U \rangle$ and $\langle V_{Karp-Luby} \rangle \approx (2-\kappa)N^2 \langle U \rangle$.

(c) $\kappa = 1$. In this case $\langle V_{Barrett} \rangle \ll N^2 \langle U \rangle$, and its exact scaling is subtle. Of course we still have $\langle V_{Karp-Luby} \rangle \approx N^2 \langle U \rangle$.

Case (a) corresponds to the exponent $p_U$ being strictly greater than $2p_T - 2$, while cases (b) and (c) correspond to $p_U = 2p_T - 2$. We believe that in fact case (c) never occurs: although it is possible to have $U = T^2/\mathcal{N}_1 \mathcal{N}_2$ for special pairs $\omega^{(1)}, \omega^{(2)}$, e.g. perpendicular rods, it seems quite implausible that such behavior could occur (even in the limit $N \to \infty$) after averaging over all pairs of SAWs.

In Section 3.9 we shall present numerical data showing clearly that $\kappa < 1$ in dimensions $d = 2, 3$; what is less clear is whether $\kappa$ is zero or nonzero. For dimension $d > 4$ we expect that $0 < \kappa < 1$. [For $d > 4$ we expect that $\langle T \rangle \sim N^2$,[16] from which it follows that $\langle U \rangle \sim N^2$ and hence $0 < \kappa \leq 1$.[17] On the other hand, we believe, as noted above, that $\kappa = 1$ is impossible.]

Assuming that $0 \leq \kappa < 1$, it follows that $\langle V_{Barrett} \rangle$ and $\langle V_{Karp-Luby} \rangle$ are both of order $N^2 \langle U \rangle$. Therefore,

$$\text{var}(\bar{Z}) \sim \langle T \rangle^2 + R^{-1} N^2 \langle U \rangle \tag{3.55}$$

---

[16] This is rigorously proven for $d \geq 6$, modulo the problem of translating generating-function results into fixed-$N$ results. See [1, Theorem 1.5.5 and Remark following it].

[17] PROOF: $\langle U \rangle \geq \langle T^2 \rangle/(N+1)^2 \geq \langle T \rangle^2/(N+1)^2$ by (3.11) and the Schwarz inequality; while $\langle U \rangle \leq \langle T \rangle$ by (3.11). Thus, if $\langle T \rangle \sim N^2$, we have also $\langle U \rangle \sim N^2$.

for both algorithms. On the other hand, the CPU times are

$$\langle T_{CPU,Barrett} \rangle \sim N + NR \qquad (3.56\text{a})$$
$$\langle T_{CPU,Karp-Luby} \rangle \sim N + N^{-1}\langle T \rangle R \qquad (3.56\text{b})$$

The variance-time products are thus

$$\langle T_{CPU,Barrett} \rangle \operatorname{var}(\bar{Z}) \sim N\left[\langle T \rangle^2 R + N^2 \langle U \rangle R^{-1} + \langle T \rangle^2 + N^2 \langle U \rangle\right] \qquad (3.57\text{a})$$

$$\langle T_{CPU,Karp-Luby} \rangle \operatorname{var}(\bar{Z}) \sim N\left[N^{-2}\langle T \rangle^3 R + N^2 \langle U \rangle R^{-1} + \langle T \rangle^2 + \langle U \rangle\langle T \rangle\right] \qquad (3.57\text{b})$$

Recalling now the inequality $T^2 \leq (N+1)^2 U$ [cf. (3.11)], we see that the dominant term for the Barrett algorithm is the fourth one, provided that $1 \lesssim R \lesssim N^2 \langle U \rangle / \langle T \rangle^2$; and in this case we have $\langle T_{CPU} \rangle \operatorname{var}(\bar{Z}) \sim N^3 \langle U \rangle$. The optimal value is at the geometric mean of this range, i.e. $R_{opt} \sim N\langle U \rangle^{1/2}/\langle T \rangle$. We conclude that a relative variance of order 1 requires a CPU time of order $N^3 \langle U \rangle/\langle T \rangle^2$. This is at least of order $N$; but it may be larger, in case $\langle U \rangle \gg \langle T^2 \rangle / N^2$ (i.e. in case $p_U > 2p_T - 2$).

Recalling next the inequality $U \leq T$ [cf. (3.11)], we see that the dominant term for the Karp-Luby algorithm is the third one, provided that $N^2 \langle U \rangle / \langle T \rangle^2 \lesssim R \lesssim N^2/\langle T \rangle$; and in this case we have $\langle T_{CPU} \rangle \operatorname{var}(\bar{Z}) \sim N\langle T \rangle^2$. The optimal value is at the geometric mean of this range, i.e. $R_{opt} \sim N^2 \langle U \rangle^{1/2}/\langle T \rangle^{3/2}$. We conclude that a relative variance of order 1 requires a CPU time of order $N$.

Comparing the analyses of the Barrett and Karp-Luby algorithms, we see that the effect of the additional randomization in the Karp-Luby algorithm is to reduce drastically the mean CPU time per iteration ($T/\mathcal{N}_2$ versus $\mathcal{N}_{min}$) while only modestly increasing the variance. The Karp-Luby algorithm is thus superior to the Barrett algorithm whenever $\langle T \rangle \ll N^2$ (as we expect occurs in dimension $d < 4$); the two algorithms are of the same order whenever $\langle U \rangle \sim \langle T \rangle \sim N^2$ (as we expect occurs in dimension $d > 4$).

## 3.9 Barrett and Karp-Luby Algorithms: Numerical Results

In this section we report results of a rather crude Monte Carlo study of $\langle T \rangle$, $\langle T^2 \rangle$, $\operatorname{var}(T)$, $\langle U \rangle$, $\langle V_{Barrett} \rangle$ and $\langle V_{Karp-Luby} \rangle$ for SAWs in dimensions $d = 2$ and $d = 3$ (taking $N_1 = N_2 = N$). The goal is to estimate the exponents $p_T$ and $p_U$ (and the various amplitudes) in the scaling relations

$$\langle T \rangle \approx A_T N^{p_T} \qquad (3.58\text{a})$$
$$\langle T^2 \rangle \approx A_{T^2} N^{2p_T} \qquad (3.58\text{b})$$
$$\operatorname{var}(T) \approx A_{var(T)} N^{2p_T} \qquad (3.58\text{c})$$
$$\langle U \rangle \approx A_U N^{p_U} \qquad (3.58\text{d})$$
$$\langle V_{Barrett} \rangle \approx A_{V_{Barrett}} N^{2+p_U} \qquad (3.58\text{e})$$
$$\langle V_{Karp-Luby} \rangle \approx A_{V_{Karp-Luby}} N^{2+p_U} \qquad (3.58\text{f})$$

In Table 1 we report our Monte Carlo estimates for 2-dimensional self-avoiding walks at $N = 100, 200, 300, 400, 500$. These were obtained by using the dimerization algorithm to generate pairs of independent SAWs $\omega^{(1)}, \omega^{(2)}$ and then using deterministic algorithm #3 (Section 3.3) to compute $T(\omega^{(1)}, \omega^{(2)})$ and $U(\omega^{(1)}, \omega^{(2)})$. Unfortunately, some of the observables lack error bars (this is our fault). The fit to $\langle T \rangle$ gives a reasonable $\chi^2$ provided that we discard the data point at the lowest value of $N$; we then get $p_T = 1.494 \pm 0.001$. This is not far from the expected value $p_T = d\nu = 3/2$; the small discrepancy can quite plausibly be attributed to corrections to scaling. (We will make a much more careful estimate of $\langle T \rangle$ and its exponent $p_T$ in Section 4.1.) The fits to $\langle T^2 \rangle$ and var($T$) are also consistent with $p_T = 3/2$, although the lack of error bars prevents making this quantitative. The fit to $\langle V_{Karp-Luby} \rangle$ gives a reasonable $\chi^2$ only if we discard the data points at the two lowest values of $N$; we then get $p_U = 1.110 \pm 0.002$. If we use the same data points in the fits to $\langle U \rangle$ and $\langle V_{Barrett} \rangle$, we get $p_U = 1.078$ and $1.152$ (unfortunately without error bars). These estimates of $p_U$ are thus in rather mediocre agreement; and it is far from clear whether $p_U$ is strictly greater than $2p_T - 2 = 1$. Another way of looking at this is to study the ratio $\kappa_N \equiv \langle T^2 \rangle/(N^2 \langle U \rangle)$, which decreases from $\approx 0.572$ at $N = 100$ to $\approx 0.485$ at $N = 500$; it is far from clear whether this ratio is tending to zero or to a nonzero value as $N \to \infty$. Clearly data at much larger values of $N$ would be needed to resolve this question definitively. Unfortunately, such data are very time-consuming to obtain, because the deterministic algorithm #3 for computing $T$ and $U$ takes a CPU time of order $N^2$. Regarding the reverse inequality, it seems clear that $p_U$ is strictly smaller than $p_T$.

In Table 2 we report our Monte Carlo estimates for 3-dimensional self-avoiding walks at $N = 100, 200, 500, 800, 1000$. Here we used the pivot algorithm combined with deterministic algorithm #3. For all observables, we can get a reasonable $\chi^2$ provided that we discard the data points at the two lowest values of $N$. The fits to $\langle T \rangle$, $\langle T^2 \rangle$ and var($T$) yield $p_T = 1.754 \pm 0.001$. Again, the errors here are purely statistical; they do not take account of systematic errors arising from corrections to scaling. We will make a much more careful estimate of $\langle T \rangle$ and $p_T$ in Section 4.2. The fits to $\langle U \rangle$, $\langle V_{Barrett} \rangle$ and $\langle V_{Karp-Luby} \rangle$ yield $p_U = 1.598 \pm 0.001$, $1.699 \pm 0.001$ and $1.630 \pm 0.001$, respectively. Again, the agreement is mediocre; and it is far from clear whether $p_U$ is strictly greater than $2p_T - 2 \approx 1.51$ (later we will see that $2p_T - 2 \approx 1.53$ is a better estimate). Otherwise put, the ratio $\kappa_N \equiv \langle T^2 \rangle/(N^2 \langle U \rangle)$ decreases from $\approx 0.661$ at $N = 100$ to $\approx 0.528$ at $N = 1000$; it is far from clear whether this ratio is tending to zero or to a nonzero value as $N \to \infty$. Again, data at much larger values of $N$ would be useful.

It is worth remarking that, in both the 2-dimensional and 3-dimensional cases, var($T$) scales the same way as $\langle T^2 \rangle$ but is only about 3% (resp. 1.5%) as big. This means that the probability distribution of $T$ is very narrow. Otherwise put, while $N$-step SAWs vary radically among themselves in size and shape, the overlap $T(\omega^{(1)}, \omega^{(2)})$ between *two* of them is remarkably constant. Presumably this is because the operation of forming $\omega^{(1)} - \omega^{(2)}$ "fills in the holes" in the individual walks: while $\omega^{(1)}$ and $\omega^{(2)}$ are fractals, $\omega^{(1)} - \omega^{(2)}$ is "semi-solid" (roughly like Swiss cheese).

# 4 Numerical Results

## 4.1 Two Dimensions

In Table 3 we present our data for $\langle R_e^2 \rangle_N$, $\langle R_g^2 \rangle_N$, $\langle R_m^2 \rangle_N$, $\langle T \rangle_{N,N}$ and the pivot-algorithm acceptance fraction $f$ for SAWs in dimension $d = 2$ in the range $100 \leq N \leq 80000$. Most of these SAWs were generated using the pivot algorithm, using either dimerization or straight rods for initialization (see Appendix B for a discussion of the adequacy of thermalization); run lengths were between $2 \times 10^6$ and $8 \times 10^6$ pivots subsequent to thermalization. However, some data at $N \leq 500$ were generated by pure dimerization (between $10^5$ and $2 \times 10^5$ independent pairs of SAWs per run). The overlap $T(\omega^{(1)}, \omega^{(2)})$ was in most cases estimated using the Karp-Luby algorithm (Section 3.7) with $100 \leq R \leq 200$; however, some runs at $N \leq 500$ used deterministic algorithm #3 (Section 3.3). Some subtleties concerning the correct determination of error bars on data generated by the pivot algorithm are discussed in Appendix C.

Early versions of our program had a bug, in which SAWs were pivoted only at sites $k > 0$ (i.e. never at the starting point). This is harmless for single SAWs, thanks to the lattice symmetries; but for pairs of SAWs it means that the relative orientation of the initial steps of the two SAWs is never altered by the algorithm. This causes a slight bias in the estimates of $\langle T \rangle$, especially for small $N$. However, we believe that this bias is completely negligible (i.e. much less than our statistical errors) for the values of $N$ treated here; we have been unable to detect any systematic differences between runs with and without this bug. Therefore, instead of throwing away the tainted data, we have simply indicated by an $^a$ in Table 3 those values of $N$ for which some (not necessarily all) of the data suffers from the bug. Caveat lector.

Table 4 shows the resulting values for the universal amplitude ratios $\langle R_g^2 \rangle / \langle R_e^2 \rangle$, $\langle R_m^2 \rangle / \langle R_e^2 \rangle$ and $\Psi$. The error bars are here determined using the triangle inequality; they are probably overestimates by a factor of $\approx 3$.

Log-log graphs of $\langle R_e^2 \rangle$, $\langle R_g^2 \rangle$, $\langle R_m^2 \rangle$, $\langle T \rangle$ and $f$ versus $N$ are so straight that there is nothing to be gained by reproducing them here. We fit each of these quantities to the Ansatz $AN^{power}$ [where power $= 2\nu$ for the three squared radii, $2\Delta_4 - \gamma$ for $\langle T \rangle$, and $-p$ for $f$] by performing weighted least-squares regressions of their logarithms against $\log N$, using the *a priori* error bars on the raw data points (Table 3) to determine both the weights and the error bars [126, Chapter 3]. As a precaution against corrections to scaling, we performed the fit with a lower cutoff $N \geq N_{min}$, and we tried all possible values of $N_{min}$. The $\chi^2$ value (sum of squares of normalized deviations from the regression line) can serve as a test of goodness of fit. Let us define the "significance level" as the probability that $\chi^2$ would exceed the observed value, assuming the correctness of the power-law model and of the raw-data error bars; this probability can be read off the $\chi^2$ distribution with $\mathcal{D} = n - 2$ degrees of freedom (DF), where $n$ is the number of data points used in the fit. An abnormally large value of $\chi^2$ (say, a level less than 5%) may indicate *either* that the pure power-law Ansatz is incorrect (e.g. due to corrections to scaling) or else that the claimed error bars on the raw data are too small; further investigation would be necessary to determine which of these is the true cause.[18] An abnormally small value of $\chi^2$ (say, a level greater than 95%) probably indicates that the claimed error bars on the raw data are too large.

The exponents estimated from fits to $\langle R_e^2 \rangle$, $\langle R_g^2 \rangle$, $\langle R_m^2 \rangle$ and $\langle T \rangle$ are plotted as a function of $N_{min}$ in Figure 1. The exponent estimated from the fit to $f$ is plotted in Figure 2. Looking at the $\chi^2$ values in these fits, we find statistically significant corrections to scaling in $\langle R_e^2 \rangle$ and $\langle R_g^2 \rangle$ only for $N_{min} \lesssim 200$, in $\langle T \rangle$ only for $N_{min} \lesssim 400$, and in $f$ only for $N_{min} \lesssim 6000$. Using in each case the next larger value of $N_{min}$, we find:

$$\begin{aligned}
\langle R_e^2 \rangle: \quad \nu &= 0.74967 \pm 0.00011 \\
A &= 0.77582 \pm 0.00138 \\
N_{min} &= 300 \\
\chi^2 &= 32.01 \text{ (34 DF, level = 57\%)}
\end{aligned}$$

$$\begin{aligned}
\langle R_g^2 \rangle: \quad \nu &= 0.74963 \pm 0.00008 \\
A &= 0.10890 \pm 0.00014 \\
N_{min} &= 300 \\
\chi^2 &= 29.01 \text{ (34 DF, level = 71\%)}
\end{aligned}$$

$$\begin{aligned}
\langle R_m^2 \rangle: \quad \nu &= 0.74972 \pm 0.00021 \\
A &= 0.34073 \pm 0.00123 \\
N_{min} &= 1000 \\
\chi^2 &= 24.72 \text{ (26 DF, level = 54\%)}
\end{aligned}$$

$$\begin{aligned}
\langle T \rangle: \quad (2\Delta_4 - \gamma)/2 &= 0.74955 \pm 0.00010 \\
A &= 1.36283 \pm 0.00213 \\
N_{min} &= 500 \\
\chi^2 &= 26.76 \text{ (32 DF, level = 73\%)}
\end{aligned}$$

$$\begin{aligned}
f: \quad p &= 0.19075 \pm 0.00046 \\
A &= 0.94095 \pm 0.00419 \\
N_{min} &= 7200 \\
\chi^2 &= 8.29 \text{ (7 DF, level = 31\%)}
\end{aligned}$$

(error bars are one standard deviation). The results are in excellent agreement with the believed exact value $\nu = 3/4$ [108,109] — although some very slight corrections to scaling clearly remain — and with the hyperscaling relation $d\nu = 2\Delta_4 - \gamma$.

For $\langle R_e^2 \rangle$, $\langle R_g^2 \rangle$, $\langle R_m^2 \rangle$ and $\langle T \rangle$, better estimates of the (nonuniversal) amplitude $A$ can be obtained by imposing the exponents $2\nu = 2\Delta_4 - \gamma = 3/2$ and simply fitting observable/$N^{3/2}$ to a constant. For all four observables, this ratio declines very slightly with $N$; by $N \approx 5000$ it is within error bars of its apparent asymptotic value. Taking $N_{min} = 5000$ we obtain:

$$\langle R_e^2 \rangle: \quad A = 0.77106 \pm 0.00034$$
$$\chi^2 = 20.58 \text{ (24 DF, level} = 66\%)$$

$$\langle R_g^2 \rangle: \quad A = 0.10817 \pm 0.00005$$
$$\chi^2 = 20.75 \text{ (24 DF, level} = 65\%)$$

$$\langle R_m^2 \rangle: \quad A = 0.33894 \pm 0.00015$$
$$\chi^2 = 21.56 \text{ (22 DF, level} = 49\%)$$

$$\langle T \rangle: \quad A = 1.35158 \pm 0.00047$$
$$\chi^2 = 22.26 \text{ (24 DF, level} = 56\%)$$

---

[18] Note also that an abnormally large *change* in $\chi^2$ as $N_{min}$ is increased by one step — that is, a drop in $\chi^2$ by an amount $\gg 1$ — signals that the data point in question differs from the regression line by several standard deviations. This could indicate that the corrections to scaling at this value of $N$ are significant, even though the *overall* $\chi^2$ — which is dominated by contributions from larger $N$ — may look reasonable.

Another approach is to fit observable/$N^{3/2}$ to $a + b(N/1000)^{-\Delta}$ with $a$, $b$ and $\Delta$ all variable.[19] For $\langle R_m^2 \rangle$ (for which we have data only at $N \geq 1000$), we are unable to obtain a decent fit with any value of $N_{min}$ available to us; the corrections to scaling at $N \geq 1000$ are too weak. For $\langle R_e^2 \rangle$, $\langle R_g^2 \rangle$ and $\langle T \rangle$ we obtain reasonable fits even for $N_{min} = 100$:

$$\langle R_e^2 \rangle: \quad a = 0.77100 \pm 0.00040$$
$$b = 0.00098 \pm 0.00050$$
$$\Delta = 0.843 \pm 0.222$$
$$\chi^2 = 35.13 \text{ (35 DF, level} = 46\%)$$

$$\langle R_g^2 \rangle: \quad a = 0.10815 \pm 0.00005$$
$$b = 0.00011 \pm 0.00004$$
$$\Delta = 0.943 \pm 0.148$$
$$\chi^2 = 28.04 \text{ (35 DF, level} = 79\%)$$

$$\langle T \rangle: \quad a = 1.35120 \pm 0.00049$$
$$b = 0.00269 \pm 0.00038$$
$$\Delta = 0.919 \pm 0.057$$
$$\chi^2 = 29.13 \text{ (35 DF, level} = 75\%)$$

Of course, the exponents $\Delta$ produced by the above fits should not be taken too seriously; they could well be phenomenological "effective" exponents that summarize the combined effects of *two or more* correction-to-scaling terms (e.g. the leading non-analytic correction $N^{-\Delta_1}$ plus the analytic correction $N^{-1}$) over some particular range of $N$. These fits are nevertheless useful in providing simple "interpolation

---

[19]The purpose of writing here $N/1000$ instead of $N$ is to reduce the correlation between the estimates of $b$ and $\Delta$. Of course, 1000 can equally well be replaced by any reasonable value which is roughly in the middle of the $N$ values represented by the data points.

formulae" that summarize our data within error bars:

$$\langle R_e^2 \rangle = 0.77100 N^{3/2} + 0.33168 N^{0.657} \quad (4.1)$$
$$\langle R_g^2 \rangle = 0.10815 N^{3/2} + 0.07653 N^{0.557} \quad (4.2)$$
$$\langle T \rangle = 1.35120 N^{3/2} + 1.53518 N^{0.581} \quad (4.3)$$

In particular, we can use these formulae to compare our data with other workers' data at different values of $N$ (but only for $N \geq 100$).

The upshot is that the corrections to scaling are quite weak in the two-dimensional self-avoiding walk; we can barely see them at our level of precision. A serious study of corrections to scaling in this model would therefore require *much* higher statistics than are available here, at large but not-too-large values of $N$ (say, $100 \lesssim N \lesssim 2000$). We have begun such a study, and will report the results separately [127].

The corrections to scaling on the amplitude ratios $\langle R_g^2 \rangle / \langle R_e^2 \rangle$, $\langle R_m^2 \rangle / \langle R_e^2 \rangle$ and $\Psi$ are even weaker than on the original observables (this is because the original corrections to scaling all have the same sign). The two ratios of radii have *no* statistically significant corrections to scaling, at our level of precision, even if one considers the true error bars to be $\approx 1/3$ of those reported in Table 4. The interpenetration ratio does have noticeable corrections to scaling at $N \lesssim 500$, but these corrections are very small: see Figure 3. Fitting the data for $N \geq N_{min}$ to a constant, we obtain

$$\langle R_g^2 \rangle / \langle R_e^2 \rangle = 0.140264 \pm 0.000073$$
$$N_{min} = 100$$
$$\chi^2 = 2.86 \text{ (37 DF, level} \approx 100\%)$$

$$\langle R_m^2 \rangle / \langle R_e^2 \rangle = 0.439605 \pm 0.000338$$
$$N_{min} = 1000$$
$$\chi^2 = 0.74 \text{ (27 DF, level} \approx 100\%)$$

$$\Psi^* = 0.66296 \pm 0.00043$$
$$N_{min} = 1000$$
$$\chi^2 = 1.45 \text{ (32 DF, level} \approx 100\%)$$

The theorists' interpenetration ratio [defined in (2.16)] is $\Psi_R^* = 0.5579 \pm 0.0006$.

## 4.2 Three Dimensions

In Table 5 we present our data for $\langle R_e^2 \rangle_N$, $\langle R_g^2 \rangle_N$, $\langle T \rangle_{N,N}$ and the pivot-algorithm acceptance fraction $f$ for SAWs in dimension $d = 3$ in the range $100 \leq N \leq 80000$. Most of these SAWs were generated using the pivot algorithm, using either dimerization or straight rods for initialization (see Appendix B for a discussion of the adequacy of thermalization); run lengths were between $2 \times 10^6$ and $4 \times 10^7$ pivots subsequent to thermalization. However, some data at $N \leq 600$ were generated by pure dimerization (between $10^5$ and $2 \times 10^5$ independent pairs of SAWs per run). The overlap $T(\omega^{(1)}, \omega^{(2)})$ was in most cases estimated using the Karp-Luby algorithm (Section 3.7) with $100 \leq R \leq 500$; however, some runs at $N \leq 1000$ used deterministic algorithm #3 (Section 3.3). Data tainted by the bug concerning $\langle T \rangle$ are again indicated by $^a$ (see Section 4.1).

Table 6 shows the results for the universal amplitude ratios $\langle R_g^2 \rangle / \langle R_e^2 \rangle$ and $\Psi$. The error bars are determined using the triangle inequality; they are probably overestimates by a factor of $\approx 3$.

We began by fitting $\langle R_e^2 \rangle$, $\langle R_g^2 \rangle$, $\langle T \rangle$ and $f$ to the pure power-law Ansatz $AN^{power}$. The exponents estimated from fits to $\langle R_e^2 \rangle$, $\langle R_g^2 \rangle$ and $\langle T \rangle$ are plotted as a function of $N_{min}$ in Figure 4. The exponent estimated from the fit to $f$ is plotted in Figure 5. *Very* strong corrections to scaling are apparent for all these observables; the exponent estimates do not appear to stabilize until one takes $N_{min} \gtrsim 10^4$ or more. Nevertheless, the estimates of $\nu$ do appear to be converging to a common value $\nu \approx 0.5876$; in particular, hyperscaling appears to be satisfied.[20]

Another view of these same fits is presented in Figure 6: here we plot the estimate of $\nu$ versus $N_{min}^{-0.5}$. The idea here is that the correction-to-scaling exponent $\Delta_1$ is predicted by RG calculations [104,105,106,90,91,107] to be in the vicinity of 0.5 (see Section 5.1.2); if this prediction is correct, then each set of estimates should fall roughly on a straight line, at least asymptotically as $N_{min} \to \infty$. The data are consistent with this prediction, but the large fluctuations make clear that it will be difficult to get accurate estimates of $\Delta_1$ by Monte Carlo.

Next we tried fitting $\langle R_e^2 \rangle$, $\langle R_g^2 \rangle$ and $\langle T \rangle$ to the Ansatz $AN^{power} + BN^{power-\Delta}$, where $power = 2\nu$ for the squared radii and $3\nu$ for $\langle T \rangle$ (we now assume hyperscaling). For each choice of fixed exponents $\nu$ and $\Delta$, we determine $A$ and $B$ by weighted least-squares and record the resulting $\chi^2$; we then ask which pairs $(\nu, \Delta)$ lead to an acceptable $\chi^2$. (Here "acceptable" is taken to mean a significance level $> 32\%$ for $\mathcal{D} = n - 2$ degrees of freedom, where $n$ is the number of data points. This corresponds roughly to confidence limits of one standard deviation.) The results for $\langle R_g^2 \rangle$ and $\langle T \rangle$ are shown in Figure 7. (The result for $\langle R_e^2 \rangle$ is similar to that for $\langle R_g^2 \rangle$, but the band of allowed values is wider.) For $N_{min} = 100$ there are *no* pairs $(\nu, \Delta)$ that are satisfactory for both $\langle R_g^2 \rangle$ and $\langle T \rangle$. This means that at such small $N$ one needs more than a single correction-to-scaling term to fit both observables in a way compatible with hyperscaling and universality. On the other hand, for larger $N_{min}$ one obtains a swath of acceptable pairs $(\nu, \Delta)$, all contained in the range $0.58715 \leq \nu \leq 0.5882$ and $0.36 \leq \Delta \leq 0.80$. From this analysis it seems difficult to obtain much precision on $\Delta$, but it does suggest that

$$\nu = 0.5877 \pm 0.0006 \qquad (4.4)$$

(subjective 68% confidence limits).

Now suppose we impose $\nu = 0.5877$ and fit observable/$N^{power}$ [where $power = 2\nu$ for the squared radii and $3\nu$ for $\langle T \rangle$] to $a + b(N/1000)^{-\Delta}$ with $a$, $b$ and $\Delta$ all variable. For $N_{min} \leq 200$, the estimates of $\Delta$ from these three fits are not consistent (this might be guessed from Figure 7). However, for $N_{min} \geq 300$ they are consistent, and yield $\Delta \approx 0.54$–$0.58$.

---

[20]Note that the rigorous inequality $d\nu \geq 2\Delta_4 - \gamma$ [see Theorem A.1 and equations (A.20)–(A.22) in Appendix A] means that the limiting value of the upper pair of curves must be above or equal to the limiting value of the lower curve.

If we now return to the Ansatz $AN^{power} + BN^{power-\Delta}$, and take $\nu = 0.5877$ and $\Delta = 0.56$, we obtain reasonable fits in $\langle R_e^2 \rangle$ for $N_{min} \gtrsim 150$, in $\langle R_g^2 \rangle$ for $N_{min} \gtrsim 400$, and in $\langle T \rangle$ for $N_{min} \gtrsim 600$. Taking $N_{min} = 600$ we obtain:

$$\begin{aligned}
\langle R_e^2 \rangle: \quad A &= 1.21667 \pm 0.00050 \\
B &= -0.48314 \pm 0.03949 \\
\chi^2 &= 13.73 \text{ (27 DF, level = 98\%)}
\end{aligned}$$

$$\begin{aligned}
\langle R_g^2 \rangle: \quad A &= 0.19455 \pm 0.00007 \\
B &= -0.11432 \pm 0.00465 \\
\chi^2 &= 13.63 \text{ (27 DF, level = 98\%)}
\end{aligned}$$

$$\begin{aligned}
\langle T \rangle: \quad A &= 0.94477 \pm 0.00028 \\
B &= 0.61687 \pm 0.01413 \\
\chi^2 &= 12.43 \text{ (27 DF, level = 99\%)}
\end{aligned}$$

At the very least, these fits provide simple interpolation formulae that summarize our data within error bars (but for $N \gtrsim 400$ only):

$$\langle R_e^2 \rangle = 1.21667 N^{1.1754} - 0.48314 N^{0.6154} \tag{4.5}$$

$$\langle R_g^2 \rangle = 0.19455 N^{1.1754} - 0.11432 N^{0.6154} \tag{4.6}$$

$$\langle T \rangle = 0.94477 N^{1.7631} + 0.61687 N^{1.2031} \tag{4.7}$$

If we take these fits seriously, we obtain the results

$$b_{R_g}^{(1)}/b_{R_e}^{(1)} = 1.48 \pm 0.18 \tag{4.8}$$

$$b_A^{(1)}/b_{R_e}^{(1)} = -1.64 \pm 0.17 \tag{4.9}$$

for the universal ratios of correction-to-scaling amplitudes.

Another way to study the corrections to scaling is to look at an amplitude ratio which is known to tend to a nonzero constant as $N \to \infty$, and fit it to $a + b(N/1000)^{-\Delta}$ with $a$, $b$ and $\Delta$ all variable. Two candidates are $\langle R_g^2 \rangle / \langle R_e^2 \rangle$ and (assuming hyperscaling) $\Psi$. Unfortunately, the corrections to scaling in $\langle R_g^2 \rangle / \langle R_e^2 \rangle$ are too weak to yield much information on $\Delta$ (see Table 7): the amplitude $b$ is within 2–3$\sigma$ of zero, and the error bars on $\Delta$ are enormous. However, we can obtain a reasonable estimate of the universal limiting value $\langle R_g^2 \rangle / \langle R_e^2 \rangle = a$. Note first that the unusually small $\chi^2$ confirms our belief that the true error bars on $\langle R_g^2 \rangle_N / \langle R_e^2 \rangle_N$ are roughly $\approx 1/3$ of those reported in Table 4. If so, then the error bars on $a$, $b$ and $\Delta$ in Table 7 should also be reduced by a factor of $\approx 3$. Making this adjustment, we conclude that

$$\frac{\langle R_g^2 \rangle}{\langle R_e^2 \rangle} = 0.1599 \pm 0.0001 \pm 0.0001 \; ; \tag{4.10}$$

here the first error bar represents systematic error due to uncontrolled corrections to scaling (subjective 68% confidence limits) and the second error bar represents the (adjusted) statistical error (classical 68% confidence limits).

The analysis of $\Psi$ is much more favorable for estimating $\Delta$, as the corrections to scaling are rather strong (Figure 8). [This is because the corrections to scaling on $\langle T \rangle$ and $\langle R_g^2 \rangle$ have opposite signs, as is evident from Figure 4.] A sample of the results is shown in Table 8. Note that the estimates of $\Psi^*$, $b$ and $\Delta$ are quite stable as $N_{min}$ is increased, and that the error bar on $\Delta$ is remarkably small. (The unusually small $\chi^2$ confirms our belief that the true error bars on $\Psi$ are roughly $\approx 1/3$ of those reported in Table 4. If so, then the error bars on $\Psi^*$, $b$ and $\Delta$ should also be reduced by a factor of $\approx 3$.) The excellent fit can be seen graphically in Figure 9; $\Psi$ is amazingly close to a linear function of $N^{-0.56}$. This confirms our expectation that the correction-to-scaling exponent $\Delta_1$ is approximately 0.5, and suggests that it may be slightly higher, around 0.56. A fair estimate would be

$$\Delta_1 \;=\; 0.56 \pm 0.03 \qquad (4.11)$$

(subjective 68% confidence limits). On the other hand, the $\Delta$ produced by the above fit could well be a phenomenological "effective" exponent that summarizes the combined effects of *two or more* correction-to-scaling terms (e.g. the leading correction $N^{-\Delta_1}$ plus the analytic correction $N^{-1}$) over some particular range of $N$. The only way to sort this out would be to use a larger $N_{min}$ together with improved statistics. In any case, we can estimate the universal limiting value

$$\Psi^* \;=\; 0.2471 \pm 0.0001 \pm 0.0002 \;. \qquad (4.12)$$

(We have again made the factor-of-3 adjustment in the statistical error bar.) The theorists' interpenetration ratio [defined in (2.16)] is $\Psi_R^* = 0.2322 \pm 0.0001 \pm 0.0003$. Since $\Psi^* > 0$, hyperscaling is satisfied.

# 5 Discussion

## 5.1 Comparison with Previous Numerical Studies

### 5.1.1 Two Dimensions

We first compared our raw data with those of other studies [128,10,129,130,127], making direct comparisons where the values of $N$ match and using the interpolation formulae (4.1)–(4.2) at other values of $N$.[21] We find excellent agreement, with one perplexing exception: when compared with the extremely precise data of [127], our values for $\langle R_e^2 \rangle$ and $\langle R_g^2 \rangle$ are about $3.5\sigma$ low at $N = 100$, and slightly low ($\lesssim 2\sigma$) at $200 \leq N \leq 500$. Interestingly, these are precisely the runs that were performed using pure dimerization, while the runs at $N > 500$ used the pivot algorithm. It is therefore conceivable that our dimerization program has a subtle bug which causes a very small error (perhaps one that decreases with $N$). However, we have been unable to find any such bug, and the small discrepancy could well be a statistical fluke.

---

[21]Warning: We believe that the theoretical premises of [130] are erroneous, in that the authors of [130] fail to distinguish correctly which quantities are universal and which are non-universal. (See Section 1.2 above.) Nevertheless, the Monte Carlo data in [130] are useful.

Our estimates of the universal ratios $\langle R_g^2 \rangle / \langle R_e^2 \rangle$ and $\langle R_m^2 \rangle / \langle R_e^2 \rangle$ agree perfectly with the best published estimates [129], and have roughly the same precision. In particular, they confirm the beautiful conformal-invariance prediction of Cardy and Saleur [131] (as corrected by [129])

$$\frac{246}{91} \frac{\langle R_g^2 \rangle}{\langle R_e^2 \rangle} - 2 \frac{\langle R_m^2 \rangle}{\langle R_e^2 \rangle} + \frac{1}{2} = 0 \ . \tag{5.1}$$

Much more precise data will be available soon [127].

To our knowledge there are no previous series-extrapolation or Monte Carlo estimates of $\Psi^*$ for two-dimensional polymers.[22] But we can compare our value with field-theoretic estimates:

Monte Carlo, simple cubic lattice, $1000 \leq N \leq 80000$ (this work):      $0.66296 \pm 0.00043$

Edwards model through order $\epsilon^2$, naive sum (Appendix D below):      $0.8243$
Edwards model through order $\epsilon^2$, exploiting $d = 1$ value (Appendix D below):      $0.6647$

This last estimate is amazingly close to the correct value; it would be interesting to know whether this close agreement is an accident.

### 5.1.2 Three Dimensions

We first compared our raw data with those of other studies [133,10,130,134,135,116,136,137], making direct comparisons where the values of $N$ match and using the interpolation formulae (4.5)–(4.6) at other values of $N$.[23] We find excellent agreement. We also compared our raw data with Nickel's interpolation/extrapolation formulae [13, equations 6 and 7], which are based on exact enumeration of short chains combined with high-precision Monte Carlo data at $N \leq 2560$. The agreement for $\langle R_e^2 \rangle$ and $\langle R_g^2 \rangle$ is excellent: most of the points differ by less than $1\sigma$, about 20% differ by between $1\sigma$ and $2\sigma$, and none differ by more than $2\sigma$. However, for $\langle T \rangle$ there are some modest discrepancies: for $N \leq 600$ and $N = 800$, our values are all between $2\sigma$ and $5.2\sigma$ lower than Nickel's; while for $N \geq 15000$, about half of our values are between $1\sigma$ and $3\sigma$ higher than Nickel's. We do not know whether either of these discrepancies is real, or what might be its cause. We suspect that it arises simply from statistical errors in Nickel's raw data for $\langle T \rangle$ (which might be several times as large as our errors); these would induce statistical errors in the coefficients of his extrapolation formula, which would in turn induce *correlated* statistical errors at nearby values of $N$.

---

[22]Table 9 (see Section 5.2 below) gives the values of $\Psi_N$ for $N \leq 7$, from exact enumeration [132]. But this series is much too short to be usefully extrapolated.

[23]Warning: Reference [116] uses an unconventional definition of $\langle R_g^2 \rangle$; it is $(1 + 1/N)^2$ times ours. Note also that the data of references [134,135,137] lack error bars.

Next we compare our estimates of the critical exponents $\nu$ and $\Delta_1$ with previous work:

    Monte Carlo, simple cubic lattice, $100 \leq N \leq 80000$ (this work):
        $\nu = 0.5877 \pm 0.0006$, $\Delta_1 = 0.56 \pm 0.03$

    Monte Carlo, simple cubic and BCC lattices, $120 \leq N \leq 2400$ [133]:
        $\nu = 0.592 \pm 0.002$

    Monte Carlo, simple cubic lattice, $200 \leq N \leq 3000$ [10]:
        $\nu = 0.592 \pm 0.002$

    Monte Carlo, simple cubic lattice, $200 \leq N \leq 7168$ [116]:
        $\nu = 0.5909 \pm 0.0003$

    Series extrapolation, various lattices [138]:
        $\nu = 0.592 \pm 0.003$

    RG, $n = 0$ field theory [104,105,106]:
        $\nu = 0.5880 \pm 0.0015$, $\Delta_1 = 0.47 \pm 0.025$

    RG, $n = 0$ field theory [107][24]:
        $\nu = 0.5872 + 0.07(g^* - 1.39) \pm 0.0004$

    Edwards model through order $z^6$ [90][25]:
        $\nu \approx 0.588$, $\Delta_1 \approx 0.473$

    Edwards model through order $z^6$ [91][25]:
        $\nu \approx 0.5886$    $\Delta_1 \approx 0.465$

(all Monte Carlo error bars are one standard deviation). It is clear in retrospect that the earlier Monte Carlo estimates [133,10,116], based on shorter walks than those used here, were biased upwards due to corrections to scaling. (This effect is seen very clearly in Figure 4. Moreover, if we truncate our own data to lie in the same range of $N$ as the previous studies, the resulting estimates of $\nu$ are almost identical to the quoted ones.) We have now done what we believe is a careful analysis of the corrections to scaling, and we have obtained reasonably good control over them. We therefore think that the current estimate is correct within its claimed error bar. (But we could be wrong!) We do not know why the series-extrapolation estimates are also high, but it could arise from the same effect; perhaps the walks probed in these analyses (up to 20–30 steps) are simply not long enough to permit an adequate analysis of corrections to scaling, even using the most sophisticated differential-approximants methods.

---

[24]The best estimate of [107] for $g^*$ is 1.39. Comparison of references [104] and [107] suggests that the uncertainty in $g^*$ may be of order 0.03–0.04. This would add an extra uncertainty of order 0.002–0.003 to $\nu$.

[25]The term of order $z^7$ has recently been obtained [107], but the extended series has not yet (to our knowledge) been analyzed.

The amazing fact is that the Monte Carlo estimates of $\nu$ have stabilized at almost exactly the value predicted by the field-theoretic calculations in their various equivalent forms ($n = 0$ field theory [104,105,106,107] or Edwards model [89,90,91,107]). The very high accuracy of the field-theoretic calculations is rather surprising to us, since this method is susceptible to serious (and quite possibly undetectable) systematic errors arising from a confluent singularity at the RG fixed point [33,139]. The weakness of this effect may be related to the apparent fact [140,139] that the confluent exponents $1/\Delta_1$ and $\Delta_2/\Delta_1$ are both very close to an integer (namely, they are $\approx 2$).

On the other hand, for the confluent exponent $\Delta_1$, the agreement between our Monte Carlo estimates and the field-theoretic predictions is not so good. Our Monte Carlo data for $\Psi$ can be fit amazingly well, all the way down to $N = 100$, by the Ansatz $\Psi = a + b/N^\Delta$ with $\Delta = 0.561 \pm 0.004$ (statistical errors only). This exponent differs considerably from the field-theoretic predictions, which all lie at $\Delta_1 \approx 0.47$. We do not know which of these estimates, if either, is the correct one — indeed, there are good reasons to be distrustful of both! On the one hand, our Monte Carlo estimate could well be a phenomenological "effective" exponent that summarizes the combined effects of *two or more* correction-to-scaling terms (e.g. the leading correction $N^{-\Delta_1}$ plus the analytic correction $N^{-1}$) over some particular range of $N$. (To test this, we would need to go to larger $N_{min}$ and obtain significantly improved statistics.) As things stand, our data do not rule out the field-theoretic prediction $\Delta_1 \approx 0.47$, provided that one includes suitable subleading correction-to-scaling terms. On the other hand, the field-theoretic prediction for $\Delta_1$ should also be taken with a grain of salt: it arises from the slope of the $\beta$-function at the fixed point $g = g^*$, and as Nickel [33,139] showed long ago, this function has, in addition to the desired term linear in $g - g^*$, also nonanalytic terms proportional to $(g^* - g)^{1/\Delta_1}$, $(g^* - g)^{\Delta_2/\Delta_1}$ and the like (see Section 5.2 below for further discussion). But the numerical methods currently employed [104,105,106,107,139,89,90,91] to extrapolate the perturbation series from $g = 0$ to $g = g^*$ assume, contrary to fact, that $\beta(g)$ is *regular* at $g = g^*$. The presence of terms $(g^* - g)^p$ with $1 < p < 2$ could well lead to systematic errors in estimates of the slope $\beta'(g^*)$ [33]. It is therefore worth considering the possibility that the true $\Delta_1$ is indeed closer to $\approx 0.56$ than to $\approx 0.47$, and that the error bars on the field-theoretic estimates may be overoptimistic.

Let us also make a brief comparison with the experimental results on polymers in a good solvent; a more detailed analysis will appear elsewhere [141]. The most systematic data on the static scaling behavior of high-molecular-weight polymers in a good solvent were obtained in the 1970's by three Japanese groups [142,143,144], using polystyrene in benzene. These data were reanalyzed, shortly after the appearance of the RG predictions for $\nu$ [145], by Cotton [146]. After making corrections for polydispersity, Cotton obtains the value $\nu = 0.586 \pm 0.004$ — in good agreement with the RG prediction, and now in good agreement with the Monte Carlo data as well. Unfortunately, we believe that this experimental value is unreliable, for the following reasons [141]:

1) The samples of Yamamoto *et al.* [142] have an unknown polydispersity, but this polydispersity is certainly not zero (as Cotton's analysis implicitly assumed).

2) The measurements of Fukuda *et al.* [143] were apparently afflicted by a serious systematic error (of magnitude ranging from 5% to 18%) arising from the way that the solution concentration was measured: see [147, footnote 19].[26]

3) The data from the three different laboratories cover almost-nonoverlapping ranges of molecular weight. As a result, the combined analysis of the three sets of data is highly susceptible to the effects of small systematic discrepancies in absolute calibration between the three laboratories, as well as to the fact that the three laboratories used slightly different temperatures (30° C for [142,143] and 25° C for [144]).

4) The raw data lack error bars. As a result, it is impossible to distinguish corrections to scaling (or systematic experimental errors) from statistical errors. This distinction is crucial to extracting a reliable value for $\nu$, as our analysis of Monte Carlo data has demonstrated.

A new generation of experiments, using modern ultra-sensitive light-scattering instrumentation [149,150,151] and an optimal statistical analysis [152], could prove to be very exciting.

Next let us compare our estimates of the universal ratio $\langle R_g^2 \rangle / \langle R_e^2 \rangle$ with previous work:

| | |
|---|---|
| Monte Carlo, simple cubic lattice, $100 \leq N \leq 80000$ (this work): | $0.1599 \pm 0.0001 \pm 0.0001$ |
| Monte Carlo, simple cubic lattice, $120 \leq N \leq 2400$ [133]: | $0.1597 \pm 0.0003$ |
| Monte Carlo, body-centered-cubic lattice, $120 \leq N \leq 2400$ [133]: | $0.1594 \pm 0.00015$ |
| Monte Carlo, simple cubic lattice, $200 \leq N \leq 3000$ [10]: | $0.1603 \pm 0.0004$ |
| Monte Carlo, simple cubic lattice, $200 \leq N \leq 7168$ [116]: | $0.1596 \pm 0.0002$ |
| Monte Carlo, simple cubic lattice, $26 \leq N \leq 3328$ [136]: | $0.16003 \pm 0.00003$ |
| Edwards model through order $z^4$ [136]: | $0.16012 \pm 0.00030$ |

(all Monte Carlo error bars are one standard deviation). The agreement is excellent.[27] It is worth noting that $\langle R_g^2 \rangle / \langle R_e^2 \rangle$ is non-monotonic as a function of $N$: for small $N$ ($\lesssim 15$), exact enumeration [154,137] shows that $\langle R_g^2 \rangle / \langle R_e^2 \rangle$ is considerably *above* its limiting value $\approx 0.1599$; but for larger $N$ (e.g. $100 \lesssim N \lesssim 1000$), our data in Table 6 show unequivocally that $\langle R_g^2 \rangle / \langle R_e^2 \rangle$ is *below* its limiting value; and our fit (Table 7) suggests that as $N \to \infty$ the approach to the limiting value is also *from below*. It would be interesting to know whether series-extrapolation techniques can "sense" this nonmonotonicity (which lies beyond the present enumerations) and predict approximately the correct limiting value.

---

[26] Unfortunately, it will not suffice simply to replace the measurements of Fukuda *et al.* [143] by those of Utiyama *et al.* [147] on a subset of their samples, because these latter measurements may suffer from a systematic error of their own, arising from the extrapolation to zero scattering angle: see [148, footnote 30].

[27] There also exists a calculation in the Edwards model to second order in $\epsilon = 4 - d$ [153], but unfortunately this expansion is too ill-behaved to be extrapolated reliably to $\epsilon = 1$ (see [10, p. 183]).

Next let us compare our estimates of the limiting interpenetration ratio $\Psi^*$ with previous work:

| | |
|---|---|
| Monte Carlo, simple cubic lattice, $100 \leq N \leq 80000$ (this work): | $0.2471 \pm 0.0001 \pm 0.0002$ |
| Monte Carlo, simple cubic lattice, $13 \leq N \leq 2560$ [13,155]: | $0.2465 \pm 0.0012$ |
| Edwards model through order $z^2$, Padé [1/1] (Appendix D below): | 0.2092 |
| Edwards model through order $\epsilon^2$, naive sum (Appendix D below): | 0.2686 |
| Edwards model through order $\epsilon^2$, exploiting $d = 1$ (Appendix D below): | 0.2486 |

(all Monte Carlo error bars are one standard deviation).[28] Our estimate is thus consistent with the very recent Monte Carlo estimate of Nickel *et al.* [13,155], but is about 4 times as precise. The Edwards-model (renormalization-group) estimates are all of the right order of magnitude, but most of them are not terribly close to the correct answer. (This is hardly surprising, in view of the very short perturbation series on which these estimates are based.) The $\epsilon$-expansion result augmented by exact information at $d = 1$ is, however, amazingly close to the correct value, both in $d = 3$ and $d = 2$. It would be useful to obtain a better understanding of whether this is a coincidence or not — perhaps by calculating the $O(\epsilon^3)$ term in $\Psi^*$.

Finally, it is worth noting that the experimental values for $\Psi^*$ also lie in the range 0.22–0.25 [11, Section 10.F] [96, Sections 2.2.5 and 2.2.6] — or more optimistically, they average to $0.245 \pm 0.005$ [146]. However, the experimental measurements of $\Psi$ are subject to all the problems noted earlier, as well as the danger of additional serious systematic errors arising from curvature in the extrapolation to zero concentration. This is a particularly severe problem for older studies which used higher concentrations in order to compensate for the less-sensitive light-scattering instrumentation then available. Again, new experiments would be highly desirable.

As explained in Section 1.2, also the ratios of correction-to-scaling amplitudes $b^{(1)}_{R_g}/b^{(1)}_{R_e}$ and $b^{(1)}_A/b^{(1)}_{R_e}$ are universal. So ideally we would like to measure these amplitude ratios, and compare them with the Edwards-model prediction $b^{(1)}_{R_g}/b^{(1)}_{R_e} = 1.249 \pm 0.035$ [136]. Unfortunately, the correction-to-scaling amplitudes are extremely sensitive to the choice of exponents $\nu$ and $\Delta_1$. Therefore, the only sensible comparison that can be made is to *impose* the same values of $\nu$ and $\Delta_1$ as are used in [136] — namely, $\nu = 0.588$ and $\Delta_1 = 0.47$ — and then perform a weighted least-squares fit to estimate the amplitudes. Unfortunately, the resulting estimates of the correction-to-scaling amplitudes are highly unstable as a function of $N_{min}$. If we take $N_{min} = 3000$, we obtain $b^{(1)}_{R_e} = -0.10 \pm 0.05$ and $b^{(1)}_{R_g} = -0.17 \pm 0.05$, hence $b^{(1)}_{R_g}/b^{(1)}_{R_e} = 1.7 \pm 1.8$ — not a very useful estimate!

---

[28]Note also that Table 9 (see Section 5.2 below) gives the values of $\Psi_N$ for $N \leq 7$, from exact enumeration [132]. But this series is much too short to be usefully extrapolated.

Finally, we can compare the estimates of the pivot-algorithm acceptance-fraction exponent $p$:

Monte Carlo, simple cubic lattice, $100 \leq N \leq 80000$ (this work):    $p \approx 0.105$ or perhaps lower

Monte Carlo, simple cubic lattice, $200 \leq N \leq 3000$ [10]:    $p = 0.1069 \pm 0.0009$
Monte Carlo, tetrahedral lattice, $49 \leq N \leq 9999$ [115]:    $p \approx 0.1146$
Monte Carlo, simple cubic lattice, $100 \leq N \leq 1600$ [135]:    $p \approx 0.103$

There is a slight discrepancy between our results and the very careful work of Zifferer [115]; we do not understand its origin. It is conceivable (though in our opinion unlikely) that the critical exponent for the pivot-algorithm acceptance fraction might vary from one three-dimensional lattice to another. More likely, the magnitude (and even the sign) of the corrections to scaling may vary radically between lattices.

## 5.2 The Sign of Approach to $\Psi^*$

For several decades, most work on the behavior of long-chain polymer molecules in dilute solution [156,157,158,11,12,96] has been based on the "two-parameter theory" in one or another of its variants: traditional (Flory-type)[29], pseudo-traditional (modified Flory-type)[30] or modern (continuous-chain-type)[31]. All two-parameter theories predict that in the limit of zero concentration, the mean-square end-to-end distance $\langle R_e^2 \rangle$, the mean-square radius of gyration $\langle R_g^2 \rangle$ and the interpenetration ratio $\Psi$ depend on the degree of polymerization $N$ (or equivalently on the molecular weight $M = N M_{monomer}$) according to

$$\langle R_e^2 \rangle = AN F_{R_e}(bN) \qquad (5.2a)$$
$$\langle R_g^2 \rangle = AN F_{R_g}(bN) \qquad (5.2b)$$
$$\Psi = F_\Psi(bN) \qquad (5.2c)$$

where $F_{R_e}, F_{R_g}, F_\Psi$ are claimed to be *universal* functions (which each specific two-parameter theory should predict), and $A$ and $b$ are *non-universal* scale factors depending on the polymer, solvent and temperature but independent of $N$. [The conventional notation is $\alpha_R^2 = F_{R_e}$, $\alpha_S^2 = 6 F_{R_g}$, $h = \alpha_S^d F_\Psi / z$ and $z = (bN)^{2-d/2}$ in spatial dimension $d$.] Moreover, virtually all the theories — and in particular the modern continuous-chain-based theories — predict that $F_\Psi$ is a monotone increasing and concave function of its argument $bN$, which approaches a limiting value $\Psi^* \approx 0.2 - 0.3$ (for $d = 3$) as $bN \to \infty$.

---

[29] See Yamakawa [157], Sections 11 and 16 (pp. 69–73 and 110–118) and parts of Sections 15, 20b and 21b (pp. 94–110, 153–164 and 167–169). See also DesCloizeaux and Jannink [12], Section 8.1 (pp. 289–313).

[30] See Yamakawa [157], most of Section 15 (pp. 94–110) and parts of Sections 20b and 21b (pp. 153–164 and 167–169). See also Domb and Barrett [159].

[31] These theories take as their starting point the Edwards model of a weakly self-avoiding continuous chain [83,84,85,86,87,11,12]. (The Edwards model is also equivalent to the continuum $\varphi^4$ field theory with $n = 0$ components.) See DesCloizeaux and Jannink [12] for a detailed treatment of the Edwards model.

But our Monte Carlo data for the self-avoiding walk (Figures 3 and 8) show precisely the opposite behavior: $\Psi$ is a *decreasing* and *convex* function of $N$, which approaches a limiting value $\Psi^* \approx 0.2471$ ($d = 3$) or $\Psi^* \approx 0.6630$ ($d = 2$) as $N \to \infty$. The same behavior was found by Nickel [13]. The decrease of $\Psi$ with $N$ is strong in $d = 3$, and weak (but noticeable) in $d = 2$.

In retrospect, this behavior is heuristically almost obvious: Short self-avoiding walks behave roughly like hard spheres, i.e. $\Psi$ is on the same order of magnitude as the hard-sphere value (2.20) [see Table 9]. On the other hand, long self-avoiding walks are fractal objects, i.e. "thinner" than hard spheres, so one expects $\Psi^* \ll \Psi_{hard-sphere}$. This is manifestly true in dimension $d \geq 4$ (where $\Psi^* = 0$) and in dimension $d = 4 - \epsilon$ [where $\Psi^* = \epsilon/8 + O(\epsilon^2)$]; it is natural to expect (and we now confirm) that it is true also in $d = 3$, and to a lesser extent in $d = 2$. Of course, in $d = 1$, SAWs *are* hard spheres (i.e. hard rods), so $\Psi^* = \Psi_{hard-sphere}$.[32] (The monotonic decrease of $\Psi^*/\Psi_{hard-sphere}$ as a function of $d$ is shown in the last line of Table 9.) If one now conjectures the simplest behavior, namely that $\Psi_N$ is a *monotonic* function of $N$, it follows that (in dimension $d \geq 2$) $\Psi_N$ must approach its limiting value $\Psi^*$ *from above*.

There is also *experimental* evidence [93,94,95,96] that for real polymers in a sufficiently good solvent, the approach to $\Psi^*$ is from above, contrary to the two-parameter theory. This behavior was considered to be a perplexing "anomalous effect", and various purported explanations were advanced [160,161,94].

The correct explanation, in our opinion, was given three years ago by Nickel [13] (see also [14,15]): theories of two-parameter type are simply wrong. Indeed, they are wrong not merely because they make incorrect predictions, but for a more fundamental reason: they purport to make universal predictions for quantities that are not in fact universal. Two-parameter theories[33] predict, among other things, that $\Psi$ is a universal function of the expansion factor $\alpha_S^2 \equiv \langle R_g^2 \rangle / \langle R_g^2 \rangle_{T_\theta}$; in particular, $\Psi$ is claimed to depend on molecular weight and temperature only through the particular combination $\alpha_S^2(M,T)$. This prediction is quite simply incorrect, both for model systems and for real polymers. Indeed, even the *sign* of the deviation from the limiting value $\Psi^*$ is not universal.

All this has a very simple renormalization-group explanation [13,14,15], so it is surprising that it was not noticed earlier. As mentioned already in Section 1.2, standard RG arguments predict, for any real or model polymer chain, the asymptotic behavior

$$\langle R_e^2 \rangle = A_{R_e} N^{2\nu}(1 + b_{R_e}^{(1)} N^{-\Delta_1} + \ldots) \qquad (5.3\text{a})$$

$$\langle R_g^2 \rangle = A_{R_g} N^{2\nu}(1 + b_{R_g}^{(1)} N^{-\Delta_1} + \ldots) \qquad (5.3\text{b})$$

$$\Psi = \Psi^*(1 + b_\Psi^{(1)} N^{-\Delta_1} + \ldots) \qquad (5.3\text{c})$$

---

[32] In $d = 1$, $\Psi_N$ is essentially constant with $N$; it differs from $\Psi^*$ only by corrections of order $1/N$ arising from the discreteness of the chain. The exact formula is $\Psi_N = (2/\sqrt{\pi})(N + \frac{1}{2})/\sqrt{N(N+2)}$.

[33] More precisely, two-parameter theories in which the scale factor $A$ is independent of temperature. In [15] these are termed "strong two-parameter theories".

as $N \to \infty$ at fixed temperature $T > T_\theta$. The critical exponents $\nu$ and $\Delta_1$ are universal. The amplitudes $A_{R_e}, A_{R_g}, b_{R_e}^{(1)}, b_{R_g}^{(1)}, b_\Psi^{(1)}$ are nonuniversal; in fact, even the *signs* of the correction-to-scaling amplitudes $b_{R_e}^{(1)}, b_{R_g}^{(1)}, b_\Psi^{(1)}$ are nonuniversal. However, the RG theory also predicts that the dimensionless amplitude *ratios* $A_{R_g}/A_{R_e}$, $\Psi^*$, $b_{R_g}^{(1)}/b_{R_e}^{(1)}$ and $b_\Psi^{(1)}/b_{R_e}^{(1)}$ are universal [13,88].

Recently, however, several papers have appeared [162,163,164,165] which attempt to explain the observed approach to $\Psi^*$ from above in terms of either an alleged "second branch" of the renormalized field theory with renormalized coupling constant $g >$ the fixed-point value $g^*$ [166,164,165] or an "extended two-parameter theory" containing an extra parameter [162,163]. We confess that we have been unable to understand the conceptual basis of these papers. Here is our attempt [14,15] to clarify what is going on:

We believe that critical phenomena are best understood in a Wilson-type (or "Wilson-de Gennes-type") renormalization-group framework [167,158]. By this we mean an RG map $\mathcal{R}$ acting on the (infinite-dimensional) space of Hamiltonians for a field theory or polymer model *with some (fixed) ultraviolet cutoff* $\Lambda < \infty$ (e.g. living on a lattice); the map $\mathcal{R}$ acts by integrating out the high-momentum (short-wavelength) degrees of freedom.[34] We wish to contrast this with a "field-theoretic" RG [69,11,12], which acts in a (finite-dimensional) space of *continuum* (renormalized) field theories or polymer models (i.e. models with the ultraviolet cutoff already *taken to infinity*) by spatial dilation. A few paragraphs from now we will explain the field-theoretic RG in terms of the Wilson RG (and we will see that the reverse is *not* possible). But for the moment let us simply proceed with the Wilson approach.

Figure 10 (taken from [14]) shows part of the (Wilson) renormalization-group flow for a cutoff field theory or polymer model in dimension $3 \leq d < 4$. Here $H^*$ is the Gaussian fixed point (which for $3 \leq d < 4$ is also the fixed point controlling the theta-solvent behavior), while $H_{GS}^*$ is the nontrivial (= good-solvent = Wilson-Fisher) fixed point. (Please ignore for now the curves $\mathcal{M}_s$, $\mathcal{M}_u$ and $\mathcal{C}$.) More precisely, Figure 10 shows the flow *on the critical surface* (corresponding to correlation length $\xi = \infty$ in the cutoff field theory, or chain length $N = \infty$ in the polymer model). Non-critical models (i.e. $\xi < \infty$ or $N < \infty$) lie above the plane of the page. Both $H^*$ and $H_{GS}^*$ have unstable (= relevant) directions coming out of the page, and these trajectories lead to the infinite-temperature fixed point $H_\infty^*$ (which has $\xi = 0$ or $N = 0$).

We must now distinguish two *very different* limiting situations in polymer theory:

(I) $N \to \infty$ at fixed temperature $T$, where either (a) $T > T_\theta$, (b) $T = T_\theta$, or (c) $T < T_\theta$.

---

[34]For example, for a continuous-space field theory with ultraviolet cutoff $\Lambda$, the map $\mathcal{R}$ could be integration over field components with momenta in the shell $\Lambda/2 < |p| \leq \Lambda$ [167]; for a lattice field theory, $\mathcal{R}$ could be passage to block spins [168]; for a discrete polymer chain, $\mathcal{R}$ could be decimation of odd-numbered sites along the chain [158].

(II) $N \to \infty$, $T \to T_\theta$ with $x \equiv N^\phi(T - T_\theta)$ fixed, where $\phi$ is a suitable *crossover exponent*.

Roughly speaking, case Ia corresponds to the good-solvent regime, while case II corresponds to the crossover regime near the theta point.[35]

Case Ia applies to any one-parameter family $\mathcal{F}$ of Hamiltonians, parametrized by the chain length $N$, that transversally intersects the critical surface (i.e. the plane of the page) at some point within the domain of attraction of $H^*_{GS}$ (i.e. anywhere to the right of $\mathcal{M}_s$). For example, the family $\mathcal{F}$ could intersect the critical surface at $P$, $Q$ or $R$ (among many other places). Then, the critical exponents and universal amplitude ratios [cf. (5.3) ff.] associated with the limit $N \to \infty$ in the family $\mathcal{F}$ are completely determined by the RG flow in an infinitesimal neighborhood of $H^*_{GS}$; they are therefore the *same* for *all* families intersecting the critical surface within the domain of attraction of $H^*_{GS}$ ("universality"). On the other hand, the nonuniversal amplitudes arise from the entire history of the RG flow that takes $\mathcal{F}$ to $H^*_{GS}$; they are therefore *different* for different families $\mathcal{F}$. For example, a family crossing the critical surface at $P$ or $R$ would approach $\Psi^*$ from below, while a family crossing at $Q$ would approach $\Psi^*$ from above. The nonuniversal amplitudes cannot be predicted except through detailed knowledge of the *microscopic* physics (i.e. of the family $\mathcal{F}$). In particular, they cannot be predicted by any "coarse-grained" theory (such as a renormalized field theory), or indeed by *any* simple mathematical model.

As will be explained below, the manifold $\mathcal{M}_u$ (which extends also out of the plane of the page) corresponds to the continuum Edwards models, with the plane of the page corresponding to $z = \infty$. It is important to note that $\mathcal{M}_u$ has *no special status* with respect to the good-solvent fixed point $H^*_{GS}$; it is merely one of many trajectories whose intersection with the critical surface is attracted to $H^*_{GS}$. The universal properties of polymer chains in a good solvent (case Ia) can indeed be extracted from the limit $z \to \infty$ of the continuum Edwards model; but they can be extracted equally well from the limit $N \to \infty$ of *any* family $\mathcal{F}$ whose intersection with the critical surface is attracted to $H^*_{GS}$. One family may be more convenient for computation than another, but all have the same conceptual status.

Let us now explain the relation between the Wilson RG and continuum field theories. Let $H^*$ be any critical fixed point (for the moment it doesn't matter whether $H^*$ is Gaussian), and let $\mathcal{M}_s$ (resp. $\mathcal{M}_u$) be the stable (resp. unstable) manifold of $H^*$.[36] Continuum limits are obtained by taking a sequence of initial Hamiltonians $H_n$ approaching the stable manifold, and rescaling lengths by suitable factors. This rescaling is equivalent to applying the map $\mathcal{R}$ a suitable number of times. The low-energy effective Hamiltonians $H_n^{eff} \equiv \mathcal{R}^n H_n$ then tend to the *unstable manifold* (see Figure 10). Continuum field theories $F$ are thus in one-to-one correspondence with Hamiltonians $H$ on the unstable manifold: the correlation functions of $F$ at momenta $|p| \leq \Lambda$ are *equal* to the correlation functions of $H$ with cutoff $\Lambda$. This point of view has been emphasized by Wilson and Kogut [167] and others [169,168].

---

[35] However, it is crucial to understand that cases I and II refer to families of *limiting paths* in the $(T, N)$-plane, not to "regions" or "domains" of the finite $(T, N)$-plane. Failure to appreciate this distinction — which may at first seem rather pedantic — can lead to apparent paradoxes regarding the sign of the correction-to-scaling terms.

[36] For simplicity let us assume that there are no marginal operators. The presence of marginal operators (such as the $\varphi^6$ operator at the Gaussian fixed point in $d = 3$) does *not* affect the fundamental conclusions, but merely induces multiplicative logarithmic corrections.

In particular, the *Gaussian* fixed point $H^*$ has (in dimension $3 \leq d < 4$) a two-dimensional unstable manifold $\mathcal{M}_u$: the two unstable directions correspond to $\varphi^2$ and $\varphi^4$ interactions (or in polymer language, a chain-length fugacity and a two-body self-interaction).[37] So $\mathcal{M}_u$ corresponds to the manifold of superrenormalizable $\varphi^4$ continuum field theories — or in polymer language, to the continuum Edwards model.

We can now understand the connection between the Wilson and field-theoretic renormalization groups.[38] The homogeneous field-theoretic RG equations [171,172,173] describe how a family of continuum field theories is mapped into itself under spatial dilation. On the other hand, the continuum field theories are in one-to-one correspondence with Hamiltonians on the unstable manifold; and this correspondence takes spatial dilation into the RG map $\mathcal{R}$. Thus, the field-theoretic RG is nothing other than the Wilson RG *restricted to the unstable manifold* and then rewritten in terms of "renormalized" parameters.

Having understood this, we can now evaluate the attempts [166,164,165] to explain the "wrong" sign of approach to $\Psi^*$ (or the analogue in liquid-gas critical points) in terms of an alleged "second branch" of the renormalized field theory, located at $g > g^*$.[39] The trouble is that, as far as we know, *no such branch exists*. As explained above, continuum field theories correspond to the unstable manifolds of critical fixed points. Thus, the alleged "second branch" could exist *only* if there were an as-yet-unknown critical fixed point $H^*_{new}$ lying somewhere to the right of $H^*_{GS}$, part of whose unstable manifold is attracted to $H^*_{GS}$. There is (to our knowledge) no evidence whatsoever for the existence of such a fixed point.[40]

---

[37] We ignore here the marginal operator $(\nabla \varphi)^2$, which corresponds to a physically trivial rescaling of field strengths.

[38] This connection was worked out by Kupiainen and Sokal in 1983, and was quite possibly known to others as well. However, to our knowledge it first appeared publicly in [14]. Very similar ideas appeared earlier in work of Hughes and Liu [170].

[39] Of course, no such explanation is needed, because we have already given a complete and straightforward explanation in terms of the Wilson RG. But there would be no harm in giving an alternate explanation of the same phenomenon — provided that this explanation is correct!

[40] The nonexistence of such a branch of *continuum* field theories seems to be recognized by Krüger and Schäfer, who note [164, abstract and p. 764] that "the strong coupling branch implicitly relies on the existence of a finite segment size" ($\simeq$ ultraviolet cutoff). But on the other hand they insist that renormalized field theory *can* be used for $g > g^*$ [165, p. 3]. We do not understand how these two statements can be reconciled. Perhaps Krüger and Schäfer want to study the *cutoff* field theories to the right of $H^*_{GS}$. But there is no distinguished one-parameter family of such theories (i.e. the putative extension of $\mathcal{M}_u$); there is simply the infinite-dimensional space of *all* cutoff theories (with complicated Hamiltonians as well as simple ones) lying to the right of $H^*_{GS}$, all of which have

The conventional field-theoretic RG approach [69,166,164,165] makes another error in assuming that the $\beta$-function describing the field-theoretic RG flow is regular at $g = g^*$. In fact, as pointed out long ago by Nickel [33,139], the $\beta$-function should be expected to contain nonanalytic terms like $(g^* - g)^{1/\Delta_1}$, $(g^* - g)^{\Delta_2/\Delta_1}$ and many others. This can easily be understood in our Wilson-type RG framework: The field-theoretic $\beta$-function describes the Wilson RG flow restricted to the unstable manifold $\mathcal{M}_u$ (and rewritten in terms of the "renormalized" parameter $g$). Now $\mathcal{M}_u$ has no special status at $H^*_{GS}$: like every other RG trajectory[41], it approaches $H^*_{GS}$ tangent to the leading irrelevant direction; but barring a miraculous coincidence, $\mathcal{M}_u$ also has nonzero components (with respect to the nonlinear scaling fields at $H^*_{GS}$) in all of the subleading irrelevant directions. This induces nonanalytic terms $(g^* - g)^{\Delta_2/\Delta_1}$, $(g^* - g)^{\Delta_3/\Delta_1}$, ... in the $\beta$-function. In addition, the analytic corrections to scaling at $H^*_{GS}$ induce nonanalytic terms like $(g^* - g)^{1/\Delta_1}$, $(g^* - g)^{2/\Delta_1}$, ... in the $\beta$-function [33].

In summary, the field-theoretic $\beta$-function is *nonanalytic as $g \uparrow g^*$*, and is (as far as we know) *not defined at all for $g > g^*$*. These facts cannot be understood by manipulation of formal expressions within renormalized field theory, but they can be understood when the renormalized field theory is placed within the Wilson-type RG context.

Finally, this framework allows us to understand [14,15] the special role played by the continuum Edwards model. This model has no special status at $H^*_{GS}$, but it does have special status at $H^*$: it is the unstable manifold $\mathcal{M}_u$. For this reason, the continuum Edwards model describes the *universal crossover scaling behavior in an infinitesimal region just above the theta temperature*, namely the limit $N \to \infty$, $T \to T_\theta$ with $x \equiv N^\phi(T - T_\theta)$ fixed $\geq 0$ (case II above), where $\phi = 2 - d/2$ for $3 < d < 4$ and $\phi = \frac{1}{2} \times \log^{-4/11}$ for $d = 3$.[42] That is, the continuum Edwards model controls the behavior of any two-parameter family $\mathcal{F}'$ of Hamiltonians that transversally intersects the critical surface in some curve $\mathcal{C}$ which in turn transversally intersects $\mathcal{M}_s$ (see Figure 10). Since the theta point is beyond the scope of the present paper, we refer the reader to [14,15] for details.

---

equivalent conceptual status.

The same objection applies to the "extended two-parameter theory" of Chen and Noolandi [162,163]. Perhaps their extra parameter is intended to correspond to the coefficient of the leading irrelevant coupling at $H^*_{GS}$; but in that case it is merely an inordinately complicated restatement of (5.3) ff., and moreover it neglects the second and higher irrelevant couplings. Of course, one could consider the manifold corresponding to the *vanishing* of the second and higher irrelevant couplings (with respect to the nonlinear scaling fields at $H^*_{GS}$); but this manifold, on the left side of $H^*_{GS}$, does *not* coincide with $\mathcal{M}_u$.

Finally, we believe that the same objection applies to the *second half* of Nickel's paper [13, after p. 1359], in connection with his "recursion model". He seems to realize this, as he describes the upper branch in his simplified recursion model as "*quasi*universal".

[41] Except for a measure-zero set of exceptional trajectories.

[42] By the latter expression we mean that in $d = 3$ the correct scaling variable is $x = N^{1/2}(\log N)^{-4/11}(T - T_\theta)$ [174,175,176].

## 5.3 Prospects for Future Work

Let us close this paper by mentioning briefly some interesting areas for future work.

Higher virial coefficients can be studied by the same methods that we have used here to study the second virial coefficient. For example, the third virial coefficient between molecules of types $i, j, k$ is [102]

$$
B_3^{(ijk)} = \frac{1}{3} \sum_{\substack{s \in S_i \\ s' \in S_j \\ s'' \in S_k}} \sum_{\substack{x' \in \mathbb{Z}^d \\ x'' \in \mathbb{Z}^d}} W_i(s) W_j(s') W_k(s'') \times
$$
$$
\left[1 - e^{-\mathcal{V}_{ij}((0,s),(x',s'))}\right] \left[1 - e^{-\mathcal{V}_{ik}((0,s),(x'',s''))}\right] \left[1 - e^{-\mathcal{V}_{jk}((x',s'),(x'',s''))}\right]
$$
(5.4)

[compare (2.13)], where the interaction energy $\mathcal{V}$ is given as usual by (2.14). The ratio $B_3^{(N,N,N)}/B_2^{(N,N)2}$ is expected to be universal for polymers in a good solvent, and it would be interesting to know its value. (At present, only a crude field-theoretic estimate is available: $B_3/B_2^2 \approx 0.517$, based on first-order perturbation theory.[43]) It should be noted that this quantity plays a crucial role in the extraction of the *second* virial coefficient from experimental light-scattering data, due to the necessity of extrapolating to zero concentration [141].

The "hit-or-miss" algorithm (Section 3.5) can easily be generalized to compute the $n^{th}$ virial coefficient, given $n$ independent SAW samples. Moreover, the efficiency should remain reasonably good whenever hyperscaling holds. We do not know whether the Barrett and Karp-Luby algorithms can be generalized to virial coefficients of order $n \geq 3$. Finally, the deterministic Fourier method (Section 3.3) can be applied to the third virial coefficient and to at least some of the graphs for virial coefficients of order $n \geq 4$ (namely, those graphs that can be decomposed into series and parallel connections). For example, to compute $B_3$ one first uses the Fourier method to compute $\rho_{12}(x)$ and thus $I_{12}(x)$, and likewise for 13 and 23; one then computes $(I_{12} * \check{I}_{13} * I_{23})(0)$ by passing to Fourier space.

Another interesting extension of our work would be to study the Domb-Joyce model [177] of weakly self-repelling walks, in which each self-intersection is penalized by a weight $e^{-\lambda}$ ($0 \leq \lambda \leq +\infty$); obviously this model interpolates between ordinary random walks ($\lambda = 0$) and SAWs ($\lambda = +\infty$). This model can be studied in two *very different* limits:

(Ia) $N \to \infty$ at fixed $\lambda > 0$.

(II) $N \to \infty$, $\lambda \to 0$ with $x \equiv \lambda N^{2-d/2}$ fixed.

Case Ia is the good-solvent regime, and the *universal* quantities should take exactly the same values as for SAWs (for any $\lambda > 0$). On the other hand, the *non-universal* quantities will manifestly be $\lambda$-dependent; we expect the approach to $\Psi^*$ to be from below for small $\lambda$, and from above for large $\lambda$. In particular, there will be some intermediate value $\lambda_*$ for which the leading correction to scaling will *vanish*. (In Figure 10 this would be achieved e.g. by a family of Hamiltonians that crosses the critical surface about halfway between $P$ and $Q$.) Of course, subleading corrections to scaling will still be present, but these will be suppressed by $N^{-\Delta_2}$, which decays much more rapidly than $N^{-\Delta_1}$. By studying the Domb-Joyce model systematically in the $(\lambda, N)$-plane, it should be possible to locate $\lambda_*$ empirically, and to exploit this knowledge to obtain estimates of $\nu$, $\Psi^*$ and other universal quantities that are less contaminated by the effects of corrections to scaling. (We thank Jim Barrett and Bernie Nickel for this observation [92].) This is in the same spirit as the analyses of series expansions by Chen, Fisher and Nickel [38,41], who found that by varying an irrelevant parameter and imposing universality, they could reduce the sensitivity of exponent estimates to the effects of corrections to scaling.

---

[43] See DesCloizeaux–Jannink [12], p. 544, equation (13.1.28). Note that in their notation $B_3/B_2^2 = \frac{4}{3} h(z)/g(z)^2$ [compare their equation (13.1.2)].

Case II in the Domb-Joyce model should be given (rather trivially) by the continuum Edwards model. This fact is hardly in doubt, but the results could serve as a useful nonperturbative check on the reliability of the extrapolations to large $z$ in the Edwards model [90,91]. This is a warm-up for the problem of the crossover scaling behavior near the theta point.

The deepest — and most difficult — extension of this work would be to SAWs with nearest-neighbor attraction, for the purpose of studying the crossover scaling behavior near the theta point. In dimension $d = 3$, the crossover scaling functions are predicted [14,15] to be given exactly, modulo two nonuniversal scale factors, by the continuum Edwards model. In dimension $d = 2$, there is as yet no theoretical prediction for the crossover scaling functions, but there are some predictions for critical exponents [178]. It is a highly nontrivial problem to develop Monte Carlo algorithms that work well near the theta point. The pivot algorithm (taken alone) does *not* work terribly well in this regime [179,180].

## A  Some Geometrical Theorems

In this appendix we prove some geometrical bounds on $T(A, B) \equiv \#(A - B)$ and $U(A, B)$ [see (3.1) and (3.9)] that played a role in Sections 2 and 3 of this paper. When averaged over pairs of independent SAWs of length $N$, these bounds show that

(a) $\langle T \rangle \leq \text{const} \times N^{d\nu}$

(b) $\langle T \rangle \geq \text{const} \times N^{2\nu}$

(c) $\langle U \rangle \geq 2 \log N$

modulo the usual assumptions on the scaling of individual SAWs. In particular, bounds (a) and (b) together prove (modulo these assumptions) hyperscaling for SAWs in dimension $d = 2$.

## A.1 Theorems and Proofs

Let us begin by defining several measures of the size of a set $A \subset \mathbb{Z}^d$:

- The *span* in the $\alpha^{th}$ coordinate direction

$$S_\alpha(A) = \max_{x \in A} \mathbf{e}_\alpha \cdot x - \min_{x \in A} \mathbf{e}_\alpha \cdot x, \tag{A.1}$$

where $\mathbf{e}_\alpha$ is the unit vector along the $\alpha$ axis ($1 \leq \alpha \leq d$).

- The *Euclidean diameter*

$$\mathrm{diam}(A) = \max_{x, x' \in A} \|x - x'\|_2, \tag{A.2}$$

where $\|\cdot\|_2$ denotes the Euclidean norm.

- The *sup-norm diameter*

$$\mathrm{diam}_\infty(A) = \max_{x, x' \in A} \|x - x'\|_\infty = \max_{1 \leq \alpha \leq d} S_\alpha(A), \tag{A.3}$$

where $\|x\|_\infty = \max_{1 \leq \alpha \leq d} |\mathbf{e}_\alpha \cdot x|$.

Let $R$ be a positive real number. A set $A \subset \mathbb{Z}^d$ is said to be $R$-*connected* if for all pairs $x, x' \in A$ there exists a sequence $x = x_0, x_1, \ldots, x_n = x'$ of points in $A$ such that $\|x_i - x_{i-1}\|_\infty \leq R$ for $1 \leq i \leq n$. A set is said to be *connected* if it is 1-connected. (Note that this allows diagonal as well as nearest-neighbor connections.)

Finally, we let $\Lambda_d(R)$ be the maximum number of lattice points in a closed Euclidean ball of radius $R$, namely

$$\Lambda_d(R) = \sup_{x_0 \in \mathbb{R}^d} \#\{x \in \mathbb{Z}^d : \|x - x_0\|_2 \leq R\}. \tag{A.4}$$

Note that $\Lambda_d(R) = \tau_d R^d + O(R^{d-1})$ as $R \to \infty$, where $\tau_d \equiv 2\pi^{d/2}/[d\Gamma(d/2)]$ is the volume of the unit ball in $\mathbb{R}^d$.

**Theorem A.1** *Let $A, B \subset \mathbb{Z}^d$. Then*

$$\#(A - B) \leq \prod_{\alpha=1}^{d} \left[ S_\alpha(A) + S_\alpha(B) + 1 \right] \tag{A.5}$$

$$\#(A - B) \leq \Lambda_d\left(\mathrm{diam}(A) + \mathrm{diam}(B)\right) \tag{A.6}$$

**Theorem A.2** *Let $A, B$ be $R$-connected subsets of $\mathbb{Z}^d$ ($d \geq 2$). Let $G_d \subset O(d)$ be the orthogonal symmetry group of $\mathbb{Z}^d$ [so that $\#(G_d) = 2^d d!$]. Then*

$$\frac{1}{\#(G_d)} \sum_{g \in G_d} \#(A - gB) \geq \frac{1}{2R^2} \mathrm{diam}_\infty(A) \mathrm{diam}_\infty(B). \tag{A.7}$$

**Theorem A.3** *Let $A, B \subset \mathbb{Z}^d$. Then*

$$U(A,B) \geq 2 \sum_{k=1}^{\mathcal{N}_{min}-1} \frac{1}{k} + \frac{\mathcal{N}_{max} - \mathcal{N}_{min} + 1}{\mathcal{N}_{min}} \tag{A.8a}$$

$$\geq \log(\mathcal{N}_{max} + 1) + \log \mathcal{N}_{min} , \tag{A.8b}$$

*where $\mathcal{N}_{min} = \min[\#(A), \#(B)]$ and $\mathcal{N}_{max} = \max[\#(A), \#(B)]$. The first inequality is the best possible in terms of $\mathcal{N}_{min}$ and $\mathcal{N}_{max}$.*

PROOF OF THEOREM (A.1). These bounds follow immediately from the fact that the set $A - B$ can be contained in a rectangular parallelopiped or a sphere of the indicated size. ∎

PROOF OF THEOREM (A.2). Let us consider the case $d = 2$ first. Fix $g \in G_2$. There exist $I \geq 1$ and $x_1, \ldots, x_I \in A$ such that $||x_i - x_{i-1}||_\infty \leq R$ for $i = 2, \ldots, I$ and $||x_I - x_1||_\infty = \text{diam}_\infty(A)$. Similarly, there exist $J \geq 1$ and $y_1, \ldots, y_J \in gB$ such that $||y_j - y_{j-1}||_\infty \leq R$ for $j = 2, \ldots, J$ and $||y_J - y_1||_\infty = \text{diam}_\infty(gB) = \text{diam}_\infty(B)$. Then Lemma 8.1 of Aizenman [111] implies that

$$\begin{aligned}\#(A - gB) &\geq \frac{1}{R^2} ||x_I - x_1||_2 \, ||y_J - y_1||_2 \, |\sin \theta(x_I - x_1, y_J - y_1)| \\ &\geq \frac{1}{R^2} \text{diam}_\infty(A) \, \text{diam}_\infty(B) \, |\sin \theta(x_I - x_1, y_J - y_1)| \end{aligned} \tag{A.9}$$

where $\theta(v, w)$ is the angle that the vector $w$ makes with respect to the vector $v$.

Let $g_+ \in G_2$ be the operator of rotation by $+\pi/2$. Then for any vectors $v$ and $w$, $\theta(v, g_+ w) = \theta(v, w) + \pi/2$, so

$$|\sin \theta(v, w)| + |\sin \theta(v, g_+ w)| = |\sin \theta(v, w)| + |\cos \theta(v, w)| \geq 1 . \tag{A.10}$$

From (A.9), we see that

$$\#(A - gB) + \#(A - g_+ gB) \geq \frac{1}{R^2} \text{diam}_\infty(A) \, \text{diam}_\infty(B) . \tag{A.11}$$

The theorem for $d = 2$ now follows by averaging (A.11) over all $g \in G_2$ and dividing both sides by 2.

Now consider the case $d \geq 3$. Fix $g \in G_d$. Choose indices $\beta$ and $\beta'$ so that $S_\beta(A) = \text{diam}_\infty(A)$ and $S_{\beta'}(gB) = \text{diam}_\infty(B)$. If $\beta \neq \beta'$, then let $P$ be the orthogonal projection of $\mathbb{Z}^d$ onto the $(\beta, \beta')$ coordinate plane. If $\beta = \beta'$, then let $\beta''$ be any element of $\{1, \ldots, d\} \setminus \{\beta\}$, and let $P$ be the orthogonal projection onto the $(\beta, \beta'')$ coordinate plane. Let $h \in G_d$ be the rotation by $+\pi/2$ in the $(\beta, \beta')$ [or $(\beta, \beta'')$] coordinate plane, which leaves all the other coordinates fixed. Note that

$Ph = g_+P$, where $g_+ \in G_2$ was defined above. We now use the result (A.11) for $d = 2$:

$$\begin{aligned}
\#(A - gB) + \#(A - hgB) &\geq \#(PA - PgB) + \#(PA - PhgB) \\
&\geq \frac{1}{R^2} \operatorname{diam}_\infty(PA) \operatorname{diam}_\infty(PgB) \\
&= \frac{1}{R^2} \operatorname{diam}_\infty(A) \operatorname{diam}_\infty(B) \,. \quad (A.12)
\end{aligned}$$

The theorem now follows by averaging, as for the $d = 2$ case. ∎

We remark that some strengthened versions of Theorem A.2, while quite plausible at first sight, are wrong. For example, we at first thought that for 1-connected sets (and in particular walks) in $\mathbb{Z}^2$,

$$\#(A - B) \geq \text{const} \times S_1(A) S_2(B) \,. \quad (A.13)$$

This is false, since if $A$ and $B$ are both $N$-step self-avoiding walks from $(0,0)$ to $(N/2, N/2)$ that stay in the diagonal strip $|x_1 - x_2| \leq 1$, then $\#(A - B) \sim N$ while $S_1(A)S_2(A) \sim N^2$. Thus one must average over rotations in some way. We had also conjectured that for any sets $A, B \subset \mathbb{Z}^2$,

$$\#(A - B) + \#(A - g_+B) \geq C \,\#\operatorname{Proj}_1(A) \,\#\operatorname{Proj}_2(B) \,, \quad (A.14)$$

where $\operatorname{Proj}_\alpha$ is the projection onto the $x_\alpha$ coordinate axis, and $g_+$ is rotation by $\pi/2$; here $C$ is a constant that is independent of the connectedness $R$. A counterexample to this conjecture was found for us by G.L. O'Brien, as follows: Let $M$ be a large integer, and let

$$A = B = \{(0,0) + k(1, M) + l(M, -1) : 0 \leq k, l < M\} \,. \quad (A.15)$$

(This is a large sparse square, slightly tilted.) Then $\#\operatorname{Proj}_1(A) = \#\operatorname{Proj}_2(B) = M^2$, but $\#(A - B) = \#(A - g_+B) \approx 4M^2$. We still do not know if (A.14) holds if its left-hand side is replaced by the left-hand side of (A.7) with $d = 2$.

PROOF OF THEOREM (A.3). Throughout this proof, we shall assume without loss of generality that $\mathcal{N}_1 \equiv \#(A) \geq \mathcal{N}_2 \equiv \#(B)$.

To show that the first inequality is the best possible, let $A$ (resp. $B$) consist $\mathcal{N}_1$ (resp. $\mathcal{N}_2$) consecutive sites on the $x_1$ axis (so $A$ and $B$ are parallel rods). Then it is easy to check that the first inequality holds as an equality.

Let $v$ be a vector in $\mathbb{R}^d$ such that the inner-product mapping $x \mapsto v \cdot x$ is one-to-one on $A \cup B$. (For example, the coordinates of $v$ could be $d$ irrational numbers that are linearly independent over the rationals.) Denote the elements of $A$ by $a_1, \ldots, a_{\mathcal{N}_1}$, ordered so that $v \cdot a_i < v \cdot a_{i+1}$ for every $i$. Similarly, denote the elements of $B$ by $b_1, \ldots, b_{\mathcal{N}_2}$, ordered so that $v \cdot b_j < v \cdot b_{j+1}$ for every $j$.

We shall use the following observation:

<u>Observation 1</u>: If $a_i - b_{\mathcal{N}_2} = a_l - b_m$ for some $i, l, m$, then $l \leq i$.

(To prove this, note that the hypothesis implies that $a_i - a_l = b_{\mathcal{N}_2} - b_m$, and so $v \cdot a_i - v \cdot a_l = v \cdot b_{\mathcal{N}_2} - v \cdot b_m \geq 0$. Therefore $v \cdot a_i \geq v \cdot a_l$, which implies $l \leq i$.) The proof of the following observation is exactly analogous:

<u>Observation 2</u>: If $a_{\mathcal{N}_1} - b_j = a_m - b_l$ for some $j, l, m$, then $l \leq j$.

We now claim that for any $i$,
$$\rho(a_i - b_{\mathcal{N}_2}) \leq i. \tag{A.16}$$

To prove this, observe that $\rho(x)$ equals the number of distinct $a_l$'s in $A$ such that $x \in a_l - B$. But if $a_i - b_{\mathcal{N}_2} \in a_l - B$, then $l \leq i$ by Observation 1. This proves (A.16). Next, we can use Observation 2 in an analogous way to prove that for any $j$,
$$\rho(a_{\mathcal{N}_1} - b_j) \leq j. \tag{A.17}$$

Now define the following subsets of $A - B$:
$$S' = \{a_i - b_{\mathcal{N}_2} : i = 1, \ldots, \mathcal{N}_2 - 1\} \tag{A.18a}$$
$$S'' = \{a_{\mathcal{N}_1} - b_j : j = 1, \ldots, \mathcal{N}_2 - 1\} \tag{A.18b}$$
$$S''' = \{a_k - b_{\mathcal{N}_2} : k = \mathcal{N}_2, \ldots, \mathcal{N}_1\} \tag{A.18c}$$

It follows from Observation 1 that $S'$ and $S''$ are disjoint, and from Observation 2 that $S''$ and $S'''$ are disjoint; trivially $S'$ and $S'''$ are disjoint. Therefore
$$\begin{aligned} U(A,B) &\geq \sum_{x \in S'} \frac{1}{\rho(x)} + \sum_{x \in S''} \frac{1}{\rho(x)} + \sum_{x \in S'''} \frac{1}{\rho(x)} \\ &\geq 2 \sum_{k=1}^{\mathcal{N}_2 - 1} \frac{1}{k} + \frac{\mathcal{N}_1 - \mathcal{N}_2 + 1}{\mathcal{N}_2}, \end{aligned} \tag{A.19}$$

where the last inequality uses (A.16), (A.17), and the trivial bound $\rho(x) \leq \mathcal{N}_2$. A simple comparison of Riemann sum to integral shows that $\sum_{k=1}^{n} k^{-1} \geq \log(n+1)$. ∎

## A.2 Application to SAWs

Now let us apply these bounds to the case in which $A$ and $B$ are independent SAWs $\omega^{(1)}, \omega^{(2)}$ of length $N$. We first note that that the spans $S_1, \ldots, S_d$ and the diameters $D$ and $D_\infty$ of either one of these SAWs are expected to scale like $N^\nu$ in the sense that
$$\langle S_1^{k_1} \cdots S_d^{k_d} D^\ell D_\infty^m \rangle \sim N^{(k_1 + \ldots + k_d + \ell + m)\nu} \tag{A.20}$$

for any exponents $k_1, \ldots, k_d, \ell, m \geq 0$. (In particular, this holds if the probability distribution of the SAW, with lengths rescaled by $N^\nu$, converges weakly to some probability measure on a space of continuum chains with respect to which the spans have finite nonzero moments.)

Assuming such scaling, it follows that $\Psi_N$ is bounded as $N \to \infty$. PROOF: By Theorem A.1 we have

$$T(\omega^{(1)}, \omega^{(2)}) \leq \prod_{\alpha=1}^{d} \left[ S_\alpha(\omega^{(1)}) + S_\alpha(\omega^{(2)}) + 1 \right] ; \qquad (A.21)$$

but from this and (A.20) it easily follows that

$$\frac{c_{N,N}}{c_N^2} = \langle T(\omega^{(1)}, \omega^{(2)}) \rangle_{N,N} \leq \text{const} \times N^{d\nu} . \qquad (A.22)$$

Q.E.D. Assuming the usual scaling (2.6) for the radius of gyration, it follows that $\Psi_N$ is bounded as $N \to \infty$.

Equation (A.20) also implies that

$$\langle T(\omega^{(1)}, \omega^{(2)}) \rangle_{N,N} \geq \text{const} \times N^{2\nu} \qquad (A.23)$$

in any dimension $d \geq 2$. This may be deduced from Theorem A.2 as follows. Since $\mathcal{S}_N$ (the set of all $N$-step SAWs which start at the origin) is invariant under lattice symmetries, we have

$$\begin{aligned}
\langle T(\omega^{(1)}, \omega^{(2)}) \rangle_{N,N} &= \frac{1}{c_N^2} \sum_{\omega^{(1)}, \omega^{(2)} \in \mathcal{S}_N} \#(\omega^{(1)} - \omega^{(2)}) \\
&= \frac{1}{c_N^2 \#(G_d)} \sum_{g \in G_d} \sum_{\omega^{(1)}, \omega^{(2)} \in \mathcal{S}_N} \#(\omega^{(1)} - g\,\omega^{(2)}) \\
&\geq \frac{1}{2 c_N^2} \sum_{\omega^{(1)}, \omega^{(2)} \in \mathcal{S}_N} \text{diam}_\infty(\omega^{(1)}) \, \text{diam}_\infty(\omega^{(2)}) \\
&= \frac{1}{2} \langle D_\infty \rangle_N^2 ,
\end{aligned} \qquad (A.24)$$

and combining this with (A.20) gives (A.23).

Lastly, we observe that the scaling assumption (A.20) implies that hyperscaling holds in two dimensions. Indeed, equations (A.22) and (A.23) together imply that

$$\langle T(\omega^{(1)}, \omega^{(2)}) \rangle_{N,N} \sim \text{const} \times N^{2\nu} \quad \text{when } d = 2. \qquad (A.25)$$

# B   Adequacy of Thermalization in the Pivot Algorithm

As pointed out in Section 2.2, the initialization of the pivot algorithm is a highly nontrivial issue. For not-too-large $N$ (up to a few thousand at least), one can use the dimerization algorithm [119,1,81] to produce a perfect equilibrium start. However, for very large $N$, dimerization is unfeasible, and it is necessary to "thermalize" the system by discarding the first $n_{disc} \gg \tau_{exp} \sim N/f$ iterations. But this is painful, because $\tau_{exp}$ is a factor $\sim N$ larger than $\tau_{int,A}$ for global observables $A$, and for very large $N$ ($\gtrsim 10^5$) the CPU time of the algorithm could end up being dominated by the thermalization. One is therefore tempted to cut corners in the choice of $n_{disc}$. In this appendix we want to illustrate how acquiescing to this temptation can lead to disaster in the form of large systematic errors.

Figure 11 shows the temporal history of our pivot-algorithm run for the 3-dimensional SAW at $N = 70000$. The initial configuration was a pair of parallel rods. We averaged the observables $R_e^2$ and $T$ over bins of $10^5$ pivot iterations each, and plotted the resulting averages for the first 40 bins. Clearly, severe initialization bias is present in $R_e^2$ until at least time $4 \times 10^5$, and in $T$ until at least time $8 \times 10^5$. (Here we used the following seat-of-the-pants criterion: if the value at bin $n$ is larger than all subsequent bins and smaller than all preceding bins, then severe initialization bias is present at time $n$.) Significant initialization bias — enough to cause a systematic error comparable to the (very small) statistical error — could well be present at times twice or three times this. So, to be safe, we discarded in this case the first $3 \times 10^6$ iterations. (The total run length was approximately $8 \times 10^6$ iterations, so a little less than half of the run was discarded.)

In general, we expect the thermalization to require a time proportional to $\tau_{exp} \sim N/f$, where $f$ is the pivot-algorithm acceptance fraction. From the run shown in Figure 11, we infer that severe initialization bias is present until at least time $\approx 3N/f$. Therefore, to be safe, we have always discarded at least $9N/f$ iterations (except of course for runs using a dimerized start). We believe that this rule is sufficiently conservative to render the systematic errors arising from inadequate thermalization much smaller than the statistical errors.

## C  Some Statistical Subtleties

Let $A_1, \ldots, A_n$ be the time series for some observable $A$ (in equilibrium). As discussed in [10, Appendix C], the error bar on the sample mean $\bar{A} \equiv n^{-1} \sum_{i=1}^n A_i$ is

$$\text{stddev}(\bar{A}) \approx \left( \frac{2\tau_{int,A} C_{AA}(0)}{n} \right)^{1/2}, \qquad (C.1)$$

where

$$C_{AA}(t) \equiv \langle A_s A_{s+t} \rangle - \langle A \rangle^2 \qquad (C.2)$$

$$\rho_{AA}(t) \equiv C_{AA}(t)/C_{AA}(0) \qquad (C.3)$$

$$\tau_{int,A} \equiv \frac{1}{2} \sum_{t=-\infty}^{\infty} \rho_{AA}(t) \qquad (C.4)$$

Since $\tau_{int,A}$ and $C_{AA}(0)$ are not known, they must be estimated from the time series: we define

$$\hat{\hat{C}}_{AA}(t) \equiv \frac{1}{n-|t|} \sum_{i=1}^{n-|t|} (A_i - \bar{A})(A_{i+|t|} - \bar{A}) \qquad (C.5)$$

$$\widehat{\widehat{\rho}}_{AA}(t) \equiv \widehat{\widehat{C}}_{AA}(t)/\widehat{\widehat{C}}_{AA}(0) \tag{C.6}$$

$$\widehat{\widehat{\tau}}_{int,A} \equiv \frac{1}{2} \sum_{t=-M}^{M} \widehat{\widehat{\rho}}_{AA}(t) \tag{C.7}$$

where we must still choose the "window" $M$.

One reasonable approach to choosing $M$ is the "automatic windowing algorithm" of [10, Appendix C]: choose $M$ to be the smallest integer such that $M \geq c\widehat{\widehat{\tau}}_{int,A}(M)$, for a suitable "window factor" $c$. If $\rho_{AA}(t)$ were roughly a pure exponential, then it would suffice to take e.g. $c \approx 5$ (since $e^{-5} < 1\%$). However, for *global* observables $A$ in the pivot algorithm, $\rho_{AA}(t)$ is expected to be *very* slowly decaying: after a brief initial decay, one expects $\rho_{AA}(t) \sim 1/t^{\approx 1}$ up until a time of order $\tau_{exp} \sim N/f$, after which time $\rho_{AA}(t)$ decays rapidly. This is the behavior exhibited by the exact solution of the pivot algorithm for *ordinary* random walk, where $\rho_{AA}(t) \sim 1/t$ in the intermediate region [10, Section 3.3]. We confirmed this behavior empirically for the *self-avoiding* walk by making an extremely long ($n = 10^8$ pivots) simulation at $N = 1000$ ($d = 2$): the sample autocorrelation functions $\widehat{\widehat{\rho}}_{AA}(t)$ for $A = R_e^2, R_g^2, T_{Karp-Luby,R=20}$ are shown in a *log-log* plot in Figure 12. For $R_e^2$, there is a wide intermediate region ($20 \lesssim t \lesssim 1000$) in which the log-log plot is roughly straight, yielding $\rho_{AA}(t) \sim 1/t^q$ with $q \approx 1.0$–1.2. For $R_g^2$ there is more curvature, but the slope $q$ is in the same ballpark. For $T_{Karp-Luby,R=20}$, in contrast, the curvature is so great that we are unable to identify an intermediate region with clear $1/t^q$ behavior. In any case, we have rough empirical confirmation for the theoretically predicted behavior; we are unable to say whether $q = 1$ or $q = 1 + p$ or something else, but clearly $q \approx 1$.

Because of this very slow decay, the automatic windowing algorithm leads to significant *underestimates* of $\tau_{int,A}$, even when the window factor $c$ is as large as 10 or 20. We therefore defined a modified estimator $\tilde{\tau}_{int,A}$ by extrapolating $\rho_{AA}(t)$ proportionally to $1/t$ in the region beyond the window, i.e. $\tilde{\tau}_{int,A} = M\widehat{\widehat{\rho}}_{AA}(M)/t$ for $t > M$, and cutting off the sum at a much larger time $M^*$ of order $\tau_{exp}$:

$$\tilde{\tau}_{int,A} = \frac{1}{2} \sum_{t=-M}^{M} \widehat{\widehat{\rho}}_{AA}(t) + M\widehat{\widehat{\rho}}_{AA}(M) \log(M^*/M) . \tag{C.8}$$

(Since $M, M^* \gg 1$, we approximated the second sum by an integral.) Here $M$ is defined, as before, by the automatic windowing algorithm. In the absence of any precise knowledge of $\tau_{exp}$, we took $M^* = N/f$; luckily $\tilde{\tau}_{int,A}$ is not very sensitive to the choice of $M^*$, because of the logarithmic dependence and because $\widehat{\widehat{\rho}}_{AA}(M) \ll 1$.

In Table 10 we show the results for a typical run ($d = 2$, $N = 80000$) for the observables $A = R_e^2, R_g^2, R_m^2, T_{Karp-Luby,R=150}$ as a function of the window factor $c$. Time is measured in units of 25 pivots. Note that:

(a) The standard windowing estimates $\widehat{\widehat{\tau}}_{int,A}$ are a factor 2–3 smaller than the modified estimates $\tilde{\tau}_{int,A}$, even at $c = 20$–30.

(b) The estimates $\hat{\tilde{\tau}}_{int,A}$ are slowly but relentlessly increasing as a function of $c$; it is perfectly plausible that they will double or triple by the time $c$ reaches infinity.

(c) The estimates $\tilde{\tau}_{int,A}$ are roughly stable as a function of $c$ as soon as $c \gtrsim 4$. However, they do show some fluctuation, because they are based on extrapolation from a single noisy point $\hat{\tilde{\rho}}_{AA}(M)$.

We therefore chose as our final estimate the average of $\tilde{\tau}_{int,A}$ for the eleven values $c = 20, 21, \ldots, 30$; by taking the average, we reduce the fluctuations mentioned in (c).

This whole procedure is, of course, inelegant and *ad hoc*. But it does work reasonably well: we expect that the estimates of $\tau_{int,A}$ are accurate to about 10%. This is not good enough for a serious study of dynamic critical behavior; but it is good enough for our present purpose, which is merely to set error bars on the static quantities $\bar{A}$. In the future we hope to devise better methods for analyzing time series with slow decay of the autocorrelation function.

# D  Remarks on the Field-Theoretic Estimates of Universal Amplitude Ratios

The critical exponents and universal amplitude ratios associated with polymers in a good solvent can be extracted from *any* family of theories which intersects transversally the domain of attraction (= stable manifold) of the good-solvent fixed point $H^*_{GS}$. *One* such family — which has no special status at $H^*_{GS}$, but is computationally convenient — is the continuum Edwards model (or what is equivalent, the $n = 0$ continuum $\varphi^4$ field theory) [12]. This model becomes critical (i.e. crosses the stable manifold of $H^*_{GS}$) when its bare coupling constant (self-avoidance parameter) $z$ tends to $\infty$.

There are two main approaches to the quantitative study of the continuum Edwards model in the limit $z \to \infty$:

- *Perturbation expansion in the coupling constant $z$ at fixed dimension $d = 3$ (or $d = 2$).* In this case the problem is to estimate the asymptotic behavior as $z \to \infty$ from the first few terms of a perturbation expansion around $z = 0$.

- *Expansion in $\epsilon = 4 - d$.* Here the critical exponents and limiting amplitude ratios, which correspond to the limit $z \to \infty$, can be obtained directly from a suitable renormalization-group analysis. The problem is then to extrapolate to $\epsilon = 1$ (or $\epsilon = 2$).

In this appendix we want to summarize the results for the universal amplitude ratios $\aleph \equiv 6\langle R_g^2\rangle/\langle R_e^2\rangle$, $\Psi^*$ and $\Psi^*_R$ which have been obtained by each of these methods, and to make some comments on their extrapolation.

1) *Perturbation expansion at fixed dimension $d = 3$.* Let $\alpha_R(z)$, $\alpha_S(z)$ and $h(z)$ be the conventional expansion and second virial factors of the continuum Edwards model. One then has

$$\aleph(z) = \alpha_S^2(z)/\alpha_R^2(z) \tag{D.1}$$

$$\Psi(z) = zh(z)/\alpha_S^d(z) \tag{D.2}$$

$$\Psi_R(z) = zh(z)/\alpha_R^d(z) \tag{D.3}$$

The known perturbation series in $d = 3$ are as follows [89,107,136,181,91]:

$$\aleph(z) = 1 - \frac{2}{35}z + \left(\frac{32}{105} - \frac{97\pi}{1296}\right)z^2$$
$$+ 0.056588 z^3 - 1.2202 z^4 + O(z^5) \tag{D.4}$$

$$\Psi(z) = z - \frac{1769 - 896\sqrt{2}}{105}z^2 + 25.5896415457 z^3 + O(z^4) \tag{D.5}$$

$$\Psi_R(z) = z - \frac{254 - 128\sqrt{2}}{15}z^2 + 26.1049923667 z^3 + O(z^4) \tag{D.6}$$

Crude estimates of $\aleph^* \equiv \aleph(\infty)$ can be obtained from the [1/1] and [2/2] Padé approximants

$$\aleph_{[1/1]}(z) = \frac{1 + \left(\frac{554}{105} - \frac{3395\pi}{2592}\right)z}{1 + \left(\frac{16}{3} - \frac{3395\pi}{2592}\right)z} \tag{D.7}$$

$$\aleph_{[2/2]}(z) = \frac{1 + 8.08283 z + 10.5132 z^2}{1 + 8.13998 z + 10.9087 z^2} \tag{D.8}$$

yielding $\aleph^*_{[1/1]} \approx 0.9531$ and $\aleph^*_{[2/2]} \approx 0.9637$. More sophisticated extrapolations have been performed by Shanes and Nickel [136], who find $\aleph^* = 0.9607 \pm 0.0018$. (Our Monte Carlo value is $0.9594 \pm 0.0012$.)

Likewise, crude estimates of $\Psi^* \equiv \Psi(\infty)$ and $\Psi_R^* \equiv \Psi_R(\infty)$ can be obtained from the [1/1] Padé approximants

$$\Psi_{[1/1]}(z) = \frac{z}{1 + \frac{1769 - 896\sqrt{2}}{105}z} \tag{D.9}$$

$$\Psi_{R,[1/1]}(z) = \frac{z}{1 + \frac{254 - 128\sqrt{2}}{15}z} \tag{D.10}$$

yielding $\Psi^*_{[1/1]} \approx 0.2092$ and $\Psi^*_{R,[1/1]} \approx 0.2055$. (Our Monte Carlo values are $0.2471 \pm 0.0003$ and $0.2322 \pm 0.0003$, respectively.)

We can also try the "direct renormalization" approach of des Cloizeaux, Conte and Jannink [90]: define the "effective exponent" $\sigma_R(z) \equiv (z/2) \, d\log \alpha_R^2(z)/dz$, which approaches the limiting value $\sigma_R^* = 2\nu - 1$ as $z \to \infty$; re-express $\Psi$ and $\Psi_R$ as functions of $\sigma_R$; and extrapolate these series to $\sigma_R = \sigma_R^*$, using the best estimate of $\nu$. If we carry out this last step by the most naive method imaginable — namely, straight evaluation of the cubic polynomial at $\sigma_R^* = 2(0.5877) - 1 = 0.1754$ — we obtain $\Psi^* = 0.2483$ and $\Psi_R^* = 0.2379$, which is not bad at all for such a short series.

2) *Expansion in $\epsilon = 4 - d$.* The limiting universal ratios $\aleph^*$, $\Psi^*$ and $\Psi_R^*$ can be evaluated in dimension $d = 4 - \epsilon$ in powers of $\epsilon$; the results are[44]:

$$\aleph^* = 1 - \frac{\epsilon}{96} - 0.030628\epsilon^2 + O(\epsilon^3) \tag{D.11}$$

$$\Psi^* = \frac{\epsilon}{8} + \frac{\epsilon^2}{16}\left(\frac{77}{48} + \log 2\right) + O(\epsilon^3) \tag{D.12}$$

$$\Psi_R^* = \frac{\epsilon}{8} + \frac{\epsilon^2}{16}\left(\frac{25}{16} + \log 2\right) + O(\epsilon^3) \tag{D.13}$$

Evaluating these at $\epsilon = 1, 2, 3$ yields

$$\aleph^* \approx 0.9590,\ 0.8567,\ 0.6931 \tag{D.14}$$
$$\Psi^* \approx 0.2686,\ 0.8243,\ 1.6672 \tag{D.15}$$
$$\Psi_R^* \approx 0.2660,\ 0.8139,\ 1.6438 \tag{D.16}$$

(We remark that the [1/1] Padé approximants for all three quantities have poles between $\epsilon = 0$ and $\epsilon = 1$, and so are unreliable for $\epsilon \geq 1$.)

Des Cloizeaux [182, footnote 7][45] has suggested to augment the $\epsilon$-expansions (D.11)–(D.13) by enforcing the known exact values at $d = 1$ ($\epsilon = 3$), which are $\aleph = 1/2$, $\Psi^* = 2/\pi^{1/2} \approx 1.1284$ [see (2.20)] and $\Psi_R^* = (2/\pi)^{1/2} \approx 0.7979$. This produces the cubic polynomials

$$\aleph^* = 1 - \frac{\epsilon}{96} - 0.030628\epsilon^2 - 0.007152\epsilon^3 \tag{D.17}$$

$$\Psi^* = \frac{\epsilon}{8} + \frac{\epsilon^2}{16}\left(\frac{77}{48} + \log 2\right) + \frac{\epsilon^3}{27}\left(\frac{2}{\pi^{1/2}} - \frac{327}{256} - \frac{9}{16}\log 2\right) \tag{D.18}$$

$$\Psi_R^* = \frac{\epsilon}{8} + \frac{\epsilon^2}{16}\left(\frac{25}{16} + \log 2\right) + \frac{\epsilon^3}{27}\left((2/\pi)^{1/2} - \frac{321}{256} - \frac{9}{16}\log 2\right) \tag{D.19}$$

Evaluating this at $\epsilon = 1, 2$ yields

$$\aleph^* \approx 0.9518,\ 0.7994 \tag{D.20}$$
$$\Psi^* \approx 0.2486,\ 0.6647 \tag{D.21}$$
$$\Psi_R^* \approx 0.2346,\ 0.5633 \tag{D.22}$$

---

[44]The expansion for $\aleph$ is from [153]; see also equation (13.1.60) of [12]. The expansion for $\Psi_R^*$ is from equation (5.35) of [182]; see also equations (12.3.102)/(13.1.11) of [12]. (In these works, $\Psi_R^*$ is called $g^*$. To establish the connection, compare equations (5.1.107) and (13.1.2) of [12].) The expansion for $\Psi^*$ can then be derived from these two. Alternatively, it can be found as the limit $\eta \to \infty$ of equation (4.15a) of [183].

[45]See also [12, pp. 541 and 557].

For $\aleph^*$ this modification has actually *worsened* the agreement with the Monte Carlo value, both in $d=3$ and in $d=2$. On the other hand, for $\Psi^*$ and $\Psi_R^*$ this "modified $\epsilon$-expansion" prediction is amazingly close to the correct value, both in $d=3$ and $d=2$. It would be useful to obtain a better understanding of whether this is a coincidence or not — perhaps by calculating the $O(\epsilon^3)$ term in $\Psi^*$.

# Acknowledgments

We wish to thank Jim Barrett, Sergio Caracciolo, Bertrand Duplantier, Michael Fisher, Peter Grassberger, Tony Guttmann, Bernie Nickel, George O'Brien, Enzo Orlandini, Andrea Pelissetto, Carla Tesi and Stu Whittington for many helpful conversations and correspondence. The computations reported here were carried out on various IBM RS-6000/320 and 320H, Sparcstation 1, Silicon Graphics "Crimson" and Convex C-210 machines at New York University.

The authors' research was supported in part by an operating grant from the Natural Sciences and Engineering Research Council of Canada (N.M.), U.S. Department of Energy contract DE-FG02-90ER40581 (A.D.S.), U.S. National Science Foundation grants DMS-8911273 and DMS-9200719 (A.D.S.), and by a New York University Research Challenge Fund grant (A.D.S.). Acknowledgment is also made to the donors of the Petroleum Research Fund, administered by the American Chemical Society, for partial support of this research under grants 21091–AC7 and 25553–AC7B–C (A.D.S.). During part of the period of research, B.L. was a American Chemical Society–Petroleum Research Fund Fellow.

# References

[1] N. Madras and G. Slade, *The Self-Avoiding Walk* (Birkhäuser, Boston–Basel–Berlin, 1993).

[2] P.G. deGennes, Phys. Lett. **A38**, 339 (1972).

[3] J. des Cloizeaux, J. Physique **36**, 281 (1975).

[4] M. Daoud, J.P. Cotton, B. Farnoux, G. Jannink, G. Sarma, H. Benoit, R. Duplessix, C. Picot and P.G. de Gennes, Macromolecules **8**, 804 (1975).

[5] V.J. Emery, Phys. Rev. **B11**, 239 (1975).

[6] C. Aragão de Carvalho, S. Caracciolo and J. Fröhlich, Nucl. Phys. **B215**[FS7], 209 (1983).

[7] R. Fernández, J. Fröhlich and A.D. Sokal, *Random Walks, Critical Phenomena, and Triviality in Quantum Field Theory* (Springer-Verlag, Berlin–Heidelberg–New York, 1992).

[8] M. Lal, Molec. Phys. **17**, 57 (1969).

[9] B. MacDonald, N. Jan, D.L. Hunter and M.O. Steinitz, J. Phys. A: Math. Gen. **18**, 2627 (1985).

[10] N. Madras and A.D. Sokal, J. Stat. Phys. **50**, 109 (1988).

[11] K.F. Freed, *Renormalization Group Theory of Macromolecules* (Wiley, New York, 1987).

[12] J. des Cloizeaux and G. Jannink, *Polymers in Solution: Their Modelling and Structure* (Oxford University Press, New York, 1990).

[13] B.G. Nickel, Macromolecules **24**, 1358 (1991).

[14] A.D. Sokal, Static scaling behavior of high-molecular-weight polymers in dilute solution: A reexamination, NYU preprint NYU–TH–93/05/01 (May 1993), hep-lat@ftp.scri.fsu.edu #9305009, rejected five times by Phys. Rev. Lett. A slightly abridged and revised version of this paper will appear in Europhys. Lett.

[15] A.D. Sokal, Fundamental problems in the static scaling behavior of high-molecular-weight polymers in dilute solution I. Critique of two-parameter theories, in preparation.

[16] B. Widom, J. Chem. Phys. **43**, 3892 (1965).

[17] M.E. Fisher, Rep. Prog. Phys. **30**, 615 (1967).

[18] G. Stell, J. Chem. Phys. **51**, 2037 (1969).

[19] N.S. Snider, J. Chem. Phys. **54**, 4587 (1971).

[20] B. Widom, Physica **73**, 107 (1974).

[21] M.E. Fisher, in *Collective Properties of Physical Systems* ($24^{th}$ Nobel Symposium), ed. B. Lundqvist and S. Lundqvist (Academic Press, New York–London, 1974).

[22] C.K. Hall, J. Stat. Phys. **13**, 157 (1975).

[23] T. Hara and G. Slade, Commun. Math. Phys. **147**, 101 (1992).

[24] T. Hara and G. Slade, Reviews in Math. Phys. **4**, 235 (1992).

[25] T. Hara and G. Slade, Commun. Math. Phys. **128**, 333 (1990).

[26] T. Hara, Prob. Th. and Rel. Fields **86**, 337 (1990).

[27] W.J. Camp, D.M. Saul, J.P. Van Dyke and M. Wortis, Phys. Rev. **B14**, 3990 (1976).

[28] G.A. Baker Jr., Phys. Rev. **B15**, 1552 (1977).

[29] B.G. Nickel and B. Sharpe, J. Phys. **A12**, 1819 (1979).

[30] J.J. Rehr, J. Phys. **A12**, L179 (1979).

[31] J. Zinn-Justin, J. Physique **40**, 969 (1979).

[32] B. Nickel, Physica **106A**, 48 (1981).

[33] B.G. Nickel, in *Phase Transitions* (1980 Cargèse lectures), ed. M. Lévy, J.-C. LeGuillou and J. Zinn-Justin (Plenum, New York–London, 1982).

[34] R. Roskies, Phys. Rev. **B23**, 6037 (1981).

[35] R.Z. Roskies, Phys. Rev. **B24**, 5305 (1981).

[36] J. Zinn-Justin, J. Physique **42**, 783 (1981).

[37] G.A. Baker Jr. and J.M. Kincaid, J. Stat. Phys. **24**, 469 (1981).

[38] J.-H. Chen, M.E. Fisher and B.G. Nickel, Phys. Rev. Lett. **48**, 630 (1982).

[39] J. Adler, M. Moshe and V. Privman, Phys. Rev. **B26**, 3958 (1982).

[40] M. Ferer and M.J. Velgakis, Phys. Rev. **B27**, 2839 (1983).

[41] M.E. Fisher and J.-H. Chen, J. Physique **46**, 1645 (1985).

[42] A.J. Guttmann, Phys. Rev. **B33**, 5089 (1986).

[43] R. Schrader and E. Tränkle, J. Stat. Phys. **25**, 269 (1981).

[44] B.A. Freedman and G.A. Baker Jr., J. Phys. **A15**, L715 (1982).

[45] M.N. Barber, R.B. Pearson, D. Toussaint and J.L. Richardson, Phys. Rev. **B32**, 1720 (1985).

[46] K. Binder, M. Nauenberg, V. Privman and A.P. Young, Phys. Rev. **B31**, 1498 (1985).

[47] A. Hoogland, A. Compagner and H.W.J. Blöte, Physica **A132**, 593 (1985).

[48] J. Glimm and A. Jaffe, Ann. Inst. Henri Poincaré **A22**, 97 (1975).

[49] R. Schrader, Phys. Rev. **B14**, 172 (1976).

[50] R. Schrader, Commun. Math. Phys. **49**, 131 (1976).

[51] A.D. Sokal, Ann. Inst. Henri Poincaré **A37**, 317 (1982).

[52] G.A. Baker Jr., in *Phase Transitions and Critical Phenomena*, Vol. 9, ed. C. Domb and J.L. Lebowitz (Academic Press, London, 1984).

[53] M.E. Fisher, in *Renormalization Group in Critical Phenomena and Quantum Fields*, ed. J.D. Gunton and M.S. Green (Temple University, Philadelphia 1974).

[54] F.J. Wegner and E.K. Riedel, Phys. Rev. **B7**, 248 (1973).

[55] S.-K. Ma, *Modern Theory of Critical Phenomena*, (Benjamin, Reading MA, 1976).

[56] D.J. Amit and L. Peliti, Ann. Phys. **140**, 207 (1982).

[57] M.E. Fisher, in *Critical Phenomena (Stellenbosch 1982)*, Lecture Notes in Physics #186, ed. F.J.W. Hahne (Springer-Verlag, Berlin–Heidelberg–New York, 1983), pp. 1–139.

[58] A.C.D. van Enter, R. Fernández and A.D. Sokal, J. Stat. Phys. **72**, 879 (1993).

[59] K. Gawedzki and A. Kupiainen, Commun. Math. Phys. **99**, 197 (1985).

[60] K. Gawedzki and A. Kupiainen, Nucl. Phys. **B257**[FS14], 474 (1985).

[61] K. Gawedzki and A. Kupiainen, Commun. Math. Phys. **102**, 1 (1985).

[62] J. Feldman, J. Magnen, V. Rivasseau and R. Sénéor, Commun. Math. Phys. **103**, 67 (1986).

[63] K. Gawedzki and A. Kupiainen, Commun. Math. Phys. **89**, 191 (1983).

[64] K. Gawedzki and A. Kupiainen, J. Stat. Phys. **35**, 267 (1984).

[65] K. Gawedzki and A. Kupiainen, Commun. Math. Phys. **106**, 533 (1986).

[66] G. Felder, Commun. Math. Phys. **102**, 139 (1985).

[67] T. Niemeijer and J.M.J. van Leuuwen, in *Phase Transitions and Critical Phenomena*, Vol. 6, ed. C. Domb and M. S. Green (Academic Press, London–New York–San Francisco, 1976).

[68] R.H. Swendsen, in *Phase Transitions* (1980 Cargèse lectures), ed. M. Lévy, J.-C. LeGuillou and J. Zinn-Justin (Plenum, New York–London, 1982).

[69] E. Brézin, J. C. Le Guillou and J. Zinn-Justin, in *Phase Transitions and Critical Phenomena*, Vol. 6, ed. C. Domb and M. S. Green (Academic Press, London–New York–San Francisco, 1976).

[70] D.S. Gaunt and A.J. Guttmann, in *Phase Transitions and Critical Phenomena*, Vol. 3, ed. C. Domb and M.S. Green (Academic Press, London, 1974).

[71] A.J. Guttmann, in *Phase Transitions and Critical Phenomena*, Vol. 13, ed. C. Domb and J.L. Lebowitz (Academic Press, London, 1989).

[72] J. Adler, M. Moshe and V. Privman, in *Percolation Structures and Processes*, ed. G. Deutscher, R. Zallen and J. Adler (Israel Physical Society, 1983).

[73] V. Privman, J. Phys. **A16**, 3097 (1983).

[74] M.N. Barber, in *Phase Transitions and Critical Phenomena*, Vol. 8, ed. C. Domb and J.L. Lebowitz (Academic Press, London, 1983).

[75] J.L. Cardy, ed., *Finite-Size Scaling* (North-Holland, Amsterdam, 1988).

[76] V. Privman, ed., *Finite Size Scaling and Numerical Simulation of Statistical Systems* (World Scientific, Singapore, 1990).

[77] A.D. Sokal, *Monte Carlo Methods in Statistical Mechanics: Foundations and New Algorithms*, Cours de Troisième Cycle de la Physique en Suisse Romande (Lausanne, June 1989).

[78] A.D. Sokal, Nucl. Phys. B (Proc. Suppl.) **20**, 55 (1991).

[79] A.D. Sokal, in *Quantum Fields on the Computer*, ed. M. Creutz (World Scientific, Singapore, 1992).

[80] S. Caracciolo, R.G. Edwards, S.J. Ferreira, A. Pelissetto and A.D. Sokal, Finite-size scaling at $\xi/L \gg 1$, in preparation.

[81] A.D. Sokal, in *Monte Carlo and Molecular Dynamics Simulations in Polymer Science*, ed. K. Binder (Oxford University Press, New York, to appear 1994).

[82] A. Baumgärtner and K. Binder, J. Chem. Phys. **71**, 2541 (1979).

[83] S.F. Edwards, Proc. Phys. Soc. London **85**, 613 (1965).

[84] S.R.S. Varadhan, appendix to article of K. Symanzik, in *Local Quantum Theory*, ed. R. Jost (Academic Press, New York–London, 1969).

[85] J. Westwater, Commun. Math. Phys. **72**, 131 (1980).

[86] J. Westwater, Commun. Math. Phys. **84**, 459 (1982).

[87] A. Bovier, G. Felder and J. Fröhlich, Nucl. Phys. **B230**[FS10], 119 (1984).

[88] V. Privman, P.C. Hohenberg and A. Aharony, in *Phase Transitions and Critical Phenomena*, Vol. 14, ed. C. Domb and J.L. Lebowitz (Academic Press, London–San Diego, 1991).

[89] M. Muthukumar and B.G. Nickel, J. Chem. Phys. **80**, 5839 (1984).

[90] J. des Cloizeaux, R. Conte and G. Jannink, J. Physique Lett. **46**, L-595 (1985).

[91] M. Muthukumar and B.G. Nickel, J. Chem. Phys. **86**, 460 (1987).

[92] A.J. Barrett and B.G. Nickel, private communication.

[93] H. Fujita and T. Norisuye, Macromolecules **18**, 1637 (1985).

[94] K. Huber and W.H. Stockmayer, Macromolecules **20**, 1400 (1987).

[95] H. Fujita, Macromolecules **21**, 179 (1988).

[96] H. Fujita, *Polymer Solutions* (Elsevier, Amsterdam, 1990).

[97] A.J. Liu and M.E. Fisher, J. Stat. Phys. **58**, 431 (1990).

[98] T. Hara, G. Slade and A.D. Sokal, J. Stat. Phys. **72**, 479 (1993).

[99] G. Slade, Commun. Math. Phys. **110**, 661 (1987).

[100] G. Slade, Ann. Probab. **17**, 91 (1989).

[101] G. Slade, J. Phys. A: Math. Gen. **21**, L417 (1988).

[102] G.E. Uhlenbeck and G.W. Ford, in *Studies in Statistical Mechanics*, Vol. 1, ed. J. de Boer and G.E. Uhlenbeck (North-Holland, Amsterdam, 1962).

[103] J. des Cloizeaux, private communication cited in E. Brézin, in *Order and Fluctuation in Equilibrium and Nonequilibrium Statistical Mechanics* (17th Solvay Conference, 1978), ed. G. Nichols, G. Dewel and J. W. Turner (Wiley-Interscience, New York, 1981).

[104] J.C. LeGuillou and J. Zinn-Justin, Phys. Rev. **B21**, 3976 (1980).

[105] J.C. LeGuillou and J. Zinn-Justin, J. Physique Lett. **46**, L-137 (1985).

[106] J.C. LeGuillou and J. Zinn-Justin, J. Physique **50**, 1365 (1989).

[107] D.B. Murray and B.G. Nickel, Revised estimates for critical exponents for the continuum $n$-vector model in 3 dimensions, University of Guelph preprint (1991).

[108] B. Nienhuis, Phys. Rev. Lett. **49**, 1062 (1982).

[109] B. Nienhuis, J. Stat. Phys. **34**, 731 (1984).

[110] D.S. McKenzie and C. Domb, Proc. Phys. Soc. (London) **92**, 632 (1967).

[111] M. Aizenman, Commun. Math. Phys. **86**, 1 (1982).

[112] F.J. Wegner, Phys. Rev. **B5**, 4529 (1972).

[113] H.E. Stanley, *Introduction to Phase Transitions and Critical Phenomena* (Oxford University Press, Oxford, 1971).

[114] N. Madras, A. Orlitsky and L.A. Shepp, J. Stat. Phys. **58**, 159 (1990).

[115] G. Zifferer, Macromolecules **23**, 3166 (1990).

[116] N. Eizenberg and J. Klafter, J. Chem. Phys. **99**, 3976 (1993).

[117] D.E. Knuth, *The Art of Computer Programming*, Vol. 3 (Addison-Wesley, Reading, Massachusetts, 1973), Section 6.4.

[118] T.H. Cormen, C.E. Leiserson and R.L. Rivest, *Introduction to Algorithms* (MIT Press/McGraw-Hill, Cambridge MA/New York, 1990), Chapter 12.

[119] K. Suzuki, Bull. Chem. Soc. Japan **41**, 538 (1968).

[120] G. Zifferer, Molec. Simul. **6**, 103 (1991).

[121] S. Redner and P.J. Reynolds, J. Phys. **A14**, 2679 (1981).

[122] A. Berretti and A.D. Sokal, J. Stat. Phys. **40**, 483 (1985).

[123] A.J. Barrett and B.G. Nickel, private communication. A precursor of this algorithm can be found in A.J. Barrett, Macromolecules **18**, 196 (1985), Section 3.

[124] R.M. Karp and M. Luby, in $24^{th}$ *IEEE Symposium on Foundations of Computer Science* (IEEE, New York, 1983), pp. 56–64.

[125] R.M. Karp, M. Luby and N. Madras, J. Algorithms **10**, 429 (1989).

[126] S.D. Silvey, *Statistical Inference* (Chapman and Hall, London, 1975).

[127] S. Caracciolo, A.J. Guttmann, B. Li, A. Pelissetto and A.D. Sokal, Correction-to-scaling exponents for two-dimensional self-avoiding walks, in preparation.

[128] D.C. Rapaport, J. Phys. **A18**, L39 (1985).

[129] S. Caracciolo, A. Pelissetto and A.D. Sokal, **A23**, L969 (1990).

[130] A.J. Barrett, M. Mansfield and B.C. Benesch, Macromolecules **24**, 1615 (1991).

[131] J.L. Cardy and H. Saleur, J. Phys. **A22**, L601 (1989).

[132] A.J. Guttmann, S. Merrilees and A.D. Sokal, unpublished (1985).

[133] D.C. Rapaport, J. Phys. **A18**, 113 (1985).

[134] J. Dayantis and J.-F. Palierne, J. Chem. Phys. **95**, 6088 (1991).

[135] L.A. Johnson, A. Monge and R.A. Friesner, J. Chem. Phys. **97**, 9355 (1992).

[136] F. Shanes and B.G. Nickel, Calculation of the radius of gyration for a linear flexible polymer chain with excluded volume interaction, J. Chem. Phys. (to appear).

[137] J. Dayantis and J.-F. Palierne, Phys. Rev. **B49**, 3217 (1994).

[138] A.J. Guttmann, J. Phys. **A22**, 2807 (1989).

[139] B.G. Nickel, Physica **A177**, 189 (1991).

[140] K.E. Newman and E.K. Riedel, Phys. Rev. **B30**, 6615 (1984).

[141] A.D. Sokal, Fundamental problems in the static scaling behavior of high-molecular-weight polymers in dilute solution II. Critical review of the experimental literature, in preparation.

[142] A. Yamamoto, M. Fujii, G. Tanaka and H. Yamakawa, Polym. J. **2**, 799 (1971).

[143] M. Fukuda, M. Fukutomi, Y. Kato and T. Hashimoto, J. Polym. Sci.: Polym. Phys. Ed. **12**, 871 (1974).

[144] Y. Miyaki, Y. Einaga and H. Fujita, Macromolecules **11**, 1180 (1978).

[145] J.-C. LeGuillou and J. Zinn-Justin, Phys. Rev. Lett. **39**, 95 (1977).

[146] J.P. Cotton, J. Physique Lett. **41**, L-231 (1980).

[147] H. Utiyama, S. Utsumi, Y. Tsunashima and M. Kurata, Macromolecules **11**, 506 (1978).

[148] B. Appelt and G. Meyerhoff, Macromolecules **13**, 657 (1980).

[149] H.R. Haller, C. Destor and D.S. Cannell, Rev. Sci. Instrum. **54**, 973 (1983).

[150] B. Chu, R. Xu, T. Maeda and H.S. Dhadwal, Rev. Sci. Instrum. **59**, 716 (1988).

[151] K.B. Strawbridge, F.R. Hallett and J. Watton, Can. J. Appl. Spectroscopy **36**, 53 (1991).

[152] A.D. Sokal, Optimal statistical analysis of static light-scattering data from dilute polymer solutions, in preparation.

[153] M. Benhamou and G. Mahoux, J. Physique Lett. **46**, L-689 (1985).

[154] C. Domb and F.T. Hioe, J. Chem. Phys. **51**, 1915 (1969).

[155] M. van Prooyen and B.G. Nickel, The second virial coefficient for self-avoiding walks on a lattice, in preparation.

[156] P.J. Flory, *Principles of Polymer Chemistry* (Cornell University Press, Ithaca, NY, 1953).

[157] H. Yamakawa, *Modern Theory of Polymer Solutions* (Harper and Row, New York, 1971).

[158] P.G. DeGennes, *Scaling Concepts in Polymer Physics* (Cornell Univ. Press, Ithaca, NY, 1979).

[159] C. Domb and A.J. Barrett, Polymer **17**, 179 (1976).

[160] J.F. Douglas and K.F. Freed, Macromolecules **18**, 201 (1985).

[161] J.F. Douglas and K.F. Freed, J. Phys. Chem. **88**, 6613 (1984).

[162] Z.Y. Chen and J. Noolandi, J. Chem. Phys. **96**, 1540 (1992).

[163] Z.Y. Chen and J. Noolandi, Macromolecules **25**, 4978 (1992).

[164] B. Krüger and L. Schäfer, Long polymer chains in good solvent: Beyond the universal limit, Universität Essen preprint (late 1993).

[165] L. Schäfer, On the sign of correction to scaling amplitudes: Field theoretic considerations and results for self repelling walks, Universität Essen preprint (1994).

[166] C. Bagnuls and C. Bervillier, Phys. Rev. **B41**, 402 (1990).

[167] K.G. Wilson and J. Kogut, Phys. Reports **12C**, 75 (1974).

[168] K. Gawędzki and A. Kupiainen, in *Critical Phenomena, Random Systems, Gauge Theories [Les Houches 1984], Part I*, ed. K. Osterwalder and R. Stora (North-Holland, Amsterdam, 1986), pp. 185–293.

[169] J. Polchinski, Nucl. Phys. **B231**, 269 (1984).

[170] J. Hughes and J. Liu, Nucl. Phys. **B307**, 183 (1988).

[171] S. Weinberg, Phys. Rev. **D8**, 3497 (1973).

[172] J.C. Collins and A.J. Macfarlane, Phys. Rev. **D10**, 1201 (1974).

[173] S.W. MacDowell, Phys. Rev. **D12**, 1089 (1975).

[174] B. Duplantier, J. Physique **43**, 991 (1982).

[175] B. Duplantier, J. Chem. Phys. **86**, 4233 (1987).

[176] B. Duplantier, Phys. Rev. **A38**, 3647 (1988).

[177] C. Domb and G.S. Joyce, J. Phys. **C5**, 956 (1972).

[178] B. Duplantier and H. Saleur, Phys. Rev. Lett. **59**, 539 (1987).

[179] S. Caracciolo, G. Ferraro, A. Pelissetto and A.D. Sokal, work in progress.

[180] E. Orlandini, M.C. Tesi and S.G. Whittington, private communication.

[181] G. Tanaka and K. Šolc, Macromolecules **15**, 791 (1982).

[182] J. des Cloizeaux, J. Physique **42**, 635 (1981).

[183] J.F. Douglas and K.F. Freed, Macromolecules **17**, 1854 (1984).

| $N$ | $\langle T\rangle$ | $\langle T^2\rangle/10^4$ | $\text{var}(T)/10^4$ | $\langle U\rangle$ | $\langle V_{Barrett}\rangle/10^4$ | $\langle V_{Karp-Luby}\rangle/10^4$ |
|---|---|---|---|---|---|---|
| 100 | 1373.53 (0.48) | 193.95 | 5.29 | 339.07 | 157.22 | 483.94 ( 0.28) |
| 200 | 3854.69 (1.39) | 1529.10 | 43.23 | 723.10 | 1435.55 | 4236.24 ( 2.58) |
| 300 | 7065.32 (2.68) | 5139.55 | 147.68 | 1124.57 | 5196.87 | 15025.26 ( 9.79) |
| 400 | 10860.24 (4.11) | 12146.26 | 351.79 | 1535.59 | 12898.02 | 36803.23 ( 24.10) |
| 500 | 15152.67 (8.32) | 23639.57 | 679.22 | 1950.72 | 26002.92 | 73527.83 ( 70.29) |

Table 1: Quantities relevant to the Barrett and Karp-Luby algorithms, as a function of $N_1 = N_2 = N$, for 2-dimensional self-avoiding walks. Error bars are one standard deviation.

| $N$ | $\langle T\rangle$ | $\langle T^2\rangle/10^4$ | $\text{var}(T)/10^4$ | $\langle U\rangle$ | $\langle V_{Barrett}\rangle/10^4$ | $\langle V_{Karp-Luby}\rangle/10^4$ |
|---|---|---|---|---|---|---|
| 100 | 3346.95 ( 0.93) | 1136.81 ( 0.63) | 16.60 ( 0.88) | 1719.60 ( 0.75) | 617.35 ( 0.15) | 2337.71 ( 0.90) |
| 200 | 11159.79 ( 7.25) | 12646.28 ( 16.27) | 192.20 ( 22.94) | 5149.61 ( 5.25) | 8158.67 ( 5.25) | 28739.30 ( 26.18) |
| 500 | 55272.12 (17.35) | 310214.61 ( 193.40) | 4713.83 ( 272.40) | 22089.93 (11.11) | 244244.80 ( 87.33) | 795935.08 ( 364.63) |
| 800 | 125986.30 (40.59) | 1611900.23 (1029.67) | 24645.50 (1451.35) | 46787.65 (24.22) | 1389999.88 ( 532.61) | 4381808.49 (2079.36) |
| 1000 | 186508.58 (61.41) | 3532249.52 (2309.75) | 53704.50 (3253.13) | 66896.98 (35.56) | 3170834.30 (1271.38) | 9855248.60 (4819.17) |

Table 2: Quantities relevant to the Barrett and Karp-Luby algorithms, as a function of $N_1 = N_2 = N$, for 3-dimensional self-avoiding walks. Error bars are one standard deviation.

| $N$ | $\langle R_e^2 \rangle$ | | $\langle R_g^2 \rangle$ | | $\langle R_m^2 \rangle$ | | $\langle T \rangle$ | | $f$ |
|---|---|---|---|---|---|---|---|---|---|
| 100 | 777.690 $\pm$ | 0.757 | 109.142 $\pm$ | 0.060 | — | | 1373.53 $\pm$ | 0.48 | — |
| 200 | 2194.000 $\pm$ | 2.180 | 307.451 $\pm$ | 0.174 | — | | 3854.69 $\pm$ | 1.39 | — |
| 300 | 4017.710 $\pm$ | 4.179 | 563.694 $\pm$ | 0.334 | — | | 7065.32 $\pm$ | 2.68 | — |
| 400 | 6185.440 $\pm$ | 6.388 | 867.673 $\pm$ | 0.512 | — | | 10860.24 $\pm$ | 4.11 | — |
| 500 | 8621.527 $\pm$ | 10.642 | 1210.549 $\pm$ | 0.948 | — | | 15154.26 $\pm$ | 7.85$^a$ | — |
| 1000 | 24459.300 $\pm$ | 42.838 | 3425.590 $\pm$ | 5.623 | 10742.500 $\pm$ | 17.119 | 42853.78 $\pm$ | 57.45 | 0.25378 $\pm$ 0.00024 |
| 1500 | 44774.300 $\pm$ | 85.796 | 6282.270 $\pm$ | 11.785 | 19684.600 $\pm$ | 34.762 | 78573.80 $\pm$ | 115.01 | 0.23452 $\pm$ 0.00024 |
| 2000 | 68929.400 $\pm$ | 131.083 | 9653.390 $\pm$ | 18.207 | 30289.000 $\pm$ | 53.735 | 120773.27 $\pm$ | 182.27 | 0.22140 $\pm$ 0.00023 |
| 2500 | 96755.100 $\pm$ | 216.912 | 13559.200 $\pm$ | 30.191 | — | | 169394.08 $\pm$ | 303.43$^a$ | — |
| 3000 | 127104.073 $\pm$ | 176.229 | 17802.697 $\pm$ | 23.913 | 55851.100 $\pm$ | 88.597 | 222357.79 $\pm$ | 246.92$^a$ | 0.20509 $\pm$ 0.00019 |
| 3500 | 159980.000 $\pm$ | 390.518 | 22420.700 $\pm$ | 52.365 | — | | 280117.27 $\pm$ | 555.52$^a$ | — |
| 4000 | 194741.264 $\pm$ | 327.675 | 27332.984 $\pm$ | 46.167 | 85730.297 $\pm$ | 178.520 | 341511.22 $\pm$ | 447.94$^a$ | 0.19376 $\pm$ 0.00023 |
| 4500 | 232822.000 $\pm$ | 662.439 | 32693.300 $\pm$ | 84.724 | — | | 408602.67 $\pm$ | 848.48$^a$ | — |
| 5000 | 272893.000 $\pm$ | 709.164 | 38287.900 $\pm$ | 106.209 | 119932.000 $\pm$ | 303.951 | 478110.68 $\pm$ | 996.75$^a$ | — |
| 5500 | 315118.000 $\pm$ | 821.224 | 44140.800 $\pm$ | 116.423 | — | | 551118.77 $\pm$ | 1220.07$^a$ | — |
| 6000 | 359467.102 $\pm$ | 613.861 | 50437.817 $\pm$ | 84.619 | 158072.674 $\pm$ | 255.803 | 630667.93 $\pm$ | 852.57$^a$ | 0.17953 $\pm$ 0.00021 |
| 6500 | 404321.000 $\pm$ | 1028.130 | 56654.600 $\pm$ | 149.773 | 177615.000 $\pm$ | 443.270 | 708214.98 $\pm$ | 1548.77$^a$ | — |
| 7000 | 450254.000 $\pm$ | 1217.260 | 63233.900 $\pm$ | 187.088 | — | | 790171.67 $\pm$ | 1863.89$^a$ | — |
| 7200 | 471499.566 $\pm$ | 762.481 | 66130.359 $\pm$ | 104.382 | 207146.970 $\pm$ | 379.611 | 825946.29 $\pm$ | 1038.17$^a$ | 0.17309 $\pm$ 0.00018 |
| 7500 | 500172.000 $\pm$ | 1396.910 | 70329.000 $\pm$ | 207.664 | 220086.000 $\pm$ | 587.985 | 878189.91 $\pm$ | 2013.06$^a$ | — |
| 8000 | 551978.094 $\pm$ | 926.915 | 77426.100 $\pm$ | 127.431 | 242583.184 $\pm$ | 384.604 | 967214.84 $\pm$ | 1288.57$^a$ | 0.16939 $\pm$ 0.00017 |
| 8250 | 577862.000 $\pm$ | 1508.860 | 80968.200 $\pm$ | 206.938 | 253902.000 $\pm$ | 602.582 | 1012461.00 $\pm$ | 2061.87$^a$ | — |
| 8400 | 593186.000 $\pm$ | 1532.640 | 83241.100 $\pm$ | 213.843 | 260645.000 $\pm$ | 621.496 | 1040352.40 $\pm$ | 2145.18$^a$ | — |
| 8650 | 620944.000 $\pm$ | 1835.070 | 87038.900 $\pm$ | 249.425 | 272775.000 $\pm$ | 753.115 | 1087701.30 $\pm$ | 2552.58$^a$ | — |
| 9000 | 657926.882 $\pm$ | 1129.894 | 92273.834 $\pm$ | 161.439 | 289167.124 $\pm$ | 464.199 | 1152906.64 $\pm$ | 1583.77$^a$ | 0.16559 $\pm$ 0.00017 |
| 9200 | 677758.000 $\pm$ | 1865.980 | 95064.200 $\pm$ | 270.506 | 298003.000 $\pm$ | 756.903 | 1189077.20 $\pm$ | 2590.80$^a$ | — |
| 9500 | 710439.000 $\pm$ | 2295.450 | 99690.100 $\pm$ | 316.909 | 312471.000 $\pm$ | 877.201 | 1247906.30 $\pm$ | 3197.34$^a$ | — |
| 9650 | 728607.000 $\pm$ | 1967.000 | 102546.000 $\pm$ | 253.146 | 320619.000 $\pm$ | 781.385 | 1279259.90 $\pm$ | 2697.09$^a$ | — |
| 9700 | 736619.000 $\pm$ | 2076.780 | 103406.000 $\pm$ | 304.886 | 324039.000 $\pm$ | 859.093 | 1293086.50 $\pm$ | 2902.13$^a$ | — |
| 9750 | 738974.000 $\pm$ | 1968.260 | 103711.000 $\pm$ | 281.340 | 325021.000 $\pm$ | 810.977 | 1297868.30 $\pm$ | 2753.30$^a$ | — |
| 9850 | 755190.000 $\pm$ | 2200.680 | 106059.000 $\pm$ | 272.353 | 331942.000 $\pm$ | 854.863 | 1322669.90 $\pm$ | 2677.55$^a$ | — |
| 9900 | 760424.333 $\pm$ | 1121.483 | 106678.709 $\pm$ | 159.471 | 334263.606 $\pm$ | 467.136 | 1332241.41 $\pm$ | 1605.57$^a$ | — |
| 10000 | 770061.308 $\pm$ | 1294.861 | 108002.619 $\pm$ | 175.686 | 338560.337 $\pm$ | 522.242 | 1350156.71 $\pm$ | 1853.36$^a$ | 0.16247 $\pm$ 0.00016 |
| 15000 | 1418675.444 $\pm$ | 1916.596 | 198953.381 $\pm$ | 265.417 | 624522.711 $\pm$ | 1018.853 | 2486038.05 $\pm$ | 2606.19 | 0.15029 $\pm$ 0.00011 |
| 20000 | 2177520.000 $\pm$ | 5305.390 | 305098.000 $\pm$ | 771.573 | 957108.000 $\pm$ | 2249.350 | 3817040.60 $\pm$ | 7883.18 | 0.14216 $\pm$ 0.00019 |
| 40000 | 6177080.000 $\pm$ | 17206.500 | 864495.000 $\pm$ | 2482.060 | 2712030.000 $\pm$ | 7031.920 | 10807802.00 $\pm$ | 23097.90 | 0.12451 $\pm$ 0.00019 |
| 60000 | 11352100.000 $\pm$ | 31723.800 | 1595490.000 $\pm$ | 4517.620 | 4991160.000 $\pm$ | 13044.200 | 19893345.00 $\pm$ | 48018.50 | 0.11573 $\pm$ 0.00016 |
| 80000 | 17390000.000 $\pm$ | 42700.700 | 2438490.000 $\pm$ | 6045.320 | 7642600.000 $\pm$ | 17427.300 | 30476824.00 $\pm$ | 58801.80 | 0.10912 $\pm$ 0.00013 |

Table 3: The results of our runs in dimension $d = 2$. Errors are $\pm$ one standard deviation. $^a$ indicates a possible minor bug in measurement; see text.

| $N$ | $\langle R_g^2 \rangle / \langle R_e^2 \rangle$ | $\langle R_m^2 \rangle / \langle R_e^2 \rangle$ | $\Psi$ |
|---:|---|---|---|
| 100 | $0.14034 \pm 0.00021$ | — | $0.66764 \pm 0.00060$ |
| 200 | $0.14013 \pm 0.00022$ | — | $0.66514 \pm 0.00062$ |
| 300 | $0.14030 \pm 0.00023$ | — | $0.66495 \pm 0.00065$ |
| 400 | $0.14028 \pm 0.00023$ | — | $0.66402 \pm 0.00064$ |
| 500 | $0.14041 \pm 0.00028$ | — | $0.66413 \pm 0.00086^a$ |
| 1000 | $0.14005 \pm 0.00048$ | $0.43920 \pm 0.00147$ | $0.66367 \pm 0.00198$ |
| 1500 | $0.14031 \pm 0.00053$ | $0.43964 \pm 0.00162$ | $0.66353 \pm 0.00222$ |
| 2000 | $0.14005 \pm 0.00053$ | $0.43942 \pm 0.00162$ | $0.66373 \pm 0.00225$ |
| 2500 | $0.14014 \pm 0.00063$ | — | $0.66277 \pm 0.00266^a$ |
| 3000 | $0.14006 \pm 0.00038$ | $0.43941 \pm 0.00131$ | $0.66262 \pm 0.00163^a$ |
| 3500 | $0.14015 \pm 0.00067$ | — | $0.66281 \pm 0.00286^a$ |
| 4000 | $0.14036 \pm 0.00047$ | $0.44023 \pm 0.00166$ | $0.66285 \pm 0.00199^a$ |
| 4500 | $0.14042 \pm 0.00076$ | — | $0.66304 \pm 0.00310^a$ |
| 5000 | $0.14030 \pm 0.00075$ | $0.43948 \pm 0.00226$ | $0.66247 \pm 0.00322^a$ |
| 5500 | $0.14008 \pm 0.00073$ | — | $0.66237 \pm 0.00321^a$ |
| 6000 | $0.14031 \pm 0.00048$ | $0.43974 \pm 0.00146$ | $0.66335 \pm 0.00201^a$ |
| 6500 | $0.14012 \pm 0.00073$ | $0.43929 \pm 0.00221$ | $0.66318 \pm 0.00320^a$ |
| 7000 | $0.14044 \pm 0.00080$ | — | $0.66293 \pm 0.00353^a$ |
| 7200 | $0.14026 \pm 0.00045$ | $0.43934 \pm 0.00152$ | $0.66260 \pm 0.00188^a$ |
| 7500 | $0.14061 \pm 0.00081$ | $0.44002 \pm 0.00240$ | $0.66245 \pm 0.00347^a$ |
| 8000 | $0.14027 \pm 0.00047$ | $0.43948 \pm 0.00143$ | $0.66273 \pm 0.00197^a$ |
| 8250 | $0.14012 \pm 0.00072$ | $0.43938 \pm 0.00219$ | $0.66338 \pm 0.00305^a$ |
| 8400 | $0.14033 \pm 0.00072$ | $0.43940 \pm 0.00218$ | $0.66304 \pm 0.00307^a$ |
| 8650 | $0.14017 \pm 0.00082$ | $0.43929 \pm 0.00251$ | $0.66297 \pm 0.00346^a$ |
| 9000 | $0.14025 \pm 0.00049$ | $0.43951 \pm 0.00146$ | $0.66285 \pm 0.00207^a$ |
| 9200 | $0.14026 \pm 0.00079$ | $0.43969 \pm 0.00233$ | $0.66358 \pm 0.00333^a$ |
| 9500 | $0.14032 \pm 0.00090$ | $0.43983 \pm 0.00266$ | $0.66409 \pm 0.00381^a$ |
| 9650 | $0.14074 \pm 0.00073$ | $0.44004 \pm 0.00226$ | $0.66182 \pm 0.00303^a$ |
| 9700 | $0.14038 \pm 0.00081$ | $0.43990 \pm 0.00241$ | $0.66341 \pm 0.00344^a$ |
| 9750 | $0.14034 \pm 0.00075$ | $0.43983 \pm 0.00227$ | $0.66390 \pm 0.00321^a$ |
| 9850 | $0.14044 \pm 0.00077$ | $0.43955 \pm 0.00241$ | $0.66161 \pm 0.00304^a$ |
| 9900 | $0.14029 \pm 0.00042$ | $0.43958 \pm 0.00126$ | $0.66253 \pm 0.00179^a$ |
| 10000 | $0.14025 \pm 0.00046$ | $0.43965 \pm 0.00142$ | $0.66321 \pm 0.00199^a$ |
| 15000 | $0.14024 \pm 0.00038$ | $0.44022 \pm 0.00131$ | $0.66291 \pm 0.00158$ |
| 20000 | $0.14011 \pm 0.00070$ | $0.43954 \pm 0.00210$ | $0.66372 \pm 0.00305$ |
| 40000 | $0.13995 \pm 0.00079$ | $0.43905 \pm 0.00236$ | $0.66324 \pm 0.00332$ |
| 60000 | $0.14055 \pm 0.00079$ | $0.43967 \pm 0.00238$ | $0.66147 \pm 0.00347$ |
| 80000 | $0.14022 \pm 0.00069$ | $0.43948 \pm 0.00208$ | $0.66305 \pm 0.00292$ |

Table 4: Universal amplitude ratios in dimension $d = 2$. Errors are ± one standard deviation, based on triangle inequality. [a] indicates a possible minor bug in measurement; see text.

| $N$ | $\langle R_e^2 \rangle$ | | $\langle R_g^2 \rangle$ | | $\langle T \rangle$ | | $f$ |
|---|---|---|---|---|---|---|---|
| 100 | $265.053 \pm$ | $0.169$ | $41.912 \pm$ | $0.017$ | $3347.31 \pm$ | $0.62$ | $0.60873 \pm 0.00012$ |
| 150 | $429.084 \pm$ | $0.392$ | $67.954 \pm$ | $0.047$ | $6756.65 \pm$ | $2.70$ | $0.58224 \pm 0.00008$ |
| 200 | $603.999 \pm$ | $0.732$ | $95.750 \pm$ | $0.069$ | $11148.05 \pm$ | $3.40$ | — |
| 300 | $976.094 \pm$ | $0.793$ | $154.937 \pm$ | $0.076$ | $22613.93 \pm$ | $4.75$ | — |
| 400 | $1372.160 \pm$ | $2.156$ | $217.745 \pm$ | $0.204$ | $37357.58 \pm$ | $14.70$ | — |
| 500 | $1786.826 \pm$ | $1.266$ | $283.943 \pm$ | $0.131$ | $55254.94 \pm$ | $11.95$ | $0.51016 \pm 0.00012$ |
| 600 | $2217.770 \pm$ | $2.568$ | $352.598 \pm$ | $0.243$ | $76076.21 \pm$ | $21.87$ | — |
| 700 | $2661.450 \pm$ | $5.081$ | $423.525 \pm$ | $0.676$ | $99752.01 \pm$ | $126.77^a$ | — |
| 800 | $3113.250 \pm$ | $3.530$ | $495.619 \pm$ | $0.346$ | $125986.30 \pm$ | $40.59$ | — |
| 1000 | $4053.528 \pm$ | $3.937$ | $645.529 \pm$ | $0.413$ | $186509.18 \pm$ | $59.65^a$ | — |
| 1200 | $5023.820 \pm$ | $9.959$ | $801.123 \pm$ | $1.321$ | $256749.19 \pm$ | $333.41^a$ | — |
| 1500 | $6545.540 \pm$ | $12.712$ | $1042.710 \pm$ | $1.896$ | $379923.40 \pm$ | $529.85$ | — |
| 1800 | $8092.940 \pm$ | $15.878$ | $1290.860 \pm$ | $2.302$ | $523305.41 \pm$ | $726.03^a$ | — |
| 2000 | $9170.600 \pm$ | $21.899$ | $1461.760 \pm$ | $3.060$ | $629686.03 \pm$ | $1061.54^a$ | — |
| 2500 | $11926.100 \pm$ | $23.465$ | $1900.610 \pm$ | $3.589$ | $931639.59 \pm$ | $1357.98^a$ | — |
| 2800 | $13632.500 \pm$ | $29.044$ | $2173.670 \pm$ | $4.100$ | $1136711.90 \pm$ | $1612.15^a$ | — |
| 3000 | $14791.272 \pm$ | $16.145$ | $2361.192 \pm$ | $2.402$ | $1285631.38 \pm$ | $1029.93^a$ | $0.41981 \pm 0.00011$ |
| 3250 | $16289.800 \pm$ | $28.272$ | $2596.720 \pm$ | $3.918$ | $1481692.70 \pm$ | $1778.01^a$ | — |
| 3500 | $17762.000 \pm$ | $37.294$ | $2836.320 \pm$ | $5.573$ | $1686417.60 \pm$ | $2668.40^a$ | — |
| 3800 | $19526.200 \pm$ | $41.537$ | $3119.110 \pm$ | $6.092$ | $1946853.70 \pm$ | $3045.89^a$ | — |
| 4500 | $23857.422 \pm$ | $38.095$ | $3812.571 \pm$ | $5.218$ | $2626477.19 \pm$ | $2858.05^a$ | — |
| 5000 | $27006.559 \pm$ | $29.538$ | $4315.483 \pm$ | $4.431$ | $3159180.21 \pm$ | $2662.07^a$ | $0.39731 \pm 0.00011$ |
| 6000 | $33467.400 \pm$ | $43.910$ | $5341.140 \pm$ | $6.595$ | $4354063.40 \pm$ | $4277.04$ | $0.38938 \pm 0.00010$ |
| 7000 | $40124.100 \pm$ | $51.919$ | $6407.520 \pm$ | $7.676$ | $5706645.60 \pm$ | $5789.41$ | — |
| 8000 | $46951.672 \pm$ | $57.753$ | $7496.853 \pm$ | $8.063$ | $7217286.50 \pm$ | $6540.09^a$ | — |
| 10000 | $61128.600 \pm$ | $88.087$ | $9749.630 \pm$ | $12.848$ | $10702551.00 \pm$ | $12094.20$ | $0.36854 \pm 0.00011$ |
| 13000 | $83031.100 \pm$ | $122.458$ | $13264.800 \pm$ | $19.003$ | $16973895.00 \pm$ | $19439.20$ | $0.35820 \pm 0.00011$ |
| 15000 | $98538.647 \pm$ | $104.138$ | $15743.182 \pm$ | $16.204$ | $21865346.33 \pm$ | $18392.31$ | $0.35308 \pm 0.00008$ |
| 23000 | $162265.000 \pm$ | $254.103$ | $25955.600 \pm$ | $38.810$ | $46375397.00 \pm$ | $59208.20$ | $0.33728 \pm 0.00011$ |
| 25000 | $179897.000 \pm$ | $269.833$ | $28727.500 \pm$ | $41.726$ | $53815101.00 \pm$ | $67927.40$ | $0.33454 \pm 0.00011$ |
| 30000 | $222088.000 \pm$ | $342.061$ | $35490.300 \pm$ | $54.269$ | $74087837.00 \pm$ | $97497.60$ | $0.32767 \pm 0.00011$ |
| 35000 | $266714.000 \pm$ | $432.848$ | $42626.800 \pm$ | $65.234$ | $97208448.00 \pm$ | $119367.00$ | $0.32255 \pm 0.00010$ |
| 40000 | $311740.821 \pm$ | $274.332$ | $49801.999 \pm$ | $41.934$ | $122939664.03 \pm$ | $87490.24$ | $0.31793 \pm 0.00006$ |
| 70000 | $602779.000 \pm$ | $1304.950$ | $96255.700 \pm$ | $196.856$ | $330049626.00 \pm$ | $550259.00$ | $0.29983 \pm 0.00013$ |
| 80000 | $704651.925 \pm$ | $599.289$ | $112663.799 \pm$ | $89.456$ | $417196845.49 \pm$ | $279956.55$ | $0.29561 \pm 0.00005$ |

Table 5: The results of our runs in dimension $d = 3$. Errors are $\pm$ one standard deviation. $^a$ indicates a possible minor bug in measurement; see text.

| $N$ | $\langle R_g^2 \rangle / \langle R_e^2 \rangle$ | $\Psi$ |
|---:|---|---|
| 100 | $0.15813 \pm 0.00016$ | $0.27693 \pm 0.00022$ |
| 150 | $0.15837 \pm 0.00025$ | $0.27077 \pm 0.00039$ |
| 200 | $0.15853 \pm 0.00031$ | $0.26710 \pm 0.00037$ |
| 300 | $0.15873 \pm 0.00021$ | $0.26323 \pm 0.00025$ |
| 400 | $0.15869 \pm 0.00040$ | $0.26100 \pm 0.00047$ |
| 500 | $0.15891 \pm 0.00019$ | $0.25924 \pm 0.00024$ |
| 600 | $0.15899 \pm 0.00029$ | $0.25794 \pm 0.00034$ |
| 700 | $0.15913 \pm 0.00056$ | $0.25691 \pm 0.00094^a$ |
| 800 | $0.15920 \pm 0.00029$ | $0.25632 \pm 0.00035$ |
| 1000 | $0.15925 \pm 0.00026$ | $0.25528 \pm 0.00033^a$ |
| 1200 | $0.15946 \pm 0.00058$ | $0.25418 \pm 0.00096^a$ |
| 1500 | $0.15930 \pm 0.00060$ | $0.25330 \pm 0.00104$ |
| 1800 | $0.15950 \pm 0.00060$ | $0.25329 \pm 0.00103^a$ |
| 2000 | $0.15940 \pm 0.00071$ | $0.25293 \pm 0.00122^a$ |
| 2500 | $0.15937 \pm 0.00061$ | $0.25240 \pm 0.00108^a$ |
| 2800 | $0.15945 \pm 0.00064$ | $0.25179 \pm 0.00107^a$ |
| 3000 | $0.15963 \pm 0.00034$ | $0.25154 \pm 0.00059^a$ |
| 3250 | $0.15941 \pm 0.00052$ | $0.25137 \pm 0.00087^a$ |
| 3500 | $0.15968 \pm 0.00065$ | $0.25062 \pm 0.00114^a$ |
| 3800 | $0.15974 \pm 0.00065$ | $0.25088 \pm 0.00113^a$ |
| 4500 | $0.15981 \pm 0.00047$ | $0.25046 \pm 0.00079^a$ |
| 5000 | $0.15979 \pm 0.00034$ | $0.25016 \pm 0.00060^a$ |
| 6000 | $0.15959 \pm 0.00041$ | $0.25040 \pm 0.00071$ |
| 7000 | $0.15969 \pm 0.00040$ | $0.24976 \pm 0.00070$ |
| 8000 | $0.15967 \pm 0.00037$ | $0.24960 \pm 0.00063^a$ |
| 10000 | $0.15949 \pm 0.00044$ | $0.24957 \pm 0.00078$ |
| 13000 | $0.15976 \pm 0.00046$ | $0.24941 \pm 0.00082$ |
| 15000 | $0.15977 \pm 0.00033$ | $0.24849 \pm 0.00059$ |
| 23000 | $0.15996 \pm 0.00049$ | $0.24896 \pm 0.00088$ |
| 25000 | $0.15969 \pm 0.00047$ | $0.24811 \pm 0.00085$ |
| 30000 | $0.15980 \pm 0.00049$ | $0.24875 \pm 0.00090$ |
| 35000 | $0.15982 \pm 0.00050$ | $0.24795 \pm 0.00087$ |
| 40000 | $0.15975 \pm 0.00028$ | $0.24832 \pm 0.00049$ |
| 70000 | $0.15969 \pm 0.00067$ | $0.24810 \pm 0.00117$ |
| 80000 | $0.15989 \pm 0.00026$ | $0.24766 \pm 0.00046$ |

Table 6: Universal amplitude ratios in dimension $d = 3$. Errors are $\pm$ one standard deviation, based on triangle inequality. $^a$ indicates a possible minor bug in measurement; see text.

| $N_{min}$ | $a$ | $b$ | $\Delta$ | $\chi^2$ |
|---|---|---|---|---|
| 100 | $0.16006 \pm 0.00027$ | $-0.00084 \pm 0.00032$ | $0.370 \pm 0.123$ | 2.20 (32 DF, level $\approx$ 100%) |
| 200 | $0.15996 \pm 0.00026$ | $-0.00072 \pm 0.00032$ | $0.466 \pm 0.227$ | 1.93 (30 DF, level $\approx$ 100%) |
| 300 | $0.15992 \pm 0.00024$ | $-0.00067 \pm 0.00031$ | $0.526 \pm 0.288$ | 1.78 (29 DF, level $\approx$ 100%) |
| 500 | $0.15985 \pm 0.00022$ | $-0.00059 \pm 0.00030$ | $0.685 \pm 0.498$ | 1.38 (27 DF, level $\approx$ 100%) |
| 1000 | $0.15986 \pm 0.00029$ | $-0.00059 \pm 0.00032$ | $0.668 \pm 0.950$ | 1.36 (23 DF, level $\approx$ 100%) |

Table 7: Fits $\langle R_g^2 \rangle / \langle R_e^2 \rangle = a + b(N/1000)^{-\Delta}$ with $a, b, \Delta$ all variable, for 3-dimensional SAWs. The true error bars are probably $\approx 1/3$ of those indicated here (see text).

| $N_{min}$ | $\Psi^* = a$ | $b$ | $\Delta$ | $\chi^2$ |
|---|---|---|---|---|
| 100 | $0.24710 \pm 0.00027$ | $0.00819 \pm 0.00029$ | $0.561 \pm 0.013$ | 3.67 (32 DF, level $\approx$ 100%) |
| 200 | $0.24705 \pm 0.00034$ | $0.00826 \pm 0.00041$ | $0.556 \pm 0.026$ | 3.57 (30 DF, level $\approx$ 100%) |
| 300 | $0.24714 \pm 0.00036$ | $0.00813 \pm 0.00046$ | $0.569 \pm 0.035$ | 3.22 (29 DF, level $\approx$ 100%) |
| 500 | $0.24717 \pm 0.00045$ | $0.00808 \pm 0.00056$ | $0.573 \pm 0.057$ | 3.06 (27 DF, level $\approx$ 100%) |
| 1000 | $0.24711 \pm 0.00063$ | $0.00815 \pm 0.00062$ | $0.562 \pm 0.105$ | 2.92 (23 DF, level $\approx$ 100%) |
| 1500 | $0.24715 \pm 0.00083$ | $0.00822 \pm 0.00089$ | $0.574 \pm 0.201$ | 2.82 (21 DF, level $\approx$ 100%) |

Table 8: Fits $\Psi = a + b(N/1000)^{-\Delta}$ with $a, b, \Delta$ all variable, for 3-dimensional SAWs. The true error bars are probably $\approx 1/3$ of those indicated here (see text).

|  | $d=1$ | $d=2$ | $d=3$ | $d=4$ |
|---|---|---|---|---|
| $\Psi_{hard-sphere}$ | 1.1284 | 4/3 | 1.6186 | 2 |
| $\Psi_{N=1}$ | 0.9772 | 0.7427 | 0.6585 | 0.6755 |
| $\Psi_{N=2}$ | 0.9974 | 0.7048 | 0.5000 |  |
| $\Psi_{N=3}$ | 1.0197 | 0.7004 | 0.4508 |  |
| $\Psi_{N=4}$ | 1.0365 | 0.6938 | 0.4216 |  |
| $\Psi_{N=5}$ | 1.0490 | 0.6918 | 0.4033 |  |
| $\Psi_{N=6}$ | 1.0586 | 0.6891 | 0.3884 |  |
| $\Psi_{N=7}$ | 1.0662 | 0.6882 | 0.3777 |  |
| $\vdots$ |  |  |  |  |
| $\Psi^*$ | 1.1284 | 0.6630 | 0.2471 | 0 |
| $\Psi^* / \Psi_{hard-sphere}$ | 1 | 0.4972 | 0.1527 | 0 |

Table 9: $\Psi_N$ for short chains (rounded to four decimal places) from exact enumerations [132], along with our Monte Carlo values for $\Psi^*$.

|  | $A = R_e^2$ | | $A = R_g^2$ | | $A = R_m^2$ | | $A = T_{K-L,R=150}$ | |
|---|---|---|---|---|---|---|---|---|
| $c$ | $\hat{\hat{\tau}}_{int,R_e^2}$ | $\widetilde{\tau}_{int,R_e^2}$ | $\hat{\hat{\tau}}_{int,R_g^2}$ | $\widetilde{\tau}_{int,R_g^2}$ | $\hat{\hat{\tau}}_{int,R_m^2}$ | $\widetilde{\tau}_{int,R_m^2}$ | $\hat{\hat{\tau}}_{int,T}$ | $\widetilde{\tau}_{int,T}$ |
| 2 | 1.198 | 5.213 | 1.873 | 11.247 | 1.455 | 8.128 | 0.914 | 5.227 |
| 3 | 1.270 | 4.667 | 2.287 | 11.465 | 1.771 | 7.413 | 1.168 | 6.564 |
| 4 | 1.363 | 4.166 | 2.641 | 11.523 | 1.898 | 7.153 | 1.350 | 7.161 |
| 5 | 1.394 | 3.837 | 2.889 | 11.489 | 1.997 | 7.117 | 1.492 | 7.470 |
| 6 | 1.447 | 3.914 | 3.074 | 11.562 | 2.108 | 6.806 | 1.613 | 7.905 |
| 7 | 1.487 | 3.609 | 3.236 | 11.918 | 2.194 | 6.547 | 1.713 | 7.925 |
| 8 | 1.518 | 3.602 | 3.374 | 12.221 | 2.241 | 6.419 | 1.851 | 8.640 |
| 9 | 1.532 | 3.526 | 3.523 | 12.023 | 2.313 | 7.317 | 1.955 | 7.801 |
| 10 | 1.555 | 3.380 | 3.648 | 12.255 | 2.374 | 7.144 | 2.051 | 8.661 |
| 11 | 1.575 | 3.257 | 3.755 | 11.394 | 2.433 | 7.482 | 2.135 | 8.936 |
| 12 | 1.600 | 4.291 | 3.833 | 11.911 | 2.484 | 7.606 | 2.215 | 9.243 |
| 13 | 1.612 | 4.265 | 3.918 | 11.300 | 2.530 | 7.276 | 2.280 | 8.321 |
| 14 | 1.632 | 3.793 | 3.999 | 12.654 | 2.571 | 7.237 | 2.341 | 8.508 |
| 15 | 1.651 | 4.066 | 4.092 | 12.356 | 2.621 | 7.376 | 2.423 | 9.482 |
| 16 | 1.672 | 4.428 | 4.168 | 13.123 | 2.653 | 6.712 | 2.476 | 8.792 |
| 17 | 1.690 | 4.196 | 4.232 | 11.731 | 2.684 | 7.210 | 2.528 | 8.719 |
| 18 | 1.708 | 4.113 | 4.303 | 12.489 | 2.710 | 6.342 | 2.591 | 9.608 |
| 19 | 1.723 | 4.005 | 4.384 | 15.178 | 2.734 | 7.033 | 2.653 | 9.815 |
| 20 | 1.736 | 4.045 | 4.446 | 13.261 | 2.770 | 7.752 | 2.691 | 9.078 |
| 21 | 1.750 | 4.531 | 4.511 | 11.689 | 2.798 | 7.535 | 2.741 | 9.663 |
| 22 | 1.763 | 4.260 | 4.545 | 9.002 | 2.828 | 7.803 | 2.785 | 7.961 |
| 23 | 1.776 | 4.432 | 4.587 | 10.681 | 2.853 | 7.397 | 2.824 | 10.257 |
| 24 | 1.787 | 3.839 | 4.623 | 12.226 | 2.883 | 6.922 | 2.873 | 9.043 |
| 25 | 1.799 | 4.198 | 4.664 | 11.623 | 2.897 | 6.365 | 2.917 | 10.258 |
| 26 | 1.813 | 3.530 | 4.708 | 12.612 | 2.913 | 7.024 | 2.958 | 9.786 |
| 27 | 1.820 | 3.389 | 4.738 | 8.981 | 2.940 | 7.839 | 2.991 | 7.904 |
| 28 | 1.827 | 4.013 | 4.761 | 11.195 | 2.963 | 8.450 | 3.024 | 9.305 |
| 29 | 1.837 | 4.626 | 4.783 | 8.318 | 2.993 | 8.515 | 3.060 | 11.170 |
| 30 | 1.845 | 4.474 | 4.806 | 11.310 | 3.020 | 9.014 | 3.119 | 13.602 |

Table 10: Standard windowing estimate $\hat{\hat{\tau}}_{int,A}$ and modified estimate $\widetilde{\tau}_{int,A}$ for the observables $A = R_e^2, R_g^2, R_m^2, T_{Karp-Luby,R=150}$, as a function of window factor $c$, for SAWs on the square lattice at $N = 80000$. Time is measured in units of 25 pivots.

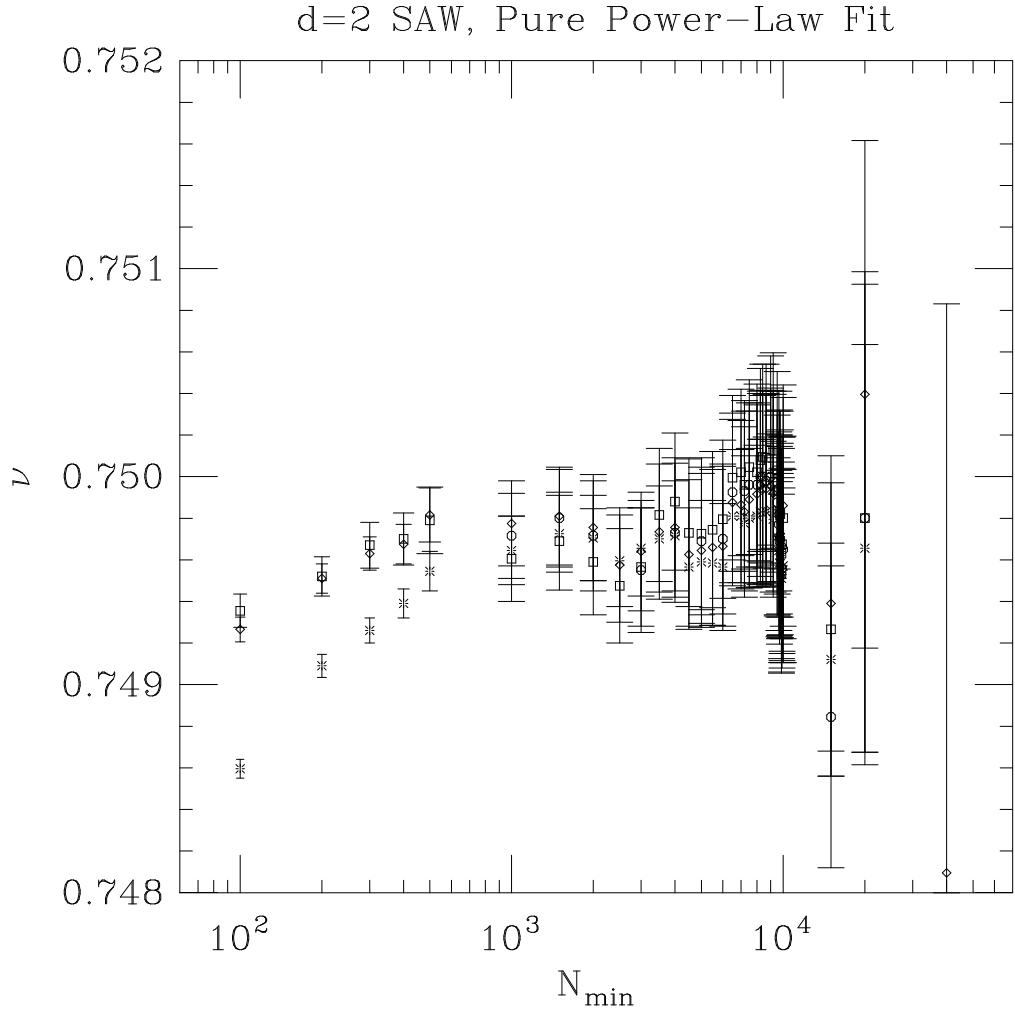

Figure 1: Estimated exponent $\nu$ from pure power-law fits to $\langle R_e \rangle^2$ ($\square$), $\langle R_g \rangle^2$ ($\diamond$) and $\langle R_m \rangle^2$ ($\circ$), and estimated exponent $(2\Delta_4 - \gamma)/2$ from pure power-law fit to $\langle T \rangle$ ($*$), plotted versus $N_{min}$. Note the good agreement with the believed exact value $\nu = 3/4$ and with the hyperscaling relation $d\nu = 2\Delta_4 - \gamma$, as soon as $N_{min} \gtrsim 1000$.

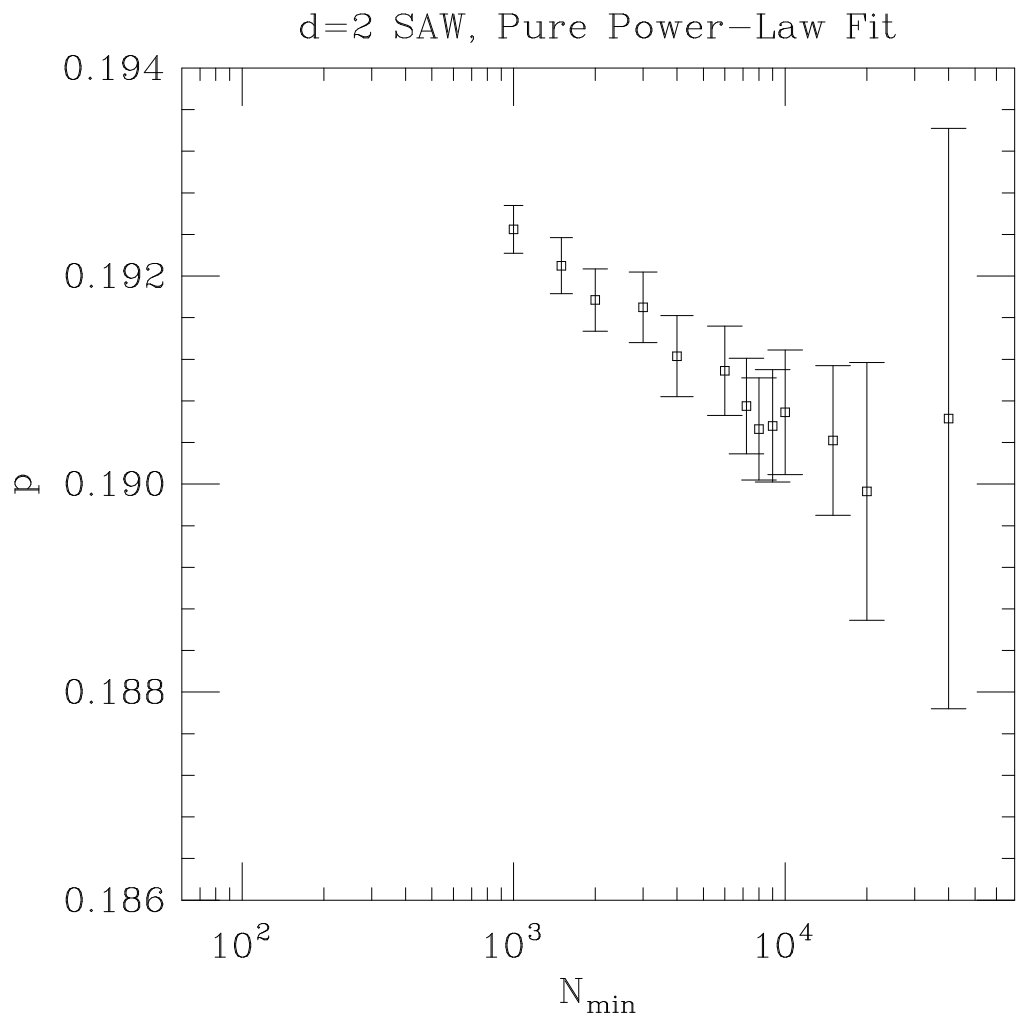

Figure 2: Estimated exponent $p$ from pure power-law fit to pivot-algorithm acceptance fraction $f$, plotted versus $N_{min}$.

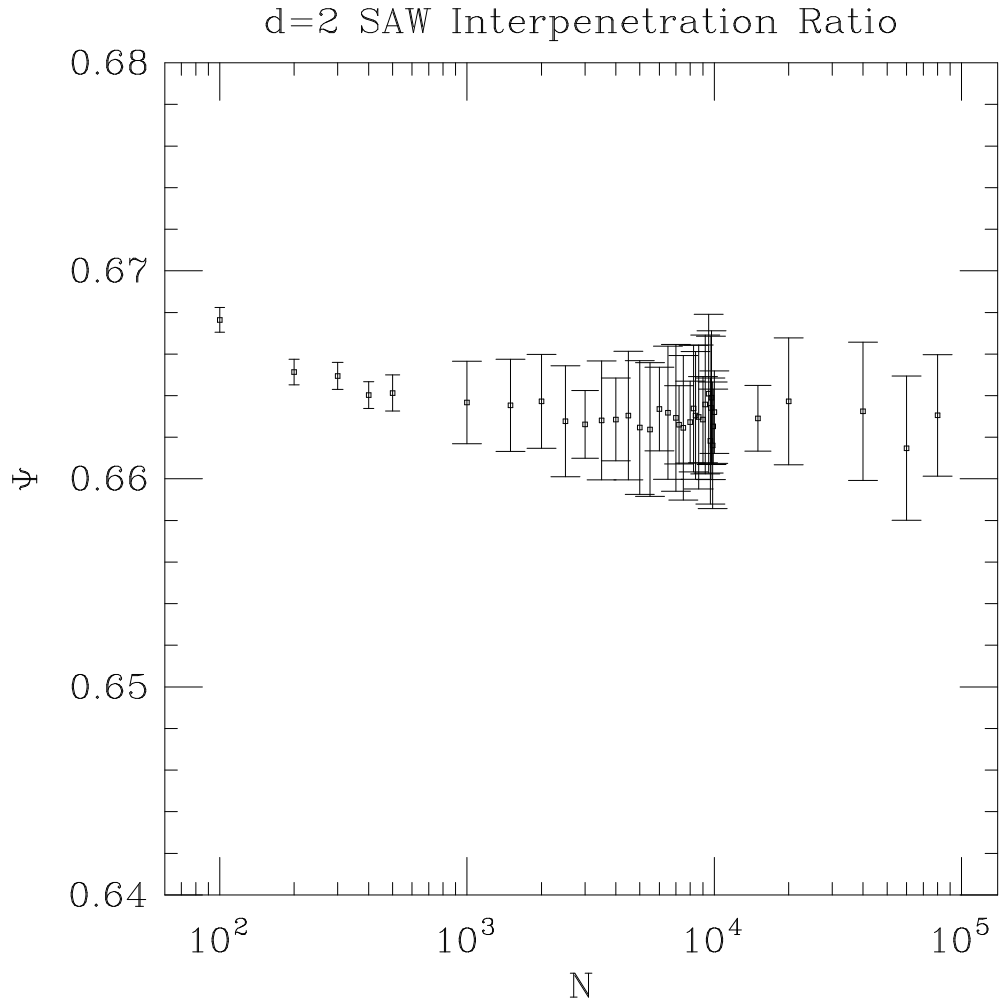

Figure 3: Interpenetration ratio $\Psi$ versus $N$, for 2-dimensional SAWs. Note that $\Psi$ varies little for $N \gtrsim 100$, and is constant within error bars for $N \gtrsim 1000$.
78

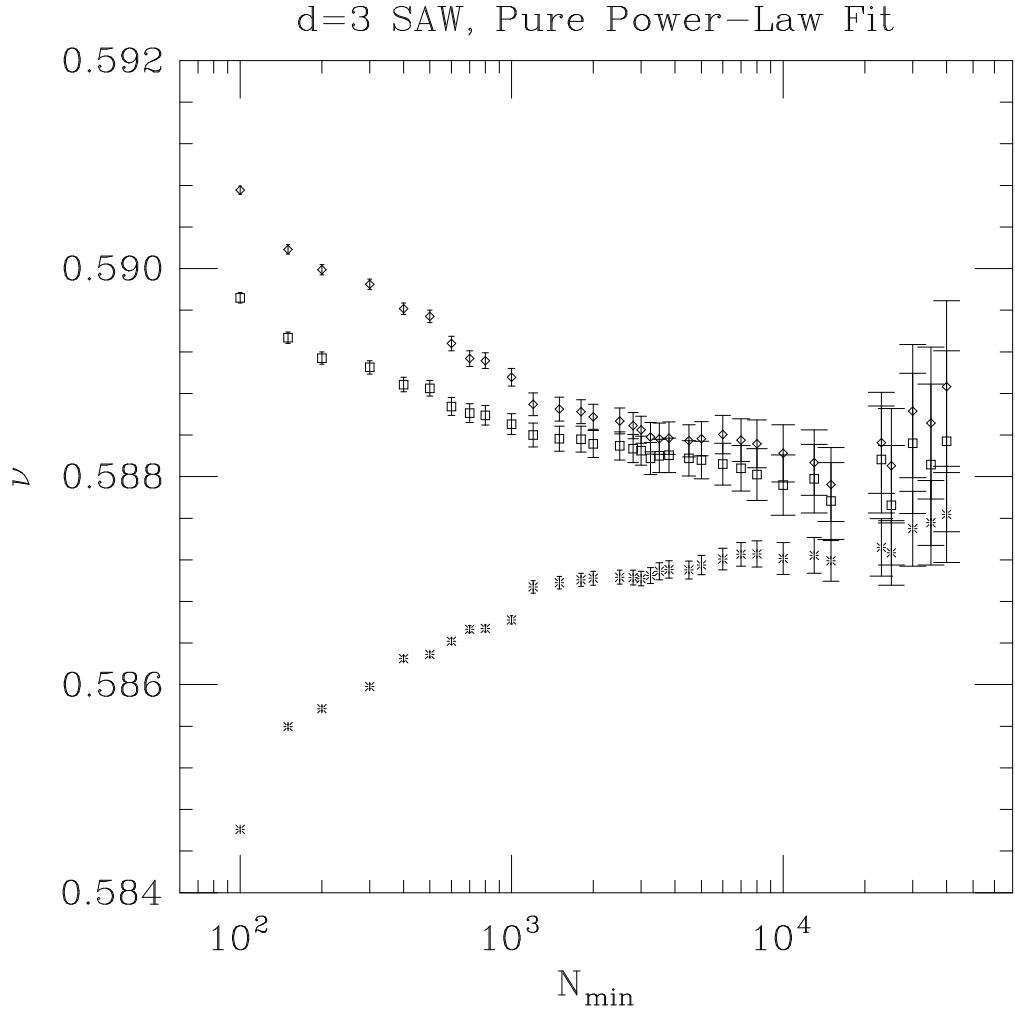

Figure 4: Estimated exponent $\nu$ from pure power-law fits to $\langle R_e \rangle^2$ ($\square$) and $\langle R_g \rangle^2$ ($\diamond$), and estimated exponent $(2\Delta_4 - \gamma)/3$ from pure power-law fit to $\langle T \rangle$ ($*$), plotted versus $N_{min}$. Note the very strong corrections to scaling, which lead to erroneous exponent estimates unless one takes $N_{min} \gtrsim 10^4$.

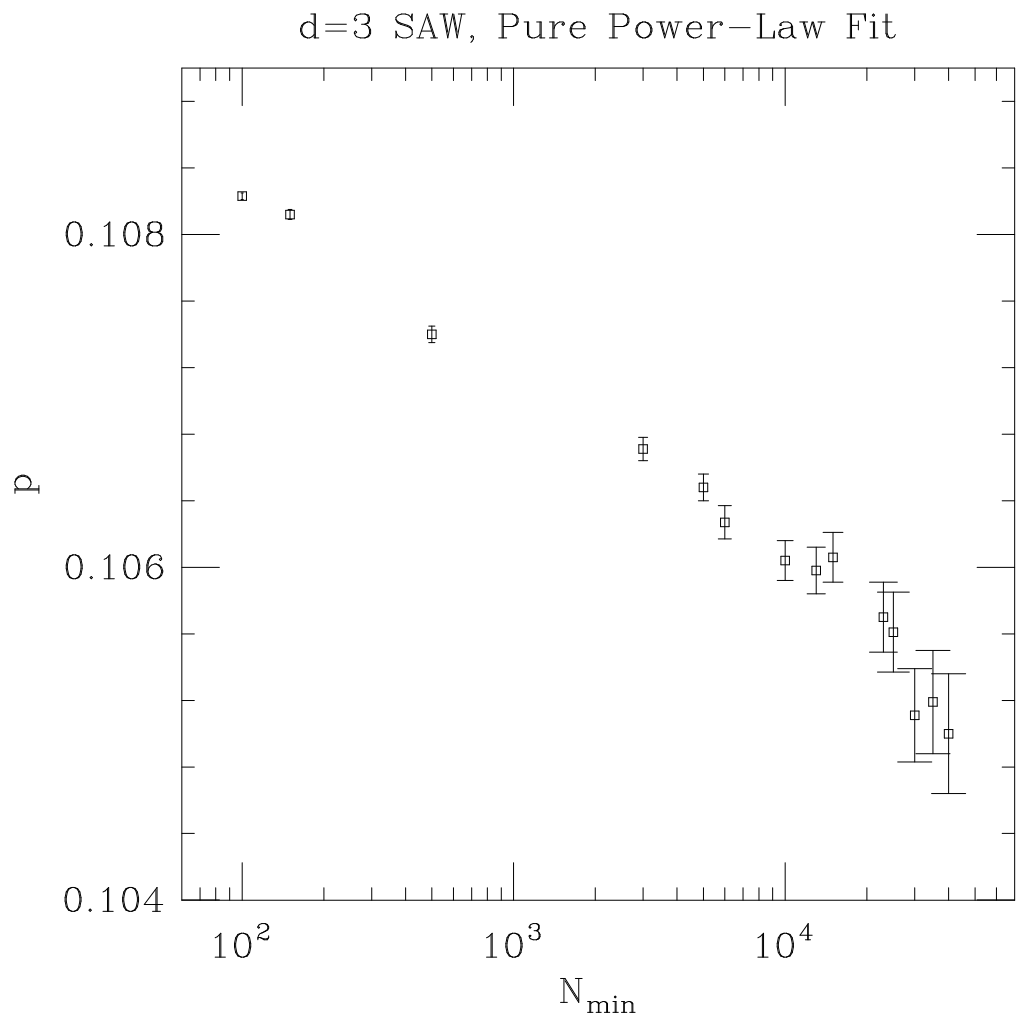

Figure 5: Estimated exponent $p$ from pure power-law fit to pivot-algorithm acceptance fraction $f$, plotted versus $N_{min}$.

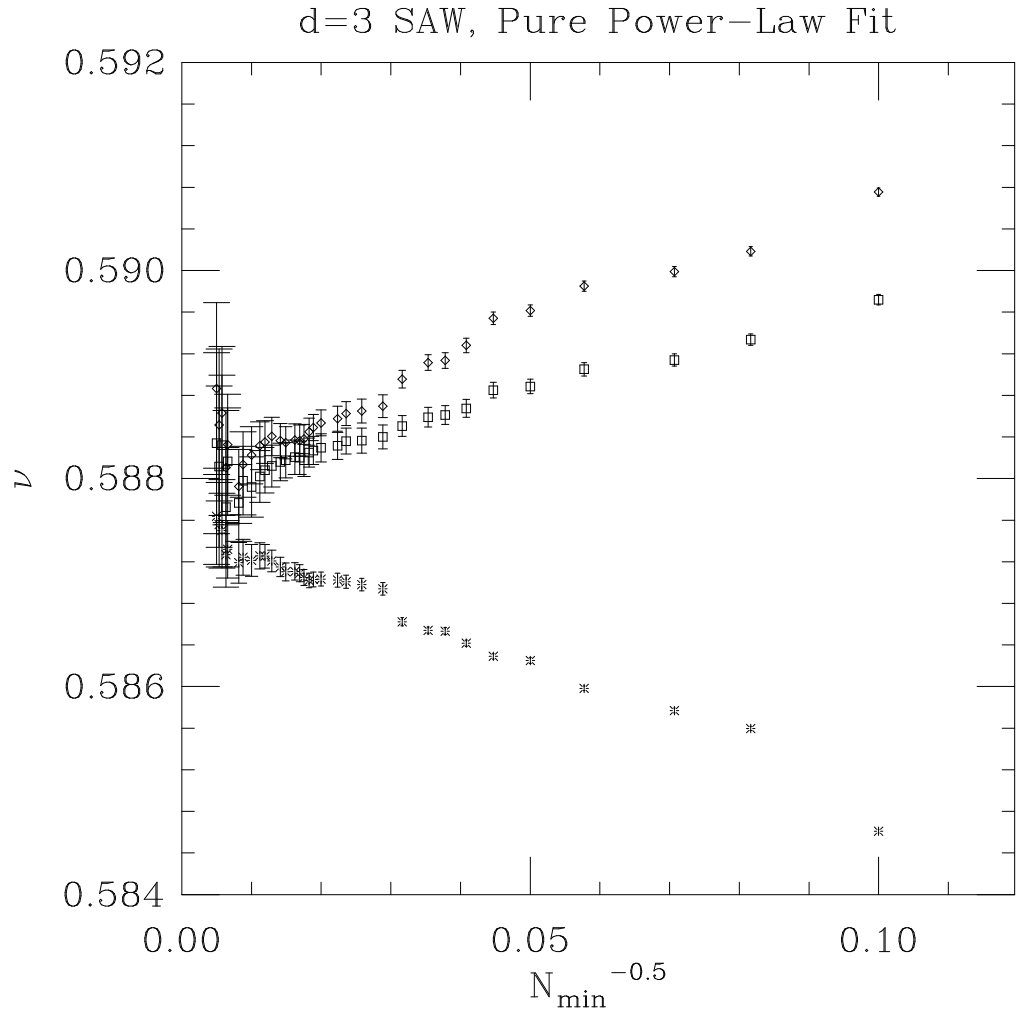

Figure 6: Estimated exponent $\nu$ from pure power-law fits to $\langle R_e \rangle^2$ ($\square$) and $\langle R_g \rangle^2$ ($\diamond$), and estimated exponent $(2\Delta_4 - \gamma)/3$ from pure power-law fit to $\langle T \rangle$ ($*$), plotted versus $N_{min}^{-0.5}$. Note the very roughly linear behavior, in agreement with the belief that $\Delta_1 \approx 0.5$.

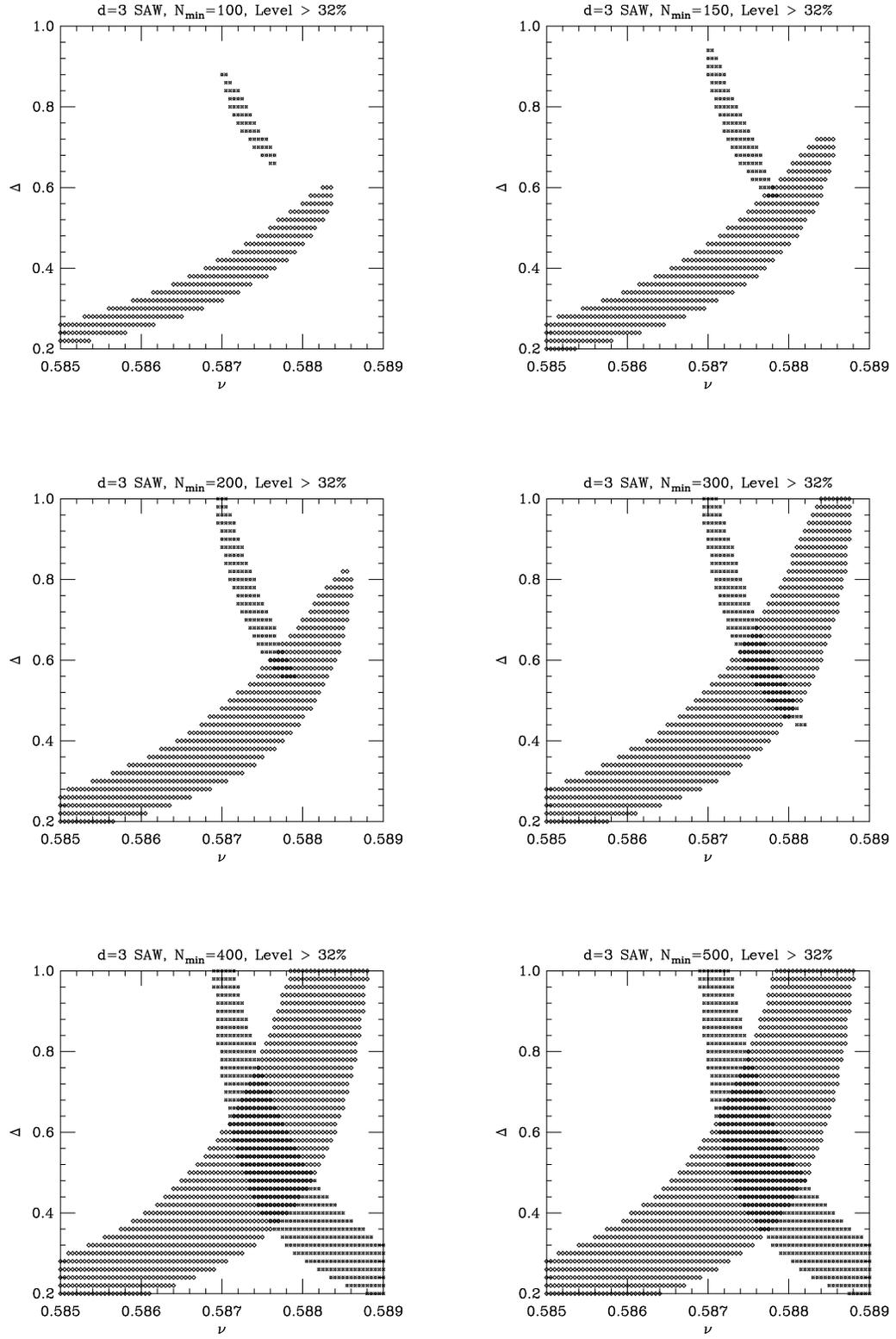

Figure 7: Pairs $(\nu, \Delta)$ for which the fit to the Ansatz $AN^{power} + BN^{power-\Delta}$ produces a $\chi^2$ acceptable at the 32% significance level (i.e. one standard deviation). Observables are $\langle R_g \rangle^2$ ($\Diamond$) and $\langle T \rangle$ ($*$).

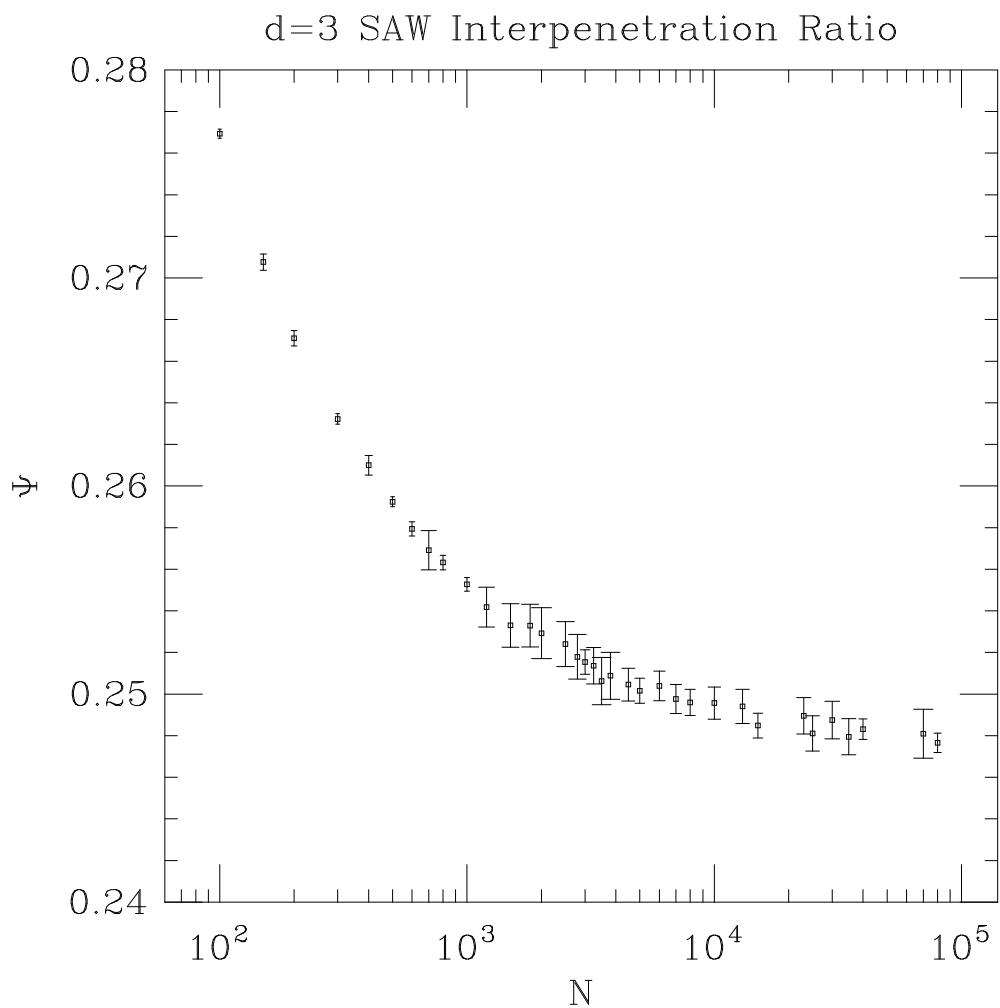

Figure 8: Interpenetration ratio $\Psi$ versus $N$, for 3-dimensional SAWs. Note that $\Psi$ is a decreasing and convex function of $N$, in flagrant disagreement with the prediction of the two-parameter renormalization-group theory.

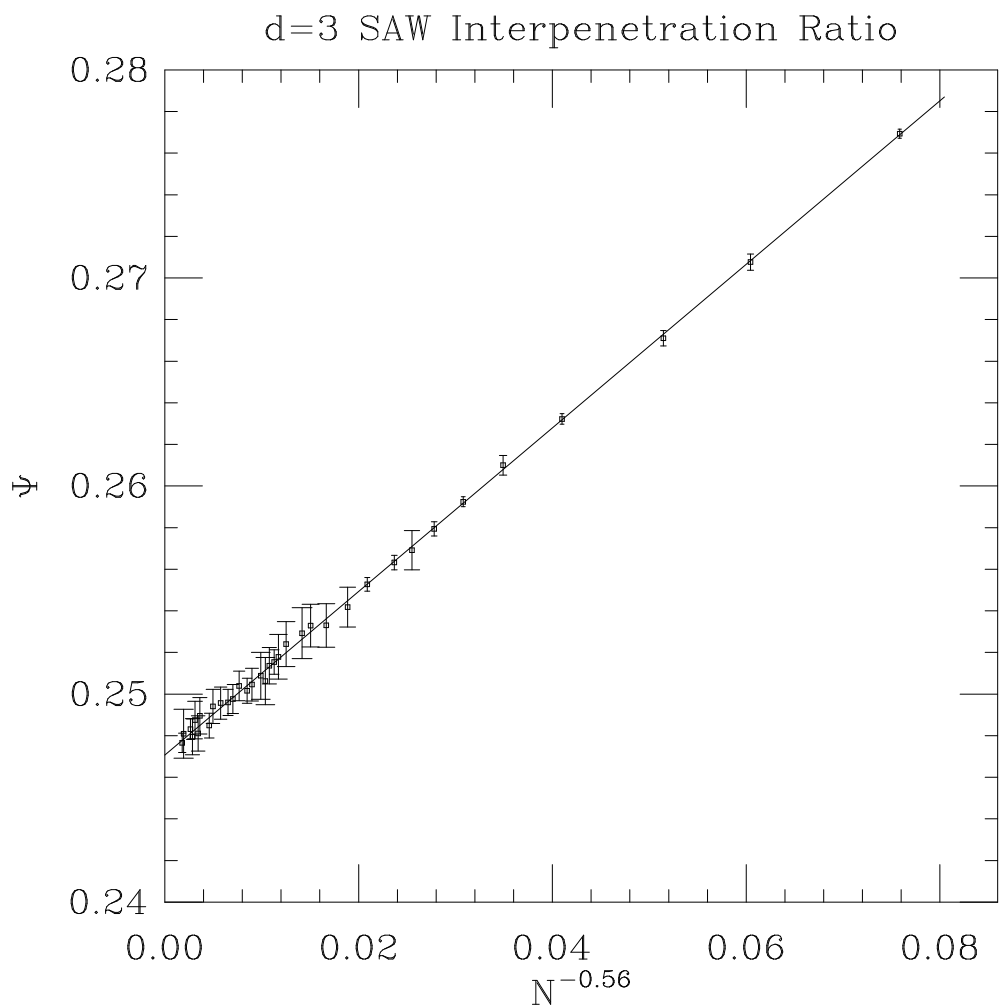

Figure 9: Interpenetration ratio $\Psi$ versus $N^{-0.56}$, for 3-dimensional SAWs. The regression line is $\Psi = 0.24707 + 0.39312/N^{0.56}$. Note the excellent linearity of the plot.

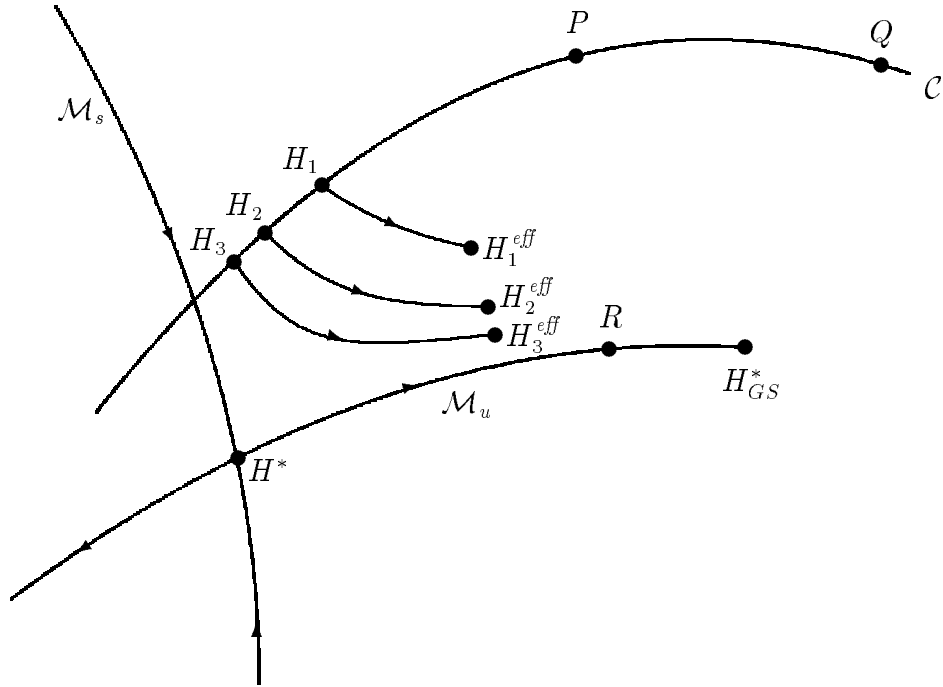

Figure 10: Wilson-de Gennes-type renormalization-group flow on the critical surface. $H^*$ (resp. $H^*_{GS}$) is the Gaussian (resp. good-solvent) fixed point. $\mathcal{M}_s$ (resp. $\mathcal{M}_u$) is the stable (resp. unstable) manifold of $H^*$. Case Ia: Models in the good-solvent regime may have correction-to-scaling amplitudes that are either negative $(P, R)$ or positive $(Q)$. Case II: The initial Hamiltonians $H_n$ approach the stable manifold, while the low-energy effective Hamiltonians $H_n^{eff} \equiv \mathcal{R}^n H_n$ approach the unstable manifold.

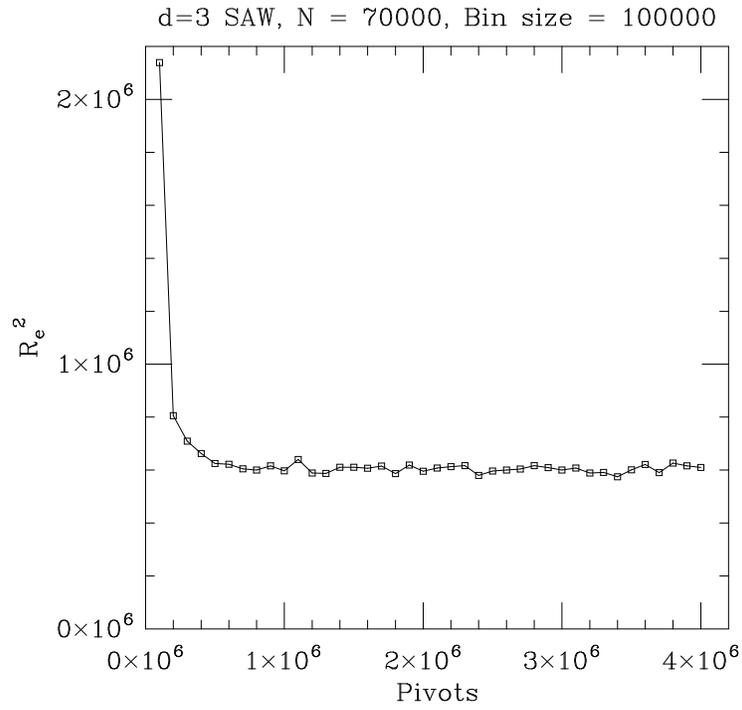

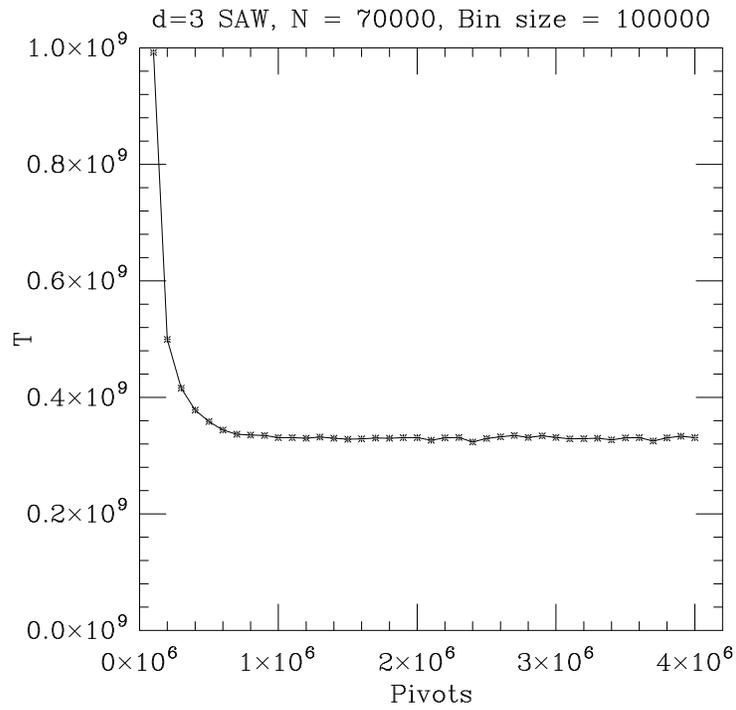

Figure 11: Averages of (a) $R_e^2$ and (b) $T$ over bins of width $10^5$ iterations, for pivot algorithm on simple cubic lattice at $N = 70000$ with parallel-rod start.

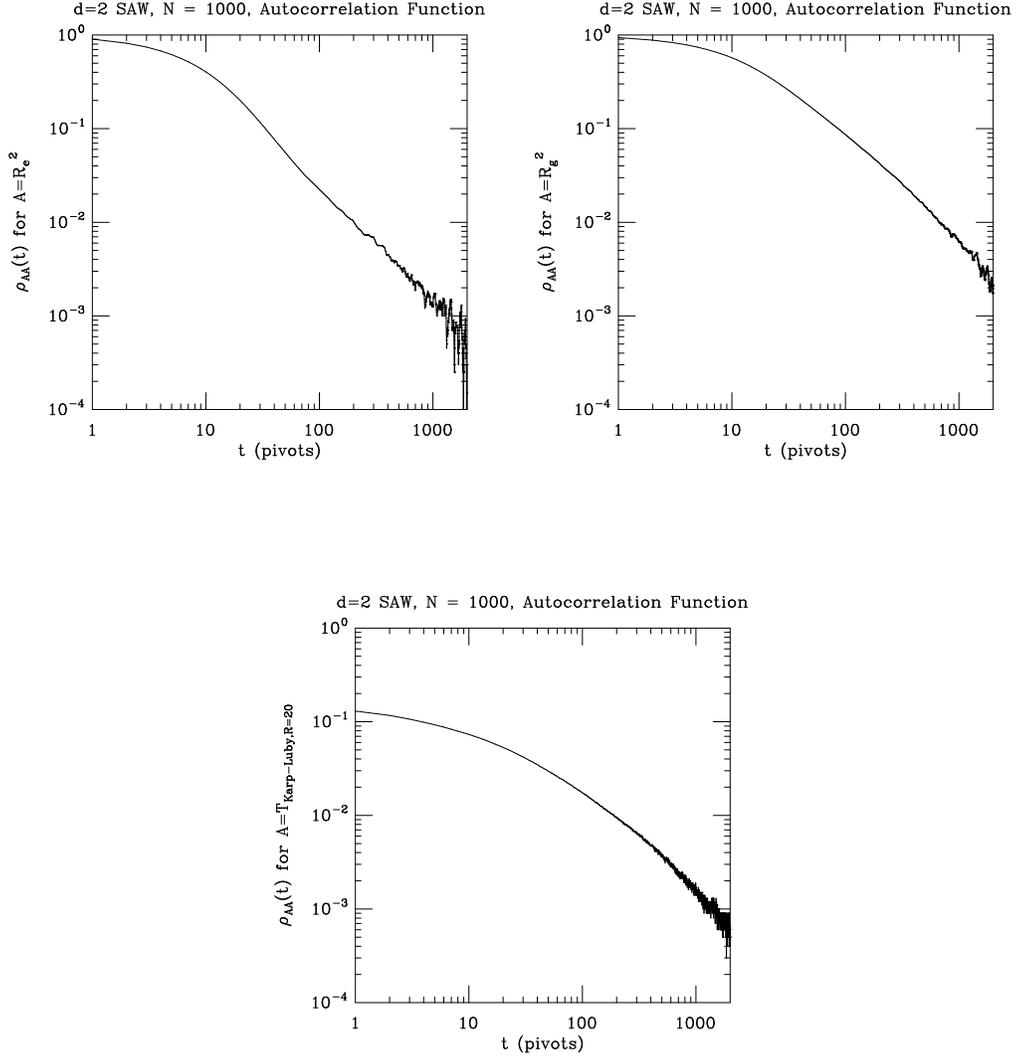

Figure 12: Log-log plot of the sample autocorrelation function $\hat{\bar{\rho}}_{AA}(t)$ for (a) $A = R_e^2$, (b) $A = R_g^2$ and (c) $A = T_{Karp-Luby, R=20}$ for pivot algorithm on square lattice at $N = 1000$.